\documentclass{article}

\usepackage{PRIMEarxiv}

\usepackage[utf8]{inputenc} %
\usepackage[T1]{fontenc}    %
\usepackage{hyperref}       %
\usepackage{url}            %
\usepackage{booktabs}       %
\usepackage{amsfonts}       %
\usepackage{nicefrac}       %
\usepackage{microtype}      %
\usepackage{lipsum}
\usepackage{fancyhdr}       %
\usepackage{graphicx}       %
\graphicspath{{media/}}     %
\usepackage{natbib}
\usepackage{amsmath}
\usepackage{titletoc} %
\usepackage{hyperref} 
\usepackage[edges]{forest}   %
\usepackage{listings}
\usepackage{xcolor}
\usepackage{amssymb}

\usepackage{booktabs}
\usepackage{makecell}

\title{BOCoDe: Engineering-Centered Benchmarking for Bayesian Optimization
\thanks{Preprint. Under Review.} 
}

\author{
  Rosen Ting-Ying Yu \\
  Massachusetts Institute of Technology \\
   \texttt{rosenyu@mit.edu} \\
  \And
  Christophe Hatterer \\
  ETH Z\"urich \\
   \texttt{christophe@hatterer.net} \\
  \And
  Advaith Narayanan \\
  Carnegie Mellon University \\
  \texttt{advaithn@andrew.cmu.edu} \\
  \And
  Cyril Picard \\
  Massachusetts Institute of Technology \\
  \texttt{cyrilp@mit.edu} \\
  \And
  Faez Ahmed \\
  Massachusetts Institute of Technology \\
  \texttt{faez@mit.edu} \\
}

\begin{document}
\maketitle

\begin{abstract}
Bayesian optimization (BO) is a sample-efficient, surrogate-based approach to black-box optimization (BBO), but its evaluation remains dominated by synthetic functions and hyperparameter optimization (HPO) tasks that are typically low-dimensional and single-objective. Engineering design poses a substantially different regime: problems are physics-based, often high-dimensional, constrained by requirements such as cost and manufacturability, and may involve multiple objectives or mixed variables. To close this benchmarking gap, we introduce BOCoDe, an open-source, PyTorch-native benchmark comprising 307 BBO problems, including 159 engineering design tasks and widely used synthetic and HPO benchmarks. Each problem includes cited provenance and machine-readable metadata that supports programmatic discovery, including by LLM-based agents, and all tasks are exposed through a unified API compatible with open-source BO libraries. We evaluate 31 BO and evolutionary algorithms across five problem classes spanning single- and multi-objective optimization, constrained and unconstrained settings, and mixed-variable search spaces. Analyses of problem structure show that engineering tasks uniquely span constrained and multi-objective settings that synthetic and HPO suites rarely cover, while embeddings from a tabular foundation model separate them most clearly from HPO tasks. Algorithm rankings also vary substantially across domains; in several problem classes, rankings obtained on standard benchmarks do not transfer to engineering tasks. BOCoDe establishes a reproducible and extensible foundation for developing and evaluating BO methods that better reflect the demands of engineering design. Code \& data can be found at \url{https://github.com/rosenyu304/BOCoDe}
\end{abstract}

\section{Introduction}
Bayesian optimization (BO) is a popular surrogate-based approach for black-box optimization (BBO), where the objective function or the constraints defining the feasible set are unknown. To model the search space, BO fits a probabilistic surrogate, usually a Gaussian process (GP), with increasing recent interest in using random forests (RFs) and tabular foundation models (TFMs) \citep{snoek2012practical,SMAC32022,muller2023pfns4bo}. Next, BO uses an acquisition function to choose the next point to evaluate, making it sample-efficient. However, the field develops BO methods yet evaluates them almost entirely on two families: synthetic analytical functions and hyperparameter optimization (HPO) tasks that are mostly single-objective (SO), rather than on the applied scientific and engineering design problems where BO is increasingly deployed \citep{liang2021benchmarking,paulson2025bayesian}.

\begin{figure}[t!]
    \centering
    \includegraphics[width=.8\columnwidth]{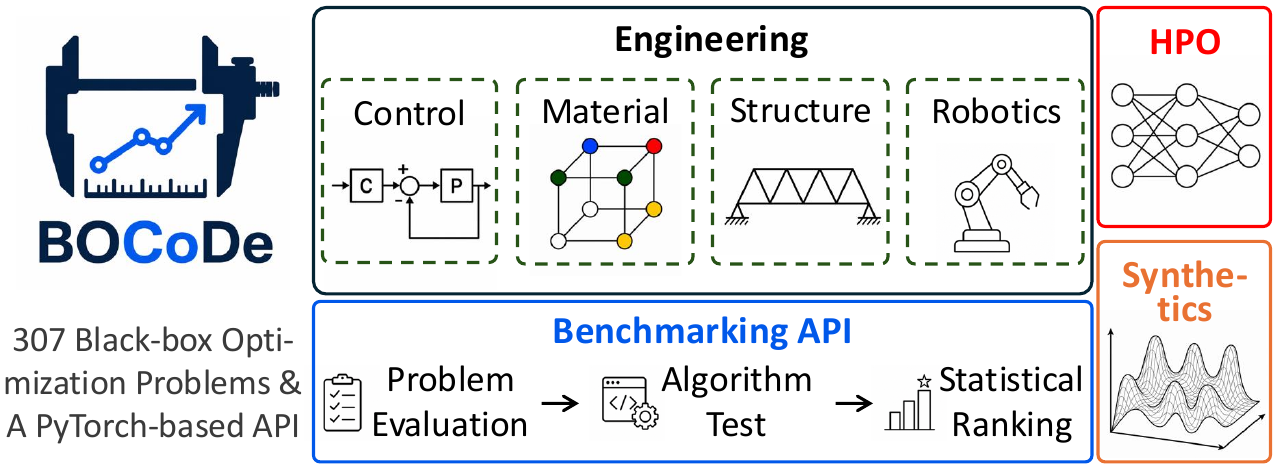}
    \caption{Overview of BOCoDe, a benchmark suite with 307 BBO problems featuring 159 engineering tasks, which aims to provide an efficient framework for solving real-world engineering design optimization problems.}
    \label{fig:Figure1}
\end{figure}

Engineering design problems occupy a different regime: typically multi-objective (MO), constrained, and high-dimensional \citep{tian2024boundary,Gourav2026MOConstrained}. These problems span one to over a thousand dimensions, up to hundreds of constraints, and up to nine objectives (Figure~\ref{fig:metadata}). Furthermore, algorithm performance could be limited by, or depend on, problem structure (e.g., SO algorithms cannot be directly deployed on MO problems). Past studies on a smaller scale show that methods that win on synthetic benchmarks do not always win on ``real'' problems (tasks reflecting real-world applications) \citep{pmlr-v133-turner21a,leriche2021revisiting}. To our knowledge, there has yet to be a study on BO's performance transfer at scale across various application domains, so method selection for physical and engineering settings rests on an untested assumption.

In fact, benchmarking BO algorithms is by itself a difficult task. Engineering, synthetic, and HPO benchmarks are scattered across incompatible codebases with conflicting dependencies and no common interface, so it is hard to compare results across papers \citep{dreczkowski2023mcbo,papenmeier2025bencher}. Fair comparison also demands aggregating performance across many runs, problems, and domains under a consistent, statistically grounded ranking protocol---something existing suites have almost never standardized. We close both gaps with \textbf{BOCoDe}, a PyTorch-native benchmark that provides 307 black-box problems, including a set of 159 engineering design problems. BOCoDe also comes with 31 reference BO and evolutionary algorithms and a statistical ranking protocol, all within a single interface, with machine-readable metadata that makes every problem programmatically selectable. Our contributions are:

\begin{enumerate}
    \item \textbf{A benchmark suite centered on engineering design problems.} We curate 159 engineering optimization problems for BOCoDe, alongside 68 synthetic and 80 HPO reference problems; these problems are constrained, multi-objective, mixed-variable, and high-dimensional far more often than those in existing suites. A structural analysis shows these problems uniquely cover the constrained, multi-objective, and mixed-variable settings existing suites omit, and a foundation-model embedding separates them from the HPO tasks that dominate BO benchmarks.
    
    \item \textbf{A unified code interface for evaluation, optimization, and ranking.} Engineering benchmarks are scattered across incompatible conventions---minimization versus maximization, and separate handling of constraints, multiple objectives, and mixed variables---so results are hard to compare across papers. BOCoDe places problems, optimizers, and a cross-domain ranking protocol behind one Python interface, and every problem has machine-readable metadata so it can be selected and run programmatically, including by LLM agents.

    \item \textbf{The first systematic evidence that BO is under-benchmarked on engineering design.} Running 31 methods across five optimization classes under a statistically grounded protocol, we find the best-ranked method changes across the engineering, synthetic, and HPO domains, so rankings from synthetic proxies do not transfer. To our knowledge, this is the first study at this scale, and it is only a starting point: closing the gap will take more problems and more algorithms, which we build BOCoDe to keep adding.

\end{enumerate}

\section{Related Work}

\subsection{Black-Box Optimization Benchmark Suites}
BBO suites exist across analytical optimization, HPO in automated machine learning (AutoML), and engineering design. Each domain has produced benchmark suites tailored to its own research gaps.

\subsubsection{Bayesian optimization benchmarks.} BO is typically benchmarked using general-purpose libraries. COCO/BBOB \citep{hansen2021coco} standardizes synthetic functions but includes no applied tasks, and BoTorch/Ax \citep{balandat2020botorch,olson2025ax} are extensible PyTorch-native frameworks whose test-problem suites are dominated by synthetic tasks. The remaining efforts are application-driven: Design-Bench \citep{trabucco2022designbench} offers only eight small-scale design tasks and provides no Bayesian optimization loop; Olympus \citep{hase2021olympus} is limited to chemistry and materials experiment planning; and BO4Mob \citep{ryu2026bomob} is a BO benchmark built around a high-dimensional urban-mobility estimation problem problem. AutoML HPO suites \citep{turner2019bayesian,eggensperger2021hpobench} advance BO for ML tuning and architecture search, yet they remain single-objective and rarely constrained. Bencher \citep{papenmeier2025bencher} instead gathers real-world problems but isolates each in its own remote environment rather than a shared, BO-native representation. However, none of them features constrained, MO, and mixed-variable problems across application domains.

\subsubsection{Evolutionary platforms.} Pymoo \citep{blank2020pymoo} and PlatEMO \citep{tian2017platemo} provide many constrained and multi-objective design problems from the evolutionary algorithm (EA) literature. However, their minimization convention and their NumPy/MATLAB, population-based interfaces do not map directly onto BO's sequential, GPU-accelerated surrogate loop without non-trivial reimplementation.

\subsection{Computational engineering design problems}
Engineering design optimization has historically been driven by evolutionary methods, so many of its standard benchmark problems originate in the EA literature, including the CEC constrained suites \citep{kumar2020cec}, the RE/CRE problems \citep{tanabe2020re}, and MODAct \citep{picard2020modact}. BO is increasingly applied to engineering problems as design evaluations and simulations become more expensive, for example in materials and experimental design \citep{liang2021benchmarking,paulson2025bayesian}, yet the problems remain isolated by domain, tied to specific solvers, and rarely exposed under a BO-native API with matched baselines. %

A further obstacle is that the same problem often appears in multiple incompatible versions, differing in search-space bounds, dimensionality, or scaling, and is frequently released without a citation to the publication or code from which it originates \citep{NEURIPS2024_fe0007fc}. This leaves users unable to identify what a problem represents or reproduce it faithfully. Thus, we propose BOCoDe, a living suite that gathers constrained, multi-objective, and mixed-variable engineering problems under a single BO-native API with reference implementations for each class.

\section{The BOCoDe Benchmark}

In the following, we detail the three main modules in BOCoDe: the optimization problem library, the algorithm suite, and the statistical ranking API.

\subsection{BOCoDe Optimization Problem Library}

\subsubsection{Optimization Definition and API.}
Every BOCoDe problem is written in a single canonical form. For a design variable $x \in \mathbb{R}^{D}$ within box bounds $x \in [x_{\text{min}},x_{\text{max}}]^D$,
\begin{equation*}
\max_{x}\; f_m(x),\quad m \in [1,M] \quad \text{s.t.}\quad g_i(x) \leq 0,\quad i \in [1,G],
\end{equation*}
with $D$ variables (dimension), $M$ objectives, and $G$ inequality constraints with feasibility as $g(x)\leq 0$. All objectives are maximized by the algorithm, following BO algorithm convention, but can be negated for minimization. We categorize the optimization problems into five classes based on the problem formulation and search space characteristics: SO and MO, unconstrained and constrained, continuous and mixed-variable problems, as shown in Table~\ref{tab:classes}. 

\begin{table}[t]
\centering
\small
\caption{Optimization problem class definition.}
\label{tab:classes}
\begin{tabular}{@{}p{0.3\linewidth}ccc@{}}
\toprule
\textbf{Class Name} & \textbf{Objective} & \textbf{Constrained} & \textbf{Continuity} \\
\midrule
SO-Unconstrained & 1 & No & Continuous\\
MO-Unconstrained & $>$1 & No & Continuous\\
SO-Constrained & 1 & Yes & Continuous\\
MO-Constrained & $>$1 & Yes & Continuous\\
SO-Mixed-Variable & 1 & No & Mixed-variable\\
\bottomrule
\end{tabular}
\end{table}

Every BOCoDe problem is a Python object with a single \texttt{evaluate(x)} method, following BoTorch/Ax test-problem style. The method takes a design variable $x$ as a PyTorch tensor and returns objective and constraint values in maximization form. This allows all optimization problems to be called through the same API (see below): 
\vspace{5em}
\begin{lstlisting}[language=Python, basicstyle={\small\ttfamily}]
import bocode
import torch
# Instantiate a benchmark problem
problem = bocode.Car()
# Read metadata
metadata = bocode.get_metadata("Car")
# Evaluate five LHS points
x = problem.sample(5, seed=0)
obj, cons = problem.evaluate(x)
\end{lstlisting}
A separate JSON metadata file records each problem's dimension, objectives, constraints, bounds, variable types, and source for programmatic selection. This structured, self-describing metadata lets both users and automated LLM agents discover, filter, and instantiate problems programmatically, supplying the type of documented schema that reliable agentic use depends on \citep{purucker2026iidgeneraltabularfoundation}. The source lives in both the metadata and the docstring, making every problem traceable to its original paper or code. 

\subsubsection{Optimization Problem Domain Analysis.} %

BOCoDe's benchmark campaign spans 307 problems across \textbf{three domain categories}: (i) \textbf{Synthetic}: $n\!=\!68$ analytical test functions from BoTorch \citep{balandat2020botorch}, (ii) \textbf{Engineering}: $n\!=\!159$ structural, mechanical, and control optimization problems drawn from the BO literature, from computational engineering design research, or from real-world engineering applications verified by three engineering design domain experts \citep{todorov2012mujoco,wang2018robotpush,kumar2020cec,tanabe2020re,picard2020modact,norheim2022mathematical,yu2025fast}, and (iii) \textbf{HPO}: $n\!=\!80$ common HPO baselines from the AutoML literature \cite{sehic2022lassobench,turner2019bayesian,eggensperger2021hpobench}. This current set of problems is a starting collection from open-source repositoriesitories rather than a complete census, and we continue to expand it.

Below, we analyze how a problem's application domain shapes its optimization characteristics from two angles: the distribution of optimization problem structural characteristics within each domain and how their problems cluster in a learned behavioral embedding. 

\begin{figure}[ht!]
\centering
\begin{minipage}[t]{0.48\textwidth}
    \centering
    \includegraphics[width=\linewidth]{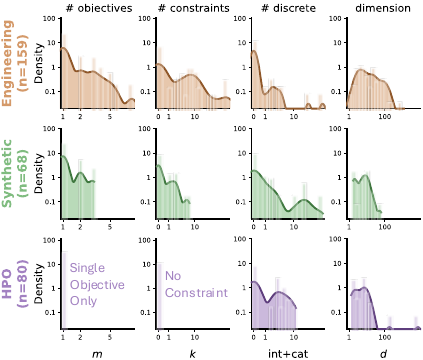}
    \caption{Distribution of problem characteristics across BOCoDe's domains. Columns: four characteristics: objectives, inequality constraints, discrete or categorical variables, and dimension. Rows: distributions of BOCoDe's three problem domains (Engineering, Synthetic, HPO).}
    \label{fig:metadata}
\end{minipage}\hfill
\begin{minipage}[t]{0.48\textwidth}
    \centering
    \includegraphics[width=\linewidth]{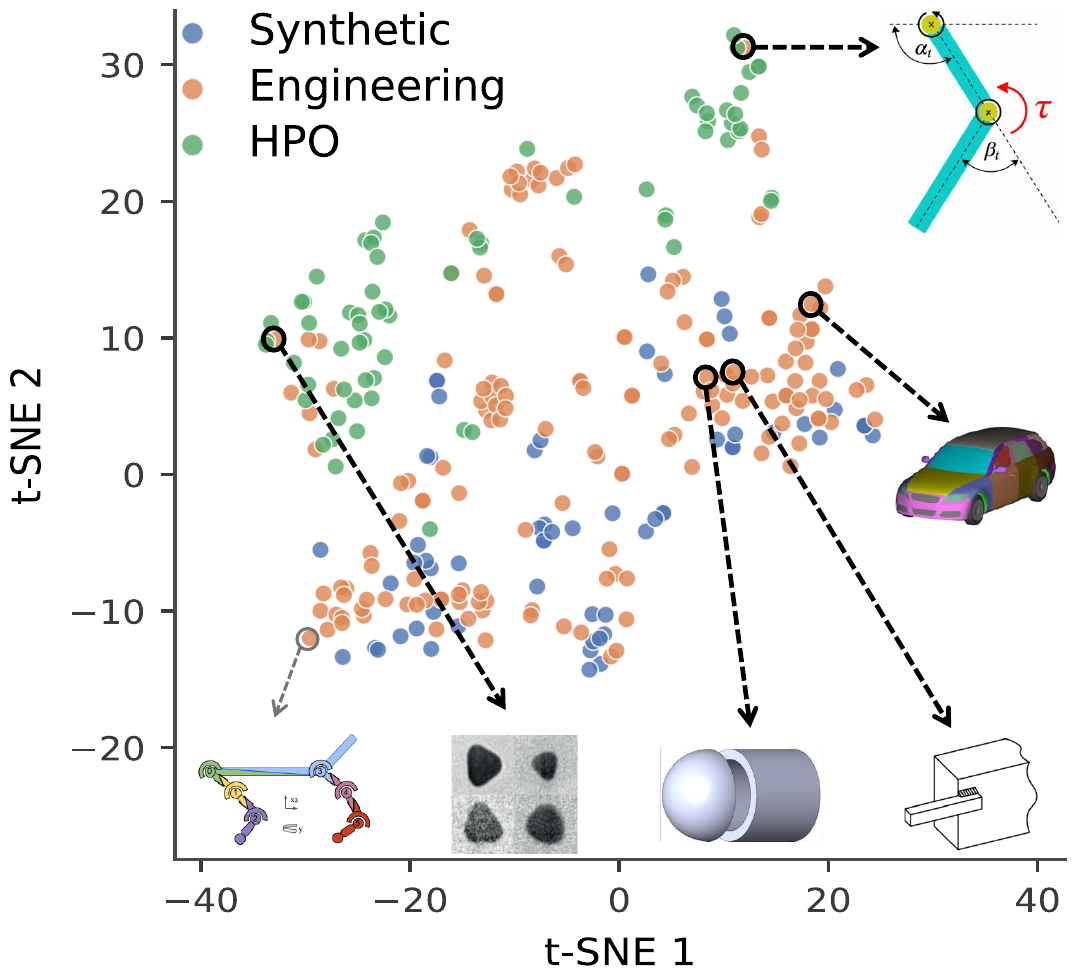}
    \caption{Two-dimensional t-SNE projection of per-problem TabPFN (v3) embeddings, colored by domain. Each point in the plot is one problem, represented by a 512-dimensional vector from TabPFN's per-problem embeddings. Nearby points have similar embeddings.}
    \label{fig:tsne}
\end{minipage}
\end{figure}

\begin{enumerate}
    \item \textbf{Optimization Problem Structural Characteristics.} Figure~\ref{fig:metadata} shows how objectives, constraints, discrete variables, and dimensions are distributed across domains. HPO problems are single-objective, and synthetic problems have fewer than five objectives. Engineering problems alone populate the constrained, MO, and mixed-variable regions and span one to over a thousand dimensions. This makes engineering tasks a broad arena for benchmarking BO across every optimization regime the field targets.
    \item \textbf{Optimization Problem Embedding.} Matching characteristic counts does not mean two problems behave alike, so we add a behavioral view. Beyond serving as a BO surrogate \citep{muller2023pfns4bo}, TabPFN produces per-point embeddings useful for clustering and visualization \citep{hollmann2025tabpfn}. Following \citep{regenwetter2026engineering}, we Latin hypercube sample (LHS) 500 points per problem, average their embeddings into one vector, and lay out all problems with t-SNE \citep{vandermaaten2008tsne}. Figure~\ref{fig:tsne} shows engineering problems in regions largely disjoint from the HPO clusters that most BO benchmarks sample, but overlap partially with synthetic clusters. In our Technical Supplement, we further demonstrate that a classifier trained on the per-problem embeddings distinguishes the domains, with the HPO problems the most separable.
\end{enumerate}

These domain differences motivate our main experiment, in which we benchmark all 31 BBO algorithms in our suite on every BOCoDe problem. In the Results section, we show that each domain favors a different algorithm, so a ranking established in one domain might not carry over to another. Per-problem information, the software code structure, and APIs are detailed in the Technical Supplement. 

\subsection{BOCoDe Algorithm Suite}

BOCoDe provides 31 BBO algorithm interface implementations: 27 BO algorithms implemented in a BoTorch-compatible interface \citep{balandat2020botorch} and 4 EA methods featured in Pymoo \citep{blank2020pymoo} that are commonly used as engineering BBO benchmarking baselines \cite{picard2020modact}, which together form an initial, non-comprehensive set for BOCoDe's collection. Table~\ref{tab:baselines} provides the full list. The algorithms are grouped by BOCoDe's five optimization problem classes, and every algorithm reads and returns the same tensors as the problems, so any method runs on any problem in its class through the single BOCoDe API:
\vspace{0.5em}
\begin{lstlisting}[language=Python, basicstyle={\small\ttfamily}]
python -m algorithms.single_obj.gp_ucb \
       --problem Car --init 10 --iters 40
python -m algorithms.multi_obj.qnehvi \
       --problem RE21 --init 10 --iters 50
\end{lstlisting}

BOCoDe also makes the surrogate a swappable component. Following the standard decomposition of BO into a surrogate model, an acquisition function, and an acquisition optimizer \citep{dreczkowski2023mcbo, balandat2020botorch}, BOCoDe fixes the acquisition and optimizer and substitutes the surrogate, supporting GP, RF, and two latest TFMs, TabPFN(v3) \citep{grinsztajn2026tabpfn3technicalreport} and TabICL(v2) \citep{qu2026tabiclv2}, due to the growing interest in TFM-based BO \citep{muller2023pfns4bo,rakotoarison2024context,yu2025gitbo}. In this study, we restrict the swap to SO problems because published TFM surrogates have so far been demonstrated only in SO BO settings; MO TFM-based BO remains an open problem. A surrogate ablation is shown in the Results \& Discussion section.

\begin{table*}[htbp!]
\centering
\small
\caption{31 Bayesian and evolutionary optimization (``Opt.'') algorithms grouped by optimization classes. This is a representative starting set rather than an exhaustive one; more methods can be added through the same API in the future.}
\label{tab:baselines}
\begin{tabular}{@{}p{0.17\textwidth}c p{0.75\textwidth}@{}}
\toprule
\textbf{Opt. Classes} & \textbf{\#} & \textbf{Opt. Algorithms} \\
\midrule
SO-Unconstrained & 9 &
\textbf{GP-UCB} \citep{srinivas2010ucb}, \textbf{GP-LogEI} \citep{ament2023logei}, \textbf{\{GP,RF,TabPFN,TabICL\}-TuRBO} \citep{eriksson2019turbo}, \textbf{BAxUS} \citep{papenmeier2022baxus}, \textbf{Vanilla-HD-BO} \citep{hvarfner2024vanilla}, \textbf{Standard-GP} \citep{xu2025standard}. \\
\addlinespace
SO-Constrained & 7 &
\textbf{CEI} \citep{gardner2014constrainedei}, \textbf{\{GP,RF,TabPFN,TabICL\}-SCBO} \citep{eriksson2021scbo}, \textbf{Penalty} \citep{Fletcher1975Penalty}, \textbf{CLF-CBO} (classifier CBO) \citep{tian2024boundary}\\
\addlinespace
MO-Unconstrained & 6 & BO methods: \textbf{qNEHVI} \citep{daulton2020qnehvi}, \textbf{qNParEGO} \citep{daulton2021qnparego}, \textbf{MESMO} \citep{Belakaria2019MESMO}, \textbf{DGEMO} \citep{Lukovic2020Diversity}. EA methods: \textbf{NSGA-II} \citep{Deb2002NSGA2}, \textbf{SPEA2} \citep{zitzler2001spea2}. \\
\addlinespace
MO-Constrained & 4 &
BO methods: \textbf{C-qNEHVI} \citep{daulton2020qnehvi}, \textbf{C-qNParEGO} \citep{daulton2021qnparego}. EA methods: \textbf{C-NSGA-II} \citep{Deb2002NSGA2,blank2020pymoo}, \textbf{C-SPEA2} \citep{zitzler2001spea2,blank2020pymoo}. \\
\addlinespace
SO-Mixed-Variable  & 5 &
\textbf{PR} \citep{daulton2022bayesian}, \textbf{Bounce} \citep{papenmeier2023bounce}, \textbf{Casmopolitan} \citep{wan2021casmopolitan}, \textbf{HEBO} \citep{cowenrivers2022hebo}, \textbf{BODi} \citep{pmlr-v206-deshwal23a}. \\
\bottomrule
\end{tabular}
\end{table*}

\subsection{BOCoDe Statistical Ranking Protocol}

Comparing algorithms across hundreds of problems raises two difficulties: each problem produces objective values of vastly different magnitudes, and each run is subject to the stochasticity of the optimizer across seeds. BOCoDe therefore ranks methods statistically. We summarize each (problem, method) pair by the median best-so-far over its seeds. We report the median rather than the mean commonly used in the BO literature, because it is what the non-parametric ranking tests below are defined on, and the median represents a realizable engineering design \citep{CARRASCO2020100665}.

Within each problem, methods are ranked ($1\!=\!$best, ties averaged) and compared by mean rank with a Friedman omnibus test, an Iman-Davenport correction, and a Nemenyi post-hoc test drawn as a critical-difference (CD) diagram \citep{JMLR:v7:demsar06a,herbold2020autorank}. Because the Nemenyi approximation weakens on small sets, for classes with fewer than 20 problems, we additionally confirm each significant pair with a Holm-corrected Wilcoxon signed-rank test and read these CD values as indicative; omnibus statistics, corrected p-values, and effect sizes are in the Technical Supplement. For each test we use only the problems on which every compared method returned a valid result, and report the effective problem count $N$ per panel in the Technical Supplement.

\section{Experiments}

\subsection{Experimental Setup}

We evaluate the 31 algorithms in BOCoDe across the 307 benchmark problems under a fixed-iteration protocol \citep{hansen2022anytime}. Each method receives $n_\text{init}=D$ LHS-sampled initial points \citep{muller2023pfns4bo} followed by a fixed budget of $250$ black-box evaluations, repeated over $25$ seeds. All methods, BO and EA, share the same initial design and the same total evaluation budget per problem. For EA baselines these $250$ evaluations are spent across generations (population $\times$ generations $=250$), so every method issues an identical number of objective calls. Per-method population sizes, full call counts, and computing infrastructure details are in the Technical Supplement.

\subsection{Metrics \& Aggregation}

We match the performance metric with each optimization class: the best-so-far objective for SO problems, normalized per problem to $[0,1]$, and hypervolume to a fixed reference point for MO problems. For constrained problems, we calculate the best feasible objective (SO-Constrained) or hypervolume (MO-Constrained) value, and document the \textit{feasibility rate} of each method: the fraction of the 25 runs that reach a feasible point within the budget \citep{yu2025fast}.

All rankings and critical-difference diagrams (Nemenyi, $\alpha\!=\!0.05$) are produced by the BOCoDe Statistical Ranking Protocol; convergence plots report the median best-so-far with a 95\% confidence-interval band, negated to follow the engineering minimization convention. Per-problem convergence analyses are in the Technical Supplement.

\section{Results \& Discussion}

\begin{figure*}[ht!]
    \centering
    \includegraphics[width=.95\linewidth]{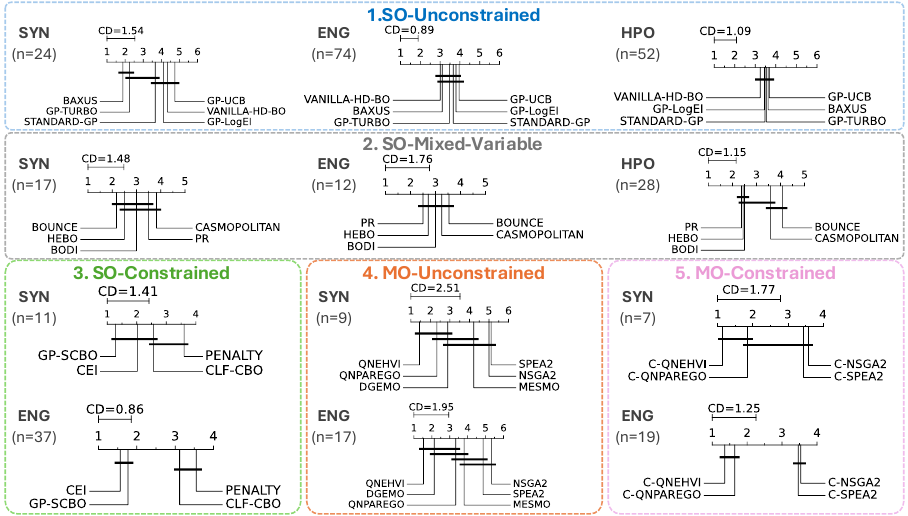}
    \caption{Statistical rankings at the final iteration, comparing GP-based methods by class and domain. Lower is better, and methods joined by a bar are statistically indistinguishable at the per-panel CD. A full per-surrogate analysis with non-GP variants is in the Technical Supplement.} %
    \label{fig:CD_Ranks}
\end{figure*}

\begin{figure*}[ht!]
    \centering
    \includegraphics[width=.95\linewidth]{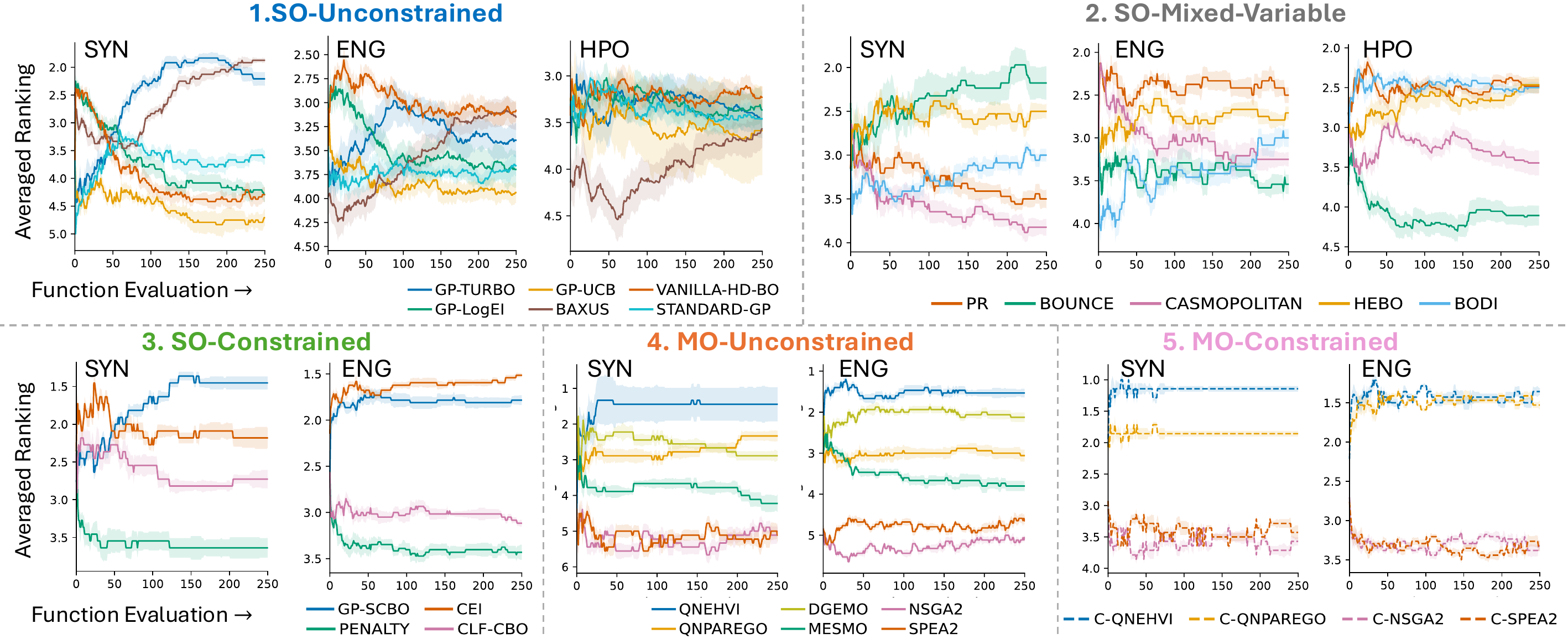}
    \caption{Rank trajectories over the 250-iteration budget, per class and domain. Lines show mean rank
and bands the 95\% CI.}
    \label{fig:rank_iterations}
\end{figure*}

We run all 31 methods on the 307 BOCoDe problems and organize the analysis around three questions: (i) whether the rankings transfer across domains (Figure~\ref{fig:CD_Ranks}), (ii) how the rankings change over the evaluation budget (Figure~\ref{fig:rank_iterations}), and (iii) whether the use of different surrogates for BO results in statistical changes in convergence behavior (Figure~\ref{fig:gp_vs_tfm}). Our main result shows that no single method dominates: for SO-Unconstrained and SO-Mixed-Variable problems, the ranking established on synthetic functions reverses on engineering problems, and even where the rank order is preserved, the set of statistically indistinguishable methods differs by domain.

\subsection{Rankings transfer unevenly across domains.} For SO-Unconstrained problems, BAxUS and GP-TuRBO lead on synthetic problems and are statistically tied there, while Vanilla-HD-BO climbs from near the bottom on synthetic problems to first on both engineering and HPO. The reshuffling, however, comes with a compression of the field: on the engineering and HPO panels, all six methods fall within a single critical difference (rank spread $0.82$ vs.\ CD $0.89$ on engineering), so no method separates significantly, whereas the synthetic panel does separate its leaders from the rest (the gap between BAxUS and Standard-GP exceeds the CD of $1.54$). SO-Mixed-Variable shows the same domain-dependence: Bounce leads on synthetic problems but is last on HPO, PR is first on engineering, and the HPO panel is led by a statistical three-way tie (PR, HEBO, BODi within $0.04$ rank). The constrained and MO classes are the stable exceptions: qNEHVI and C-qNEHVI lead every domain, and CEI and GP-SCBO swap the top spot on SO-Constrained within one CD. Yet, that stability is in the ranking only. Feasibility itself diverges by domain: every method almost always reaches feasibility on synthetic problems (0.94--1.00), but the feasibility rate drops to 0.51--0.91 across SO-Constrained methods on engineering ones (Table~\ref{tab:feasrate_combined_syneng}), so the synthetic ceiling hides constraint-handling gaps that engineering exposes.

\paragraph{Discussion.}This indicates that engineering problems are needed to evaluate BO methods for open and mixed-variable search spaces, where synthetic rankings do not carry over. Reporting only the top method also understates the result, since which methods are statistically tied is itself domain-dependent. Part of the cross-domain difference may reflect imbalanced problem counts: some classes contribute many engineering problems but far fewer synthetic ones, so the wider CD on the smaller sets can mask real gaps. Expanding the underrepresented domains is therefore needed in future work to settle borderline comparisons.

\subsection{Rankings change over the iteration budget.} The rank trajectories in Figure~\ref{fig:rank_iterations} show that this domain-dependence is also a function of the budget. Method orderings are not fixed within a run: trust-region (TR) methods such as GP-TuRBO and BAxUS start mid-pack and climb as the budget grows, while the standard GP acquisitions lead early and are overtaken, and the confidence bands separate only in later iterations. The budget at which a method leads, and how sharply the domains distinguish methods, therefore shift over the run.

\paragraph{Discussion. }Methods that favor early exploration and methods that favor late trust-region refinement appear complementary, and when each regime wins across the engineering, synthetic, and HPO domains is an open question, with budget-aware or switching strategies a natural response.

\begin{figure*}[htbp!]
    \centering
    \includegraphics[width=\linewidth]{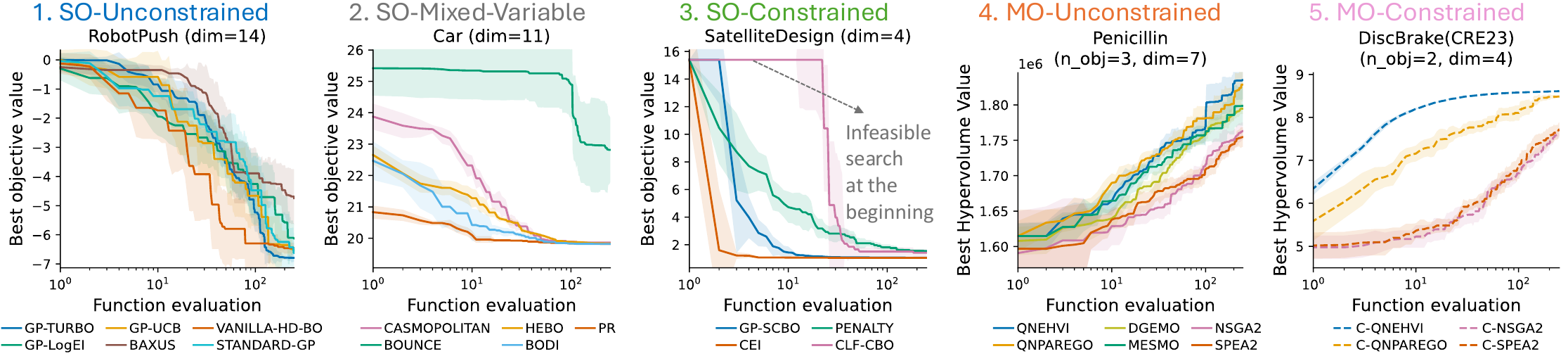}
    \caption{Example convergence for one engineering problem in each class; constrained panels show only feasible values. Full results for all 307 problems appear in Technical Supplement.}
    \label{fig:convergence_example}
\end{figure*}

\subsection{Surrogate choice changes convergence behavior.} 
\begin{figure}[htbp!]
\centering
\includegraphics[width=.5\columnwidth]{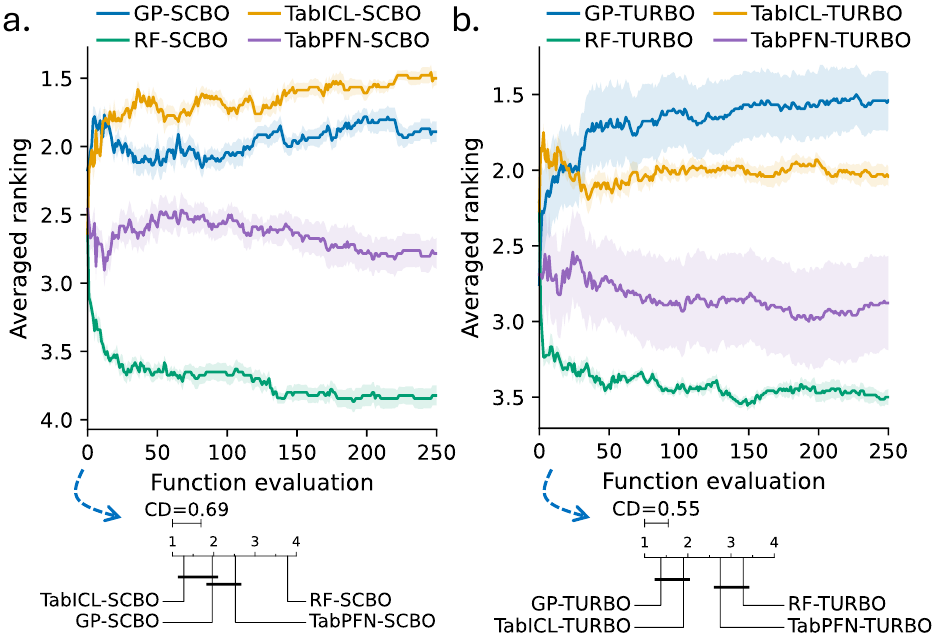}
\caption{Effect of the surrogate under (a) SCBO and (b) TuRBO on engineering problems, holding acquisition and optimizer fixed. Averaged rank over evaluations with final CD diagrams; bar-connected methods are statistically tied.}
\label{fig:gp_vs_tfm}
\end{figure}

When the acquisition and the acquisition optimizer are fixed and only the surrogate is varied (Figure~\ref{fig:gp_vs_tfm}), the surrogate produces statistically distinguishable convergence on engineering problems. GP-based variants remain in the top group, and the latest TabICL is statistically tied with the GP in both settings (TabICL-SCBO vs.\ GP-SCBO, CD\,=\,0.69; TabICL-TuRBO vs.\ GP-TuRBO, CD\,=\,0.38), while the latest TabPFN and a random-forest surrogate rank below both. The gap is between models rather than versions, since both foundation models are the most recent releases, and to our knowledge TabICL has not previously been used as a BO surrogate.

\paragraph{Discussion.}A TFM can thus be competitive with a GP as a BO surrogate on engineering problems, which makes foundation-model surrogates a concrete direction worth studying alongside GPs rather than in place of them \cite{yu2025fast}. The open questions are why TabICL is competitive while TabPFN is not, despite both being tabular foundation models, and whether this holds in the high-dimensional and constrained regimes where surrogate modeling is hardest.

\begin{table}[t]
\centering
\small
\caption{Feasibility rate for the SO-Constrained class.}
\label{tab:feasrate_combined_syneng}
\begin{tabular}{lcc}
\toprule
\textbf{Method} & \textbf{Synthetic} & \textbf{Engineering} \\
\midrule
\quad GP-SCBO & 1.00 & 0.91 \\
\quad CEI & 1.00 & 0.80 \\
\quad PENALTY & 0.96 & 0.64 \\
\quad CLF-CBO & 0.94 & 0.51 \\
\bottomrule
\end{tabular}
\end{table}

\section{Conclusions, Limitations, and Future Work}

BOCoDe is a BO benchmark of 307 problems under a single API---159 computational engineering design problems alongside synthetic and HPO ones---with 31 reference methods across five optimization classes. Our main finding is that engineering design is a different regime, not just a harder one: the best method on synthetic problems can rank last on engineering ones, at least for the unconstrained and mixed-variable classes. The same holds for the surrogate: on engineering problems a TFM statistically ties the GP. Neither is visible from synthetic benchmarks, so engineering must be benchmarked on its own. BOCoDe's coverage is still incomplete, most notably for multi-objective mixed-variable problems and BO methods that are not featured in the ML literature. Additionally, a natural next step is extending swappable foundation-model surrogates to multi-objective BO, which current TFM methods do not yet support. Robust constraint handling on engineering problems also remains open: penalty- and classifier-based methods reach feasibility on only half to two-thirds of runs. We will maintain BOCoDe as an open-source PyPI package after acceptance and extend it with larger problems linked to numerical solvers for structural or fluid-dynamics problems and with new BO methods, and we welcome community contributions.

\section*{Acknowledgments}
The authors thank Kailey Epstein, Janet Qian, and Lyle Regenwetter for their help in curating engineering problems from the literature and conducting code reviews. 

\bibliographystyle{unsrt}  
\bibliography{references}

\newpage
\appendix

\newpage
\startcontents[appendices]  %
\phantomsection

\section*{Table of Contents for Appendices}
\printcontents[appendices]{}{1}{}  %

\newpage

\section{Domain Distinguishability in the Embedding Space} \label{app:cluster_classifier}

\begin{figure}[h!]
    \centering
    \includegraphics[width=.5\linewidth]{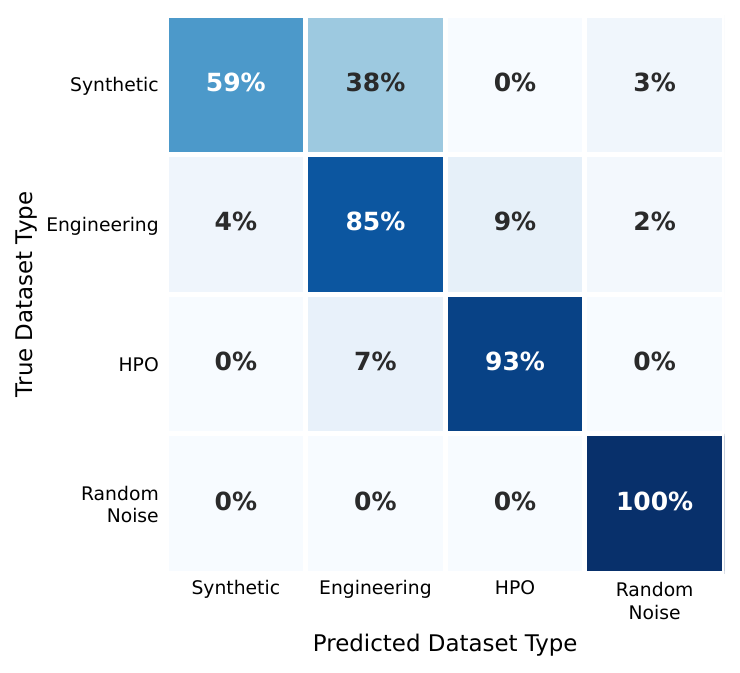}
    \caption{Confusion matrix of a classifier predicting a problem's domain from its TabPFN embedding (row-normalized recall). A random-noise control with no structure is separated perfectly, confirming the embedding captures genuine problem structure. Engineering and HPO are highly distinguishable; synthetic problems overlap the engineering region (38\% predicted as engineering).}
    \label{fig:confusion}
\end{figure}

To quantify the separation seen in the t-SNE layout (main paper's Figure 3), we train an XGBoost \citep{XGBoost2016} classifier to predict a problem's domain from its per-problem TabPFN embedding. Its confusion matrix is shown in Figure~\ref{fig:confusion}. As a control, we add a class of random-noise datasets with no structure, which are recovered with 100\% accuracy, confirming that the embedding encodes genuine problem structure rather than noise. Engineering and HPO problems are highly distinguishable (85\% and 93\% accuracy), whereas synthetic problems are the least distinguishable (59\%) and are most often mistaken for engineering ones (38\%). This matches the t-SNE view: engineering occupies a well-defined region largely disjoint from HPO, while a portion of the synthetic problems spans the engineering region.

\section{Repository Structure and Problem Catalog} \label{app:repo}

The BOCoDe registry catalogs the \textbf{307} base problems evaluated in the main paper's benchmark campaign---159 engineering, 68 synthetic, and 80 HPO---together with additional dimension and scalarization variants of these problems that were not run in this campaign. Figure~\ref{fig:Structure} shows the layout of the \texttt{bocode} Python package, and Figure~\ref{fig:tree_algorithms} the \texttt{algorithms} directory holding the reference baselines.

\begin{figure*}[tp]
\centering
\footnotesize
\begin{forest}
for tree={
  font=\ttfamily,
  grow'=0,
  child anchor=west,
  parent anchor=south,
  anchor=west,
  calign=first,
  edge path={
    \noexpand\path [draw, \forestoption{edge}]
    (!u.south west) +(4pt,0) |- (.child anchor)\forestoption{edge label};
  },
  before typesetting nodes={
    if n=1
      {insert before={[,phantom]}}
      {}
  },
  fit=band,
  before computing xy={l=0.8em},
  s sep=2pt,
  l sep=8pt,
}
[bocode/
  [\_\_init\_\_.py]
  [base.py]
  [registry.py]
  [transforms.py]
  [\_registry\_data.json]
  [opt\_problems/ {\color{gray}(problem implementations)}
    [engineering/]
    [cec2020\_rw/]
    [reproblems/]
    [modact/]
    [control/]
    [materials/]
    [hpo/]
    [nas/]
    [synthetic/]
    [synthetic\_constrained/]
    [synthetic\_mixed/]
    [\_vendor/]
    [{\color{gray}\dots}]
  ]
  [opt\_problems\_metadata/ {\color{gray}(one JSON metadata record per problem)}
    [Car.json]
    [CRE21.json]
    [{\color{gray}\dots}]
  ]
]
\end{forest}
\caption{Structure of the \texttt{bocode} Python package.}
\label{fig:Structure}
\end{figure*}

\begin{figure*}[tp]
\centering
\footnotesize
\begin{forest}
for tree={
  font=\ttfamily,
  grow'=0,
  child anchor=west,
  parent anchor=south,
  anchor=west,
  calign=first,
  edge path={
    \noexpand\path [draw, \forestoption{edge}]
    (!u.south west) +(5pt,0) |- (.child anchor)\forestoption{edge label};
  },
  before typesetting nodes={
    if n=1
      {insert before={[,phantom]}}
      {}
  },
  fit=band,
  before computing xy={l=1em},
}
[algorithms/ {\color{gray}(reference BO baselines; research scripts, not shipped)}
  [\_bo\_utils.py {\color{gray}}]
  [\_tfm\_utils.py {\color{gray}}]
  [\_tabicl\_utils.py {\color{gray}}]
  [single\_obj/ {\color{gray}}]
  [single\_obj\_constrained/ {\color{gray}}]
  [multi\_obj/ {\color{gray}}]
  [multi\_obj\_constrained/ {\color{gray}}]
  [single\_obj\_mixed\_variable/ {\color{gray}}]
]
\end{forest}
\caption{Structure of the \texttt{algorithms} directory holding the reference BO baselines.}
\label{fig:tree_algorithms}
\end{figure*}

\section{The BOCoDe API} \label{app:APIs}

This section details the programming interfaces summarized in the main text: the problem API every benchmark exposes, the transforms that derive new runnable problems from existing ones, the machine-readable metadata, and the adapters that put third-party optimizer packages on BOCoDe's shared protocol.

\subsection{Problem API}

Every BOCoDe problem derives from a common base class. The registry hands out problems by name and answers queries over the metadata described below:

\vspace{2cm}
\begin{lstlisting}[language=Python, basicstyle={\small\ttfamily}]
problem = bocode.Car()               
    # or get_problem("Car")
bocode.list_problems(                
    # all filters optional
    application="Engineering",       
    # domain, case-insensitive
    num_objectives=1,
    constrained=True,
    input_type="mixed",   
    # or "continuous" / "discrete"
    scalable=False)       
    # fixed- vs. user-set dimension
bocode.get_single_objective_constrained()  
    # one class's members
meta = bocode.get_metadata("Car")    
    # JSON record
\end{lstlisting}
\noindent

Filters combine conjunctively, and each of the five optimization classes has its own selector.

The initial design is reproducible and type-aware: continuous dimensions get a Latin-hypercube sampled (LHS) design, integer dimensions get integers, and categorical dimensions draw from their allowed levels, so one call serves continuous and mixed-variable problems alike. During optimization, every algorithm in this study proposes in the unit cube; the problem maps each proposal back to native bounds and valid levels:
\vspace{0.5em}
\begin{lstlisting}[language=Python, basicstyle={\small\ttfamily}]
X0 = problem.sample(5, seed=0)       
    # type-aware LHS design
X  = problem.enforce_variable_types(
    problem.scale(X_unit))
\end{lstlisting}

Evaluation follows the maximization convention of the main text: a constraint counts as satisfied when its value is $\le 0$, and the evaluator concatenates the CEC2020 suite's separate equality and inequality blocks, so every problem presents a single constraint tensor of width $G$:
\vspace{0.5em}
\begin{lstlisting}[language=Python, basicstyle={\small\ttfamily}]
obj_value, cons_value = problem.evaluate(X)           
    # shapes (n, M), (n, G)
\end{lstlisting}

\subsection{Transforms}

Adding a raw objective to a raw constraint violation (or to another
objective) lets whichever term has the larger magnitude dominate, so each
wrapper min--max normalizes its terms to $[0,1]$ over a Latin-hypercube
sample of the base problem (256 points by default), following standard
practice \citep{marler2004survey,tessema2009adaptive}. Inside a benchmark
run it calibrates on the already-evaluated initial design instead, so
construction costs no evaluations beyond the shared budget.
\vspace{0.5em}
\begin{lstlisting}[language=Python, basicstyle={\small\ttfamily}]
mo   = bocode.CRE21() # multi-obj., constrained
so   = bocode.ScalarizedProblem(mo) # -> single-objective
flat = bocode.PenalizedProblem(so) # -> unconstrained
cont = bocode.ContinuousRelaxation(bocode.Car()) # -> continuous
\end{lstlisting}

\begin{enumerate}
    \item \textbf{Scalarization (multi- to single-objective).} Combines the
    $m$ normalized objectives into one, with weights defaulting to
    $1/m$. Constraints pass through unchanged.
    \item \textbf{Penalization (constrained to unconstrained).} Maximizes
    $\mathrm{obj}_{\mathrm{norm}} - w \cdot \mathrm{viol}_{\mathrm{norm}}$
    with $w{=}1$ by default, where the violation is $\sum_i \max(0, g_i)$.
    A feasible point therefore scores its plain normalized objective.
    \item \textbf{Continuous relaxation (mixed to continuous).} Drops the
    integer and categorical restrictions of a mixed-variable problem.
    Dataset problems with a fixed candidate pool are rejected, since they
    admit no relaxation.
\end{enumerate}

The Problem Directory's penalty-scalarized entries apply the same idea to 44 constrained problems, twelve of them multi-objective, re-exposing each as single-objective and unconstrained.
The wrapper takes the equal-weight mean of the objectives, normalizes each
by a precomputed per-objective scale shipped with the package, and
subtracts the total constraint violation. The scales are fixed problem
constants rather than sample estimates, so every entry is a deterministic
problem definition.

\subsection{Problem Metadata}
Every registered problem ships one JSON record. The package loads these
records without importing the problem's code, so a query touches no
optional dependency, and a batch script can select problems from the
records alone, without instantiating a single one. The record for the car
side-impact problem, abridged:

\vspace{1.5mm}
\begin{lstlisting}[basicstyle={\small\ttfamily}]
{
"name": "Car",
"real_name": "Car Side Impact Design
(single-objective, 11 mixed variables)",
"application": "Engineering",
"suite": "Engineering (standalone)",
"num_objectives": 1,
"num_constraints": 10,
"dim": 11,
"input_type": "mixed",
"scalable": false,
"bounds": [[0.5, 1.5], [0.45, 1.35], ...],
"variable_types": ["continuous", ...,
[0.192, 0.345], ...],
"convex": "unknown",
"np_hard": "unknown",
"f_opt": null,
"source": "Gandomi AH, Yang XS, Alavi AH 
(2011) Mixed variable structural 
optimization using firefly algorithm. 
Computers & Structures 
89(23-24):2325-2336"
}
\end{lstlisting}
\vspace{1.5mm}

\subsection{Optimizer Adapters} %

Every baseline ships as a runnable module behind one command-line pattern, where the
module path names the optimization class and the method:

\vspace{1.5mm}
\begin{lstlisting}[language=bash]
python -m algorithms.<class>.<method> \
    --problem <Problem> --init <number> --iters <number> --seed <number>
\end{lstlisting}
\vspace{1.5mm}

For example: 

\vspace{1.5mm}
\begin{lstlisting}[language=bash]
python -m algorithms.single_obj.gp_ucb \
    --problem Car --init 20 --iters 40 --seed 42
\end{lstlisting}
\vspace{1.5mm}

\noindent
The shared runner splits the budget into an initial design, dimension-scaled
when no size is given. It issues one proposal per iteration, resumes interrupted
runs from checkpoints, and writes the same per-iteration trace for every
method: best value so far (minimized) and wall time. Third-party packages run behind thin
adapters that hand each package the same initial design and budget every
other method receives and record results in BOCoDe's conventions;
Appendix~\ref{app:baselines} documents the per-method choices, including
HEBO's sign convention and Bounce's single-call budget.

\section{Algorithms: Implementation Details}
\label{app:baselines} %

Table~\ref{tab:baselines_appendix} summarizes the implementation source, surrogate, acquisition, and
license of every baseline; the bullets below give the full hyperparameter settings and any
deviations from the reference implementations. The interchangeable-surrogate variants
introduced in this work (the RF-, TabPFN- and TabICL- rows) reuse the machinery of the
corresponding baseline with only the surrogate replaced, and are described in the main text.

\begin{table*}[tp]
\centering
\small
\caption{Overview of the reference algorithms (surrogate-swap variants share one row; the
four surrogates expand the 26 rows to the 31 algorithm entries of the main paper's Table~2).
Full hyperparameter settings and deviations are given in the per-method bullets below.}
\label{tab:baselines_appendix}
\begin{tabular}{@{}p{3.2cm}lp{2.8cm}p{2.7cm}p{3cm}c@{}}
\toprule
\textbf{Method} & \textbf{Opt.\ class} & \textbf{Surrogate} & \textbf{Acquisition / Selection} & \textbf{Source} & \textbf{License} \\
\midrule
GP-LogEI & SO-Unc & GP & LogEI & \cite{balandat2020botorch,ament2023logei} & MIT \\
GP-UCB & SO-Unc & GP & UCB ($\beta{=}2$) & \cite{balandat2020botorch,srinivas2010ucb} & MIT \\
\{GP,RF,TabPFN,TabICL\}-TuRBO & SO-Unc & swappable & trust-region Thompson sampling & \cite{eriksson2019turbo} & MIT \\
BAxUS & SO-Unc & GP (nested subspaces) & trust-region Thompson sampling & \cite{papenmeier2022baxus} & MIT \\
Vanilla-HD-BO & SO-Unc & GP (dim.-scaled prior) & qLogNEI & \cite{hvarfner2024vanilla} & --- \\
Standard-GP & SO-Unc & GP (robust $\ell_0$ init) & UCB ($\beta{=}1.5$) & \cite{xu2025standard} & --- \\
\midrule
CEI & SO-Con & GP per output & constrained LogEI & \cite{gardner2014constrainedei} & MIT \\
\{GP,RF,TabPFN,TabICL\}-SCBO & SO-Con & swappable & constrained Thompson sampling & \cite{eriksson2021scbo} & MIT \\
Penalty & SO-Con & GP & LogEI on penalized objective & \cite{Fletcher1975Penalty} & --- \\
CLF-CBO & SO-Con & GP + GP classifier & qLogEI $\times$ feasibility & \cite{tian2024boundary} & MIT \\
\midrule
qNEHVI & MO-Unc & GP per objective & qLogNEHVI & \cite{daulton2020qnehvi} & MIT \\
qNParEGO & MO-Unc & GP per objective & qLogNParEGO & \cite{daulton2021qnparego,knowles2006parego} & MIT \\
MESMO & MO-Unc & GP per objective & lower-bound max-value entropy & \cite{Belakaria2019MESMO,tu2022jes} & MIT \\
DGEMO & MO-Unc & GP per objective & diversity-guided batch HVI & \cite{Lukovic2020Diversity} & MIT \\
NSGA-II / SPEA2 & MO-Unc & --- (EA) & genetic operators & \cite{Deb2002NSGA2,zitzler2001spea2,blank2020pymoo} & Apache-2.0 \\
\midrule
C-qNEHVI & MO-Con & GP per output & feasibility-weighted qLogNEHVI & \cite{daulton2020qnehvi} & MIT \\
C-qNParEGO & MO-Con & GP per output & feasibility-weighted scalarization & \cite{daulton2021qnparego,knowles2006parego} & MIT \\
C-NSGA-II / C-SPEA2 & MO-Con & --- (EA) & constraint domination & \cite{Deb2002NSGA2,zitzler2001spea2,blank2020pymoo} & Apache-2.0 \\
\midrule
PR & SO-Mixed & GP & LogEI (probabilistic reparam.) & \cite{daulton2022bayesian} & MIT \\
Bounce & SO-Mixed & GP (nested embedding) & trust-region Thompson sampling & \cite{papenmeier2023bounce} & MIT \\
Casmopolitan & SO-Mixed & GP (mixed kernel) & interleaved trust-region search & \cite{wan2021casmopolitan} & --- \\
HEBO & SO-Mixed & input-warped GP & MACE ensemble & \cite{cowenrivers2022hebo} & MIT \\
BODi & SO-Mixed & GP (dictionary embedding) & LogEI (alternating search) & \cite{pmlr-v206-deshwal23a} & --- \\
\bottomrule
\end{tabular}
\end{table*}

All algorithms optimize over the unit cube in BOCoDe's maximization convention (constraints
feasible when $c \le 0$), start from a Latin-hypercube initial design of $n_\text{init}$
points, and run $250$ sequential ($q = 1$) iterations over $25$ seeds. We set $n_\text{init}=D$ (\texttt{dim}) problem dimension as a default.  Restarts that redraw an initial design (TuRBO, SCBO) are charged
against the same budget. We use BoTorch v0.18.1, GPyTorch v1.15.2, PyTorch v2.11.0 and pymoo
v0.6.2. Unless stated otherwise, ``BoTorch defaults'' means a single-task GP with input
normalization and output standardization fit by the exact marginal log-likelihood, and an
acquisition optimizer with $10$ restarts and $512$ raw samples. All links were last accessed on
July 25th, 2026.

\subsection{Single-Objective Unconstrained}

\begin{itemize}

\item \textbf{GP-LogEI}: From BoTorch's getting-started guide
\citep{balandat2020botorch} (\url{https://botorch.org/docs/getting_started},
MIT), with the LogEI acquisition \citep{ament2023logei}.

\item \textbf{GP-UCB}: The same loop with BoTorch's upper-confidence-bound
acquisition \citep{balandat2020botorch,srinivas2010ucb}; we set $\beta = 2$ on
the posterior standard deviation (passed to BoTorch as $4$ under its
$\mu + \sqrt{\beta}\sigma$ parameterization).

\item \textbf{GP-TuRBO}: TuRBO-1 \citep{eriksson2019turbo}, from the official
repository (\url{https://github.com/uber-research/TuRBO}, MIT) and BoTorch's
tutorial
(\url{https://github.com/pytorch/botorch/blob/main/tutorials/turbo_1/turbo_1.ipynb},
MIT). We use the reference constants: trust-region side $0.8$ initially with
bounds $[0.5^7, 1.6]$, failure tolerance $\lceil \max(4/q, d/q) \rceil$, a
candidate set of $\min(5000, \max(2000, 200d))$ Sobol points, per-coordinate
perturbation probability $\min(20/d, 1)$, and a local ARD Mat\'ern-5/2 GP with
lengthscales in $[0.005, 4]$ and noise in $[10^{-8}, 10^{-3}]$. The repository
and the tutorial disagree on the success tolerance ($3$ vs.\ $10$); we follow
the paper's $3$.

\item \textbf{BAxUS}: From BoTorch's tutorial \citep{papenmeier2022baxus}
(\url{https://github.com/pytorch/botorch/blob/main/tutorials/baxus/baxus.ipynb},
MIT), cross-checked against the official repository
(\url{https://github.com/LeoIV/BAxUS}). For the failure tolerance we follow
the official repository's rule rather than the tutorial's, whose cap can halve
the trust region after a single failed evaluation at BAxUS's small starting
dimensionalities.

\item \textbf{Vanilla-HD-BO}: From \citep{hvarfner2024vanilla}
(\url{https://github.com/hvarfner/vanilla_bo_in_highdim}). We set the paper's
dimension-scaled lengthscale and noise priors explicitly on an ARD RBF GP,
since a library default would silently substitute different ones, and keep the
reference's sample-around-best $\sigma = 0.1$; BoTorch's default $10^{-3}$
would disable that feature.

\item \textbf{Standard-GP}: From \citep{xu2025standard}
(\url{https://github.com/XZT008/Standard-GP-is-all-you-need-for-HDBO}). The
reference's lengthscale box and its $\sqrt{d}$ initialization become
inconsistent once $d \gtrsim 900$; for the one such problem in our suite
(LassoLeukemia, $d = 7129$) we widen the box's upper bound to $3\sqrt{d}$
rather than clamping the initialization that defines the method.

\end{itemize}

\subsection{Single-Objective Constrained}

\begin{itemize}

\item \textbf{CEI}: From BoTorch's closed-loop tutorial
(\url{https://botorch.org/docs/tutorials/closed_loop_botorch_only}, MIT),
realizing \citep{gardner2014constrainedei} in the numerically stable log form
\citep{ament2023logei}. Before any feasible point exists the acquisition
reduces to the probability of feasibility \citep[Eq.~9]{gelbart2014unknown}.

\item \textbf{SCBO}: From BoTorch's tutorial \citep{eriksson2021scbo}
(\url{https://botorch.org/docs/tutorials/scalable_constrained_bo}, MIT) and
the paper's Appendix~C. The paper and the tutorial disagree on both
tolerances; we take the success tolerance from the paper and the failure
tolerance from the tutorial.

\item \textbf{Penalty}: A static exterior penalty \citep{Fletcher1975Penalty},
for which no reference implementation exists; we fold the constraints into the
objective and hand the result to the same GP-LogEI loop as above, so the only
difference from CEI and SCBO is the constraint handling. The penalized
objective is $F(x) = \tilde{f}(x) - \rho\,\tilde{v}(x)$ with total violation
$v(x) = \sum_i \max(0, c_i(x))$, where $\tilde{f}$ and $\tilde{v}$ are
min--max normalized on the run's own initial design; because both terms are
normalized we set $\rho = 1$, exposed as a flag for sweeping. We report the
best feasible raw objective, as for the other constrained methods.

\item \textbf{CLF-CBO}: From the BoTorch community notebook
``Classifier-based Constrained BO''
(\url{https://github.com/meta-pytorch/botorch/blob/main/notebooks_community/clf_constrained_bo/clf_constrained_bo.ipynb},
MIT), the classifier-based approach to unknown constraints of which
\citep{tian2024boundary} is a recent instance. Before any feasible point
exists the notebook's masked incumbent is undefined; we fall back to the worst
observed objective, which keeps the acquisition positive so the feasibility
weighting steers the search.

\end{itemize}

\subsection{Multi-Objective Unconstrained}

\begin{itemize}

\item \textbf{qNEHVI}: From BoTorch's multi-objective tutorial
\citep{daulton2020qnehvi}
(\url{https://botorch.org/docs/tutorials/multi_objective_bo}, MIT), in the log
form \citep{ament2023logei}. The hypervolume reference point is inferred once
from the initial design and held fixed. We fit and optimize in a per-objective
standardized frame, which leaves dominance unchanged, because the log box
decomposition underflows when objectives differ by orders of magnitude; the
reported hypervolume is computed in the raw frame.

\item \textbf{qNParEGO}: From the same tutorial
\citep{daulton2021qnparego,knowles2006parego}; acquisition settings and the
standardized frame are as for qNEHVI.

\item \textbf{MESMO}: BoTorch's lower-bound multi-objective max-value entropy
search, implementing \citep{Belakaria2019MESMO} with the estimator of
\citep{tu2022jes}; original code at \url{https://github.com/belakaria/MESMO}.
We use the diagonal-covariance estimator rather than BoTorch's full-covariance
default: it matches MESMO's Eq.~6, and the default returns NaNs on RE61 and
CRE51 and requests $19.9$~GiB on RE91.

\item \textbf{DGEMO}: Adapted from \citep{Lukovic2020Diversity} (official
repository: \url{https://github.com/yunshengtian/DGEMO}, MIT); the greedy
diversity-aware batch selection is ported from the reference code. NSGA-II
over the GP posterior mean (population $100$, $50$ generations) replaces the
reference's Pareto discovery, and KMeans replaces its graph-cut clustering,
which requires the pygco C++ wrapper. DGEMO consumes the shared budget in
batches of $5$; the per-evaluation incumbent trace and the shared
reference-point rule keep it comparable at equal evaluation budget.

\item \textbf{NSGA-II and SPEA2}: pymoo's official implementations
\citep{Deb2002NSGA2,zitzler2001spea2,blank2020pymoo} unchanged
(\url{https://pymoo.org}, Apache-2.0). Our only settings put pymoo on the BO
baselines' budget: population size $n_\text{init}$ and
$\lceil (n_\text{init} + 250) / n_\text{init} \rceil + 2$ generations, from
which only the first $n_\text{init} + 250$ evaluated points, in evaluation
order, are used. Objectives are negated at the boundary since BOCoDe maximizes
and pymoo minimizes; hypervolume uses the same reference point and
partitioning as the BO baselines.

\end{itemize}

\subsection{Multi-Objective Constrained}

\begin{itemize}

\item \textbf{C-qNEHVI}: From BoTorch's constrained multi-objective tutorial
\citep{daulton2020qnehvi}
(\url{https://botorch.org/docs/tutorials/constrained_multi_objective_bo},
MIT); we report the running feasible hypervolume.

\item \textbf{C-qNParEGO}: From the same tutorial
\citep{daulton2021qnparego,knowles2006parego}. The constrained tutorial
normalizes the Chebyshev scalarization with the raw observations while the
unconstrained one uses the GP posterior mean; we use the posterior mean in
both ParEGO rows so they differ only in constraint handling.

\item \textbf{C-NSGA-II and C-SPEA2}: The same pymoo algorithms, defaults, and
budget rule as above \citep{Deb2002NSGA2,zitzler2001spea2,blank2020pymoo}.
BOCoDe and pymoo share the $c \le 0$ feasibility convention, so constraints
pass through unchanged; we report the running feasible hypervolume.

\end{itemize}

\subsection{Single-Objective Mixed-Variable}

\begin{itemize}

\item \textbf{PR}: From \citep{daulton2022bayesian}
(\url{https://github.com/facebookresearch/bo_pr}, MIT), with the paper's
Table~2 reparameterizations, a single-task GP, and LogEI as the inner
acquisition.

\item \textbf{Bounce}: The authors' official package v0.1.0 unmodified
\citep{papenmeier2023bounce} (\url{https://github.com/LeoIV/bounce}, MIT); we
set only the batch size to $1$ and the evaluation budget to
$n_\text{init} + 250$. Every discrete dimension, integer as well as
categorical, maps to a categorical parameter over level indices, because the
official initial sampler does not support ordinal parameters; this drops the
ordering of integer levels but keeps them inside the combinatorial embedding
the method is built on.

\item \textbf{Casmopolitan}: From \citep{wan2021casmopolitan}
(\url{https://github.com/xingchenwan/Casmopolitan}). Integer dimensions are
treated as continuous with Garrido-Merch\'an rounding inside the Mat\'ern
kernel, as in the official code.

\item \textbf{HEBO}: The official package v0.3.6 unmodified through its
ask/tell API \citep{cowenrivers2022hebo}
(\url{https://github.com/huawei-noah/HEBO}, MIT). Instead of HEBO's own
quasi-random warm-up we replay BOCoDe's shared initial design, so every method
starts from the same design; discrete parameters carry the index into the
dimension's allowed values, so proposals decode to a valid level without
rounding.

\item \textbf{BODi}: From \citep{pmlr-v206-deshwal23a}
(\url{https://github.com/aryandeshwal/BODi}), with the paper's dictionary
embedding, kernel, and alternating acquisition search.

\end{itemize}

\subsection{Non-GP Surrogates}
\label{app:sub:nonGP}

The RF, TabPFN, and TabICL rows of Table~\ref{tab:baselines} replace only the
surrogate inside their host loop (TuRBO or SCBO); the trust-region logic,
candidate generation, restart rule, and budget stay untouched, so differences
between rows of one host are attributable to the surrogate, with the random
forest as the classical non-GP control. Every surrogate meets the same
contract: a predictive mean and uncertainty per candidate, plus posterior
samples for Thompson-sampling selection.

\begin{itemize}

\item \textbf{TabPFN}: Upstream TabPFN~v3 (8.0.6)
\citep{hollmann2025tabpfn,grinsztajn2026tabpfn3technicalreport}, whose
bar-distribution readouts supply the mean, variance, and acquisition values
directly. Checkpoints are downloaded from Hugging Face under a Prior-Labs
license token; an environment variable pins a local checkpoint file, which
skips the token and fixes the exact weights a batch run used. The code
carries the Prior Labs license (Apache~2.0 with an attribution requirement);
the TabPFN-3 weights carry a separate non-commercial license on Hugging Face.

\item \textbf{TabICL}: TabICL~v2 (version 2 or later, BSD-3-Clause)
\citep{qu2026tabiclv2} through its native regressor. The uncertainty signal is
a $64$-point equal-probability quantile grid whose points are the midpoints of
equal-mass bins, so averaging over the grid gives an unbiased Monte-Carlo
estimate of any expectation; EI, quantile-UCB, and Thompson samples all read
off it. We set the ensemble size to one to match the TabPFN wrapper's cost
profile.

\item \textbf{Random forest (RF)}: scikit-learn's random-forest regressor with
$100$ trees, refit from scratch each iteration; the predictive mean and
uncertainty are the mean and cross-tree standard deviation of the ensemble.
The forest provides marginal $(\mu, \sigma)$ per candidate and no joint
posterior, so Thompson sampling draws each candidate independently from its
own $\mathcal{N}(\mu, \sigma)$ and takes the argmax, repeating without
replacement for a batch.

\end{itemize}

\section{Experiment Setup Details}

\paragraph{Budget and protocol.} Every run starts from a Latin-hypercube initial design
of $n_\text{init}$ points and adds $250$ sequential ($q = 1$) proposals, repeated over
$25$ seeds. BOCoDe's shared dimension-scaled rule of $n_\text{init}=D$. The rule
overrides each method's own initial-design default, so no method receives a larger
budget, and restarts that redraw an initial design (TuRBO, SCBO) draw against the same
budget. Compute limits set the $250$-iteration cap: the campaign spans roughly
$36{,}000$ runs, and the per-problem convergence figures in the Full Results section
show method rankings stabilize well before the cap on most problems. Longer horizons
are future work.

\paragraph{Software.} We use BoTorch v0.18.1, GPyTorch v1.15.2, PyTorch v2.11.0, and
pymoo v0.6.2.

\paragraph{Evolutionary baselines.} The size of the pymoo population $n_{\text{pop}}$ is equal to $n_\text{init}$,
with $\lceil (n_\text{init} + 250) / n_\text{init} \rceil + 2$ generations. The first $n_\text{init} + 250$ evaluated points, in evaluation order, enter the
analysis. This is the same-function-evaluation convention for comparing BO with EAs
\citep{Lukovic2020Diversity,daulton2020qnehvi}. For the multi-objective problems below
$d = 10$, the large majority, this yields a population of $20$, the value Ablation~2
validates against $10$, $40$, and $100$.

\subsection{Evaluation Accounting} Each method family's function-evaluation count 
per run is shown in  Table~\ref{tab:call_counts}. Every family consumes the same total, 
$n_{\text{init}} + 250$. The families differ only in where the initial design 
comes from and how mid-run restarts are launched.

\begin{table}[ht!]
\centering
\footnotesize
\caption{Function-evaluation accounting per run and method family. Every family
consumes the same total budget of $n_\text{init} + 250$ evaluations.}
\label{tab:call_counts}
\begin{tabular}{@{}p{0.30\columnwidth}p{0.58\columnwidth}@{}}
\toprule
\textbf{Family} & \textbf{Evaluation accounting} \\
\midrule
BoTorch loops & shared LHS design ($n_\text{init}$), then $250$ proposals, $q{=}1$ \\
Trust-region (TuRBO, SCBO, BAxUS, Casmopolitan) & as above; restart-triggered redraws
of the design are charged against the same $250$ \\
Bounce & native initialization of matched size $n_\text{init}$, then proposals up to
the shared total \\
HEBO & shared LHS design replayed through ask/tell, then $250$ proposals \\
pymoo EAs & population $n_\text{init}$,
$\lceil (n_\text{init}{+}250)/n_\text{init} \rceil {+} 2$ generations; the first
$n_\text{init}{+}250$ evaluations, in order, enter the analysis \\
\bottomrule
\end{tabular}
\end{table}

\paragraph{Hardware.} The experiments run in parallel on a shared cluster, one job per
(problem, method, seed) triple, each allocated identical resources: 8 CPU cores of an
Intel Xeon Platinum 8480+ node and one NVIDIA H200 GPU. Preempted jobs resume from
checkpoints rather than restarting.

\section{Detailed Statistical Analysis}
\label{app:stats}

\begin{figure*}[tp]
    \centering
    \includegraphics[width=1\linewidth]{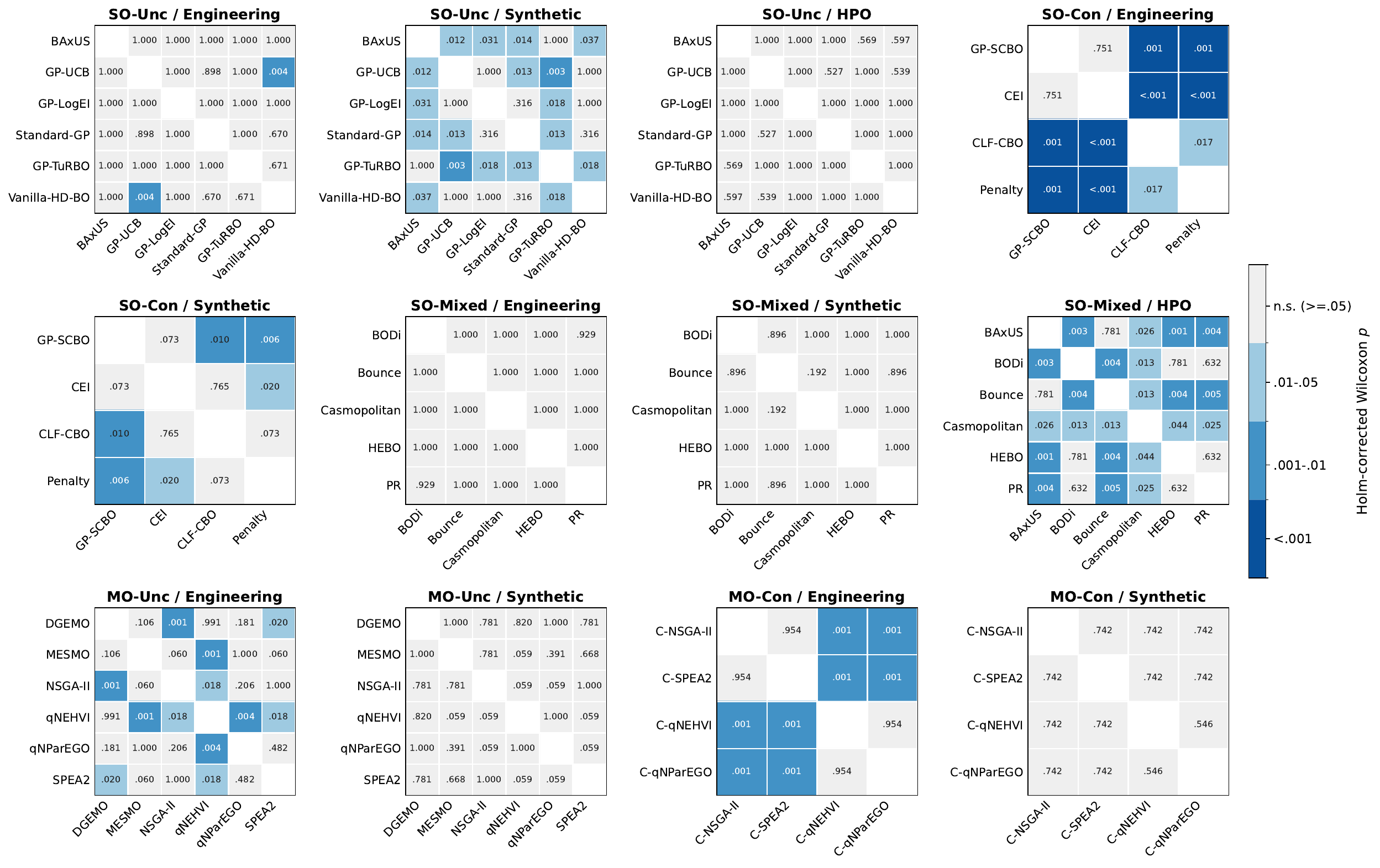}
    \caption{Pairwise Holm-corrected Wilcoxon signed-rank $p$-values on the complete-case paired subset, one panel per optimization class and domain. Blue cells are significant ($p<0.05$; darker $=$ smaller $p$), and gray cells are not; each pair is tested on the problems both methods solve, with the Holm correction applied across the panel's pairs. ``Unc'': Unconstrained; ``Con'': Constrained; ``Mixed'': Mixed-variable.}
    \label{fig:T3_matrix_TRUE}
\end{figure*}

\begin{figure*}[tp]
    \centering
    \includegraphics[width=1\linewidth]{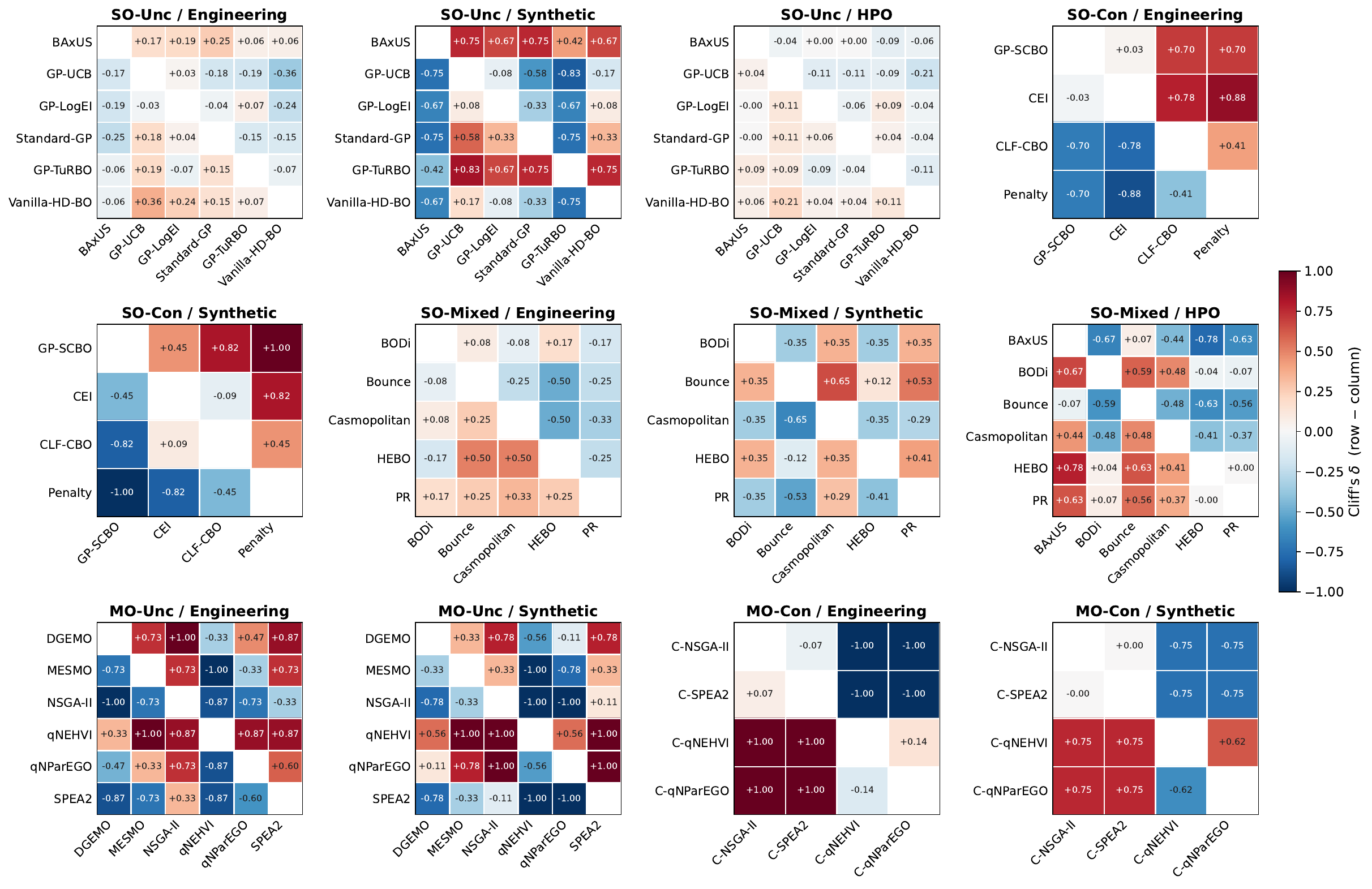}
    \caption{Matched-pairs effect sizes (Cliff's $\delta$, row minus column; $+1$ means the row method dominates) for the same panels. Red cells favor the row method, blue the column; magnitude reads with the pairwise significance in Fig.~\ref{fig:T3_matrix_TRUE}. ``Unc'': Unconstrained; ``Con'': Constrained; ``Mixed'': Mixed-variable.}
    \label{fig:T4_matrix_TRUE}
\end{figure*}

This appendix expands the per-class rankings of the main paper into the full statistical evidence behind the critical-difference (CD) diagrams. 
For every optimization class and domain, we summarize over the $25$ seeds of each (problem, method) ``panel" pair to its median final performance: 
the best-so-far objective value, or the dominated hypervolume for the MO classes. We then rank the methods within each problem 
($1{=}$ best, ties averaged) and report each method's mean rank over the panel's complete-case problems
---those on which every compared method returned a valid result---so the ranks, the CD diagrams, and the rank-vs-iteration trajectories 
of the main paper all summarize the same quantities (Tables~\ref{tab:T2} and~\ref{tab:T2_MO}; 
the effective problem count $N$ per panel is in Table~\ref{tab:T5}). 
Significance is assessed with a Friedman test (Iman--Davenport correction) over the same complete-case problems (Table~\ref{tab:T1}), 
Holm-corrected pairwise Wilcoxon signed-rank tests (Figure~\ref{fig:T3_matrix_TRUE}), and matched-pairs effect sizes 
(Cliff's $\delta$, Figure~\ref{fig:T4_matrix_TRUE}). Following \citep{JMLR:v7:demsar06a}, the CD diagrams (Table~\ref{tab:T5}) 
are the Nemenyi construction on these mean ranks; on the small panels ($N<20$) we read them as indicative and rest significance on the Holm-Wilcoxon tests. 
Cross-domain rank transfer is measured by Kendall's $\tau$ and Spearman's $\rho$ with a bootstrap (Tables~\ref{tab:T6} and~\ref{tab:T9}). 
The non-GP surrogate variants (RF, TabPFN, TabICL) enter the surrogate ablation, not the per-class rankings.

\paragraph{Friedman tests.} For each panel we rank the methods within every problem ($1{=}$ best) 
and test them with the Friedman test using the Iman--Davenport $F$-correction \citep{friedman1937use,iman1980approximations}, 
as recommended for comparing methods across many problems \citep{JMLR:v7:demsar06a}. The $p$-value is this $F$'s tail 
probability under equal ranking, and a panel is significant at $p<0.05$. All panels are significant except 
SO-Mixed-Variable/Engineering ($p=0.40$) and SO-Unconstrained/HPO ($p=0.35$). This aligns with our observations: 
on the mixed-variable engineering problems the methods swap ranks from problem to problem, and on the 
HPO problems the acquisition functions all reach comparable tuning accuracy, so no systematic ordering 
emerges. On SO-Unconstrained/Engineering the omnibus is only marginal ($p=0.028$): the Friedman test 
detects that \emph{some} ordering exists, yet no single pair separates by the Nemenyi criterion, 
which is consistent with the compressed mean-rank spread on that panel.

\paragraph{Pairwise tests and effect sizes.}
For significant panels we identify the differing pairs with Holm-corrected paired Wilcoxon signed-rank tests and size them with Cliff's $\delta$. Wilcoxon sums the signed ranks of the per-problem paired differences. Holm compares the sorted $p$-values against $\alpha/(m-i+1)$ to cap the family-wise error at $0.05$, and $\delta=\tfrac{\#(x>y)-\#(x<y)}{n_x n_y}$ gives the net dominance rate in $[-1,1]$. A pair is called different when its Holm-adjusted $p<0.05$, with $|\delta|$ reporting the margin. The engineering panels resolve cleanly (Fig.~\ref{fig:T3_matrix_TRUE}): qNEHVI beats the evolutionary baselines and MESMO but ties DGEMO; C-qNEHVI and C-qNParEGO beat the constrained EAs; GP-SCBO and CEI beat Penalty and tie each other---all with large $|\delta|$ (Fig.~\ref{fig:T4_matrix_TRUE}). The small synthetic panels often yield no significant pair despite a significant Friedman test, a power limit rather than true parity.

\paragraph{Critical differences.}
Table~\ref{tab:T5} lists the CD values of the diagrams, $\text{CD} = q_\alpha \sqrt{k(k+1)/6N}$ at $\alpha = 0.05$. 
The CD shrinks on the large engineering panels ($0.86$ for SO-Constrained/Engineering and $0.89$ for SO-Unconstrained/Engineering) 
and grows on the small synthetic and multi-objective panels (up to $2.51$ for MO-Unconstrained/Synthetic), 
so the latter establish differences less readily. A tight CD is necessary but not sufficient for separation: 
on SO-Unconstrained/Engineering the mean-rank spread ($0.82$) falls below even this small CD, so the panel remains statistically tied.

\paragraph{Rank transfer across domains.}
Table~\ref{tab:T6} reports Kendall's $\tau$ and Spearman's $\rho$ between the per-method mean-rank vectors of each domain pair, 
with bootstrap $95\%$ confidence intervals over problems. To ask whether a method ordering learned in one domain 
carries to another we correlate the two domains' rank vectors within each class. Kendall's $\tau=\tfrac{C-D}{\binom{k}{2}}$ 
is the normalized excess of concordant over discordant method pairs and Spearman's $\rho=1-\tfrac{6\sum d_i^2}{k(k^2-1)}$ 
the rank-correlation. Both run from $-1$ (reversed) to $+1$ (identical), with $0$ meaning the orderings are unrelated. 
We read transfer off the bootstrap $95\%$ CI and its $p$: a CI excluding $0$ indicates a reliable correlation. 
Two patterns emerge. First, in the single-objective classes no domain pair transfers reliably: the synthetic-to-engineering 
correlations are weak and inconsistent in sign (SO-Unconstrained $\tau=+0.47$, SO-Mixed-Variable $\tau=-0.20$, neither significant), 
and the engineering-to-HPO agreement, though positive 
(SO-Mixed-Variable $\tau=+0.80$, with $\rho$ significant at $p=0.04$; SO-Unconstrained $\tau=+0.20$), 
does not reach significance in $\tau$. Second, the constrained and multi-objective classes transfer moderately and consistently 
($\tau$ from $+0.67$ to $+0.73$, with MO-Unconstrained engineering-to-synthetic $\rho=+0.89$ at $p=0.02$). 
With at most $k=6$ methods per rank vector these tests have little power, so we read the table as trends rather than as proof: 
cross-domain orderings correlate at best moderately, which is consistent with our main finding that the engineering 
panels are governed by problem-to-problem heterogeneity rather than by a stable global ordering.

\paragraph{Robustness to problem families.}
Many engineering problems come in closely related variants (the RE family, the MODAct series). 
Table~\ref{tab:T9} repeats the rank-transfer analysis with the bootstrap resampled at the level of source families, 
so related variants count as one unit. The correlations barely move (SO-Mixed-Variable engineering-to-HPO stays $\tau=+0.80$), 
so the pattern is not an artifact of near-duplicate problems.

\paragraph{Summary.}
The tests support three claims of the main paper: (i) on panels with sufficient power, 
the leading methods sit clearly ahead of the baselines, with large $|\delta|$ on every 
engineering comparison above; (ii) method orderings do not transfer reliably across domains 
in the single-objective classes, so a ranking established on synthetic functions carries 
no significant information about the engineering ranking, while the constrained and 
multi-objective classes are moderately stable; and (iii) on the largest engineering 
panels the field is statistically tied on average because no method sustains an 
advantage across problems, so per-problem results carry more information than any 
single global ranking. Where a panel lacks power or a domain overlap is too small to test, 
we say so rather than infer equivalence.

\begin{table*}[tp]\centering\small
\caption{Per-panel Friedman test (with the Iman--Davenport correction) for each optimization class and domain. A significant $p$ indicates that at least one method differs on the panel; which methods differ is given by the pairwise tests (Figure~\ref{fig:T3_matrix_TRUE}). $N$, $k$: the complete-case problems and methods of the panel---the same block used by the mean ranks and CD diagrams (Table~\ref{tab:T5}).}
\label{tab:T1}\begin{tabular}{ll rr rr}\toprule
Class & Domain & $N$ & $k$ & Iman--Davenport $F$ & Friedman $p$ \\ \midrule
SO-Mixed-Variable & Engineering & 12 & 5 & 1.04 & 3.98e-01 \\
SO-Mixed-Variable & Synthetic & 17 & 5 & 4.06 & 5.40e-03 \\
SO-Mixed-Variable & HPO & 28 & 5 & 16.94 & 8.32e-11 \\
SO-Unconstrained & Engineering & 72 & 6 & 2.54 & 2.80e-02 \\
SO-Unconstrained & Synthetic & 24 & 6 & 14.89 & 2.95e-11 \\
SO-Unconstrained & HPO & 52 & 6 & 1.13 & 3.46e-01 \\
SO-Constrained & Engineering & 30 & 4 & 40.74 & 1.53e-16 \\
SO-Constrained & Synthetic & 11 & 4 & 11.01 & 4.87e-05 \\
MO-Unconstrained & Engineering & 15 & 6 & 24.18 & 4.79e-14 \\
MO-Unconstrained & Synthetic & 9 & 6 & 14.68 & 3.55e-08 \\
MO-Constrained & Engineering & 16 & 4 & 67.09 & 1.21e-16 \\
MO-Constrained & Synthetic & 7 & 4 & 34.83 & 1.05e-07 \\
\bottomrule\end{tabular}\end{table*}

\begin{table*}[p]
\centering\footnotesize\setlength{\tabcolsep}{3pt}
\begin{minipage}[t]{0.49\textwidth}
\centering
\caption{Single-objective classes: mean within-problem rank per method ($1{=}$ best), matching the CD diagrams. Methods are ranked within each problem and averaged (arithmetic mean) over the complete-case problems of the panel (those on which every compared method returned a valid result; $N$ in Table~\ref{tab:T5}).}
\label{tab:T2}
\begin{tabular}{lll r}\toprule
Class & Domain & Method & mean rank \\ \midrule
SO-Mixed-Variable & Engineering & PR & 2.50 \\
 &  & HEBO & 2.71 \\
 &  & BODi & 3.00 \\
 &  & Casmopolitan & 3.25 \\
 &  & Bounce & 3.54 \\
\midrule
SO-Mixed-Variable & Synthetic & Bounce & 2.18 \\
 &  & HEBO & 2.50 \\
 &  & BODi & 3.00 \\
 &  & PR & 3.50 \\
 &  & Casmopolitan & 3.82 \\
\midrule
SO-Mixed-Variable & HPO & HEBO & 2.46 \\
 &  & PR & 2.48 \\
 &  & BODi & 2.50 \\
 &  & Casmopolitan & 3.45 \\
 &  & Bounce & 4.11 \\
\midrule
SO-Unconstrained & Engineering & Vanilla-HD-BO & 3.12 \\
 &  & BAxUS & 3.12 \\
 &  & GP-TuRBO & 3.41 \\
 &  & Standard-GP & 3.67 \\
 &  & GP-LogEI & 3.74 \\
 &  & GP-UCB & 3.94 \\
\midrule
SO-Unconstrained & Synthetic & BAxUS & 1.88 \\
 &  & GP-TuRBO & 2.25 \\
 &  & Standard-GP & 3.67 \\
 &  & GP-LogEI & 4.12 \\
 &  & Vanilla-HD-BO & 4.33 \\
 &  & GP-UCB & 4.75 \\
\midrule
SO-Unconstrained & HPO & Vanilla-HD-BO & 3.16 \\
 &  & GP-LogEI & 3.42 \\
 &  & Standard-GP & 3.47 \\
 &  & GP-TuRBO & 3.57 \\
 &  & BAxUS & 3.67 \\
 &  & GP-UCB & 3.70 \\
\midrule
SO-Constrained & Engineering & CEI & 1.57 \\
 &  & GP-SCBO & 1.77 \\
 &  & CLF-CBO & 3.12 \\
 &  & Penalty & 3.55 \\
\midrule
SO-Constrained & Synthetic & GP-SCBO & 1.36 \\
 &  & CEI & 2.36 \\
 &  & CLF-CBO & 2.64 \\
 &  & Penalty & 3.64 \\
\bottomrule
\end{tabular}
\end{minipage}\hfill
\begin{minipage}[t]{0.49\textwidth}
\centering
\caption{Multi-objective classes: mean within-problem rank per method ($1{=}$ best), matching the CD diagrams. Methods are ranked within each problem and averaged (arithmetic mean) over the complete-case problems of the panel ($N$ in Table~\ref{tab:T5}).}
\label{tab:T2_MO}
\begin{tabular}{lll r}\toprule
Class & Domain & Method & mean rank \\ \midrule
MO-Unconstrained & Engineering & qNEHVI & 1.53 \\
 &  & DGEMO & 2.13 \\
 &  & qNParEGO & 3.33 \\
 &  & MESMO & 3.80 \\
 &  & SPEA2 & 4.87 \\
 &  & NSGA-II & 5.33 \\
\midrule
MO-Unconstrained & Synthetic & qNEHVI & 1.44 \\
 &  & qNParEGO & 2.33 \\
 &  & DGEMO & 2.89 \\
 &  & MESMO & 4.22 \\
 &  & NSGA-II & 5.00 \\
 &  & SPEA2 & 5.11 \\
\midrule
MO-Constrained & Engineering & C-qNEHVI & 1.44 \\
 &  & C-qNParEGO & 1.56 \\
 &  & C-NSGA-II & 3.47 \\
 &  & C-SPEA2 & 3.53 \\
\midrule
MO-Constrained & Synthetic & C-qNEHVI & 1.14 \\
 &  & C-qNParEGO & 1.86 \\
 &  & C-SPEA2 & 3.43 \\
 &  & C-NSGA-II & 3.57 \\
\bottomrule
\end{tabular}

\vspace{1.2em}
\caption{Critical-difference computation: $\text{CD}=q_\alpha\sqrt{k(k+1)/6N}$, $\alpha=0.05$. $k$ is the panel's methods and $N$ its complete-case problems (every method valid), matching the CD diagrams and the mean ranks of Tables~\ref{tab:T2} and~\ref{tab:T2_MO}.}
\label{tab:T5}\begin{tabular}{ll rrr r}\toprule
Class & Domain & $q_{\alpha}$ & $k$ & $N$ & CD \\ \midrule
SO-Mixed-Variable & Engineering & 2.728 & 5 & 12 & 1.761 \\
SO-Mixed-Variable & Synthetic & 2.728 & 5 & 17 & 1.479 \\
SO-Mixed-Variable & HPO & 2.728 & 5 & 28 & 1.153 \\
SO-Unconstrained & Engineering & 2.850 & 6 & 72 & 0.889 \\
SO-Unconstrained & Synthetic & 2.850 & 6 & 24 & 1.539 \\
SO-Unconstrained & HPO & 2.850 & 6 & 52 & 1.046 \\
SO-Constrained & Engineering & 2.569 & 4 & 30 & 0.856 \\
SO-Constrained & Synthetic & 2.569 & 4 & 11 & 1.414 \\
MO-Unconstrained & Engineering & 2.850 & 6 & 15 & 1.947 \\
MO-Unconstrained & Synthetic & 2.850 & 6 & 9 & 2.513 \\
MO-Constrained & Engineering & 2.569 & 4 & 16 & 1.173 \\
MO-Constrained & Synthetic & 2.569 & 4 & 7 & 1.773 \\
\bottomrule\end{tabular}
\end{minipage}
\end{table*}

\begin{table*}[tp]
\centering\footnotesize\setlength{\tabcolsep}{3pt}
\caption{Rank-transfer correlations between domains (per class). Kendall $\tau$/Spearman $\rho$ between method mean-rank vectors (complete-case, matching the CD diagrams); bootstrap 95\% CI over problems. Rows with few common methods ($k\le6$) and non-significant $p$ indicate a consistent \emph{trend} rather than a significant correlation; the cross-class pattern is the robust signal.}
\label{tab:T6}
\begin{tabular}{ll r rl rl}\toprule
Class & Domains & $k$ & $\tau$ & 95\% CI ($p$) & $\rho$ & 95\% CI ($p$) \\ \midrule
SO-Mixed-Variable & Eng$\leftrightarrow$Syn & 5 & -0.20 & [-0.74,+0.40] (0.82) & -0.30 & [-0.82,+0.46] (0.62) \\
SO-Mixed-Variable & Eng$\leftrightarrow$HPO & 5 & +0.80 & [-0.20,+1.00] (0.08) & +0.90 & [-0.30,+1.00] (0.04) \\
SO-Mixed-Variable & Syn$\leftrightarrow$HPO & 5 & +0.00 & [-0.60,+0.40] (1.00) & -0.10 & [-0.70,+0.40] (0.87) \\
SO-Unconstrained & Eng$\leftrightarrow$Syn & 6 & +0.47 & [-0.20,+0.83] (0.27) & +0.43 & [-0.20,+0.89] (0.40) \\
SO-Unconstrained & Eng$\leftrightarrow$HPO & 6 & +0.20 & [-0.33,+0.73] (0.72) & +0.43 & [-0.49,+0.89] (0.40) \\
SO-Unconstrained & Syn$\leftrightarrow$HPO & 6 & -0.33 & [-0.73,+0.47] (0.47) & -0.14 & [-0.83,+0.60] (0.79) \\
SO-Constrained & Eng$\leftrightarrow$Syn & 4 & +0.67 & [+0.33,+1.00] (0.33) & +0.80 & [+0.40,+1.00] (0.20) \\
MO-Unconstrained & Eng$\leftrightarrow$Syn & 6 & +0.73 & [+0.47,+1.00] (0.06) & +0.89 & [+0.66,+1.00] (0.02) \\
MO-Constrained & Eng$\leftrightarrow$Syn & 4 & +0.67 & [+0.33,+1.00] (0.33) & +0.80 & [+0.60,+1.00] (0.20) \\
\bottomrule\end{tabular}

\vspace{1.2em}
\caption{Family-clustered rank-transfer (per class): Table~\ref{tab:T6} with the bootstrap resampling by source family (e.g.\ all RE$xx$ as one unit).}
\label{tab:T9}
\begin{tabular}{ll r rl rl}\toprule
Class & Domains & $k$ & $\tau$ & 95\% CI ($p$) & $\rho$ & 95\% CI ($p$) \\ \midrule
SO-Mixed-Variable & Eng$\leftrightarrow$Syn & 5 & -0.20 & [-0.60,+0.40] (0.82) & -0.30 & [-0.80,+0.50] (0.62) \\
SO-Mixed-Variable & Eng$\leftrightarrow$HPO & 5 & +0.80 & [-0.20,+1.00] (0.08) & +0.90 & [-0.30,+1.00] (0.04) \\
SO-Mixed-Variable & Syn$\leftrightarrow$HPO & 5 & +0.00 & [-0.60,+0.40] (1.00) & -0.10 & [-0.70,+0.40] (0.87) \\
SO-Unconstrained & Eng$\leftrightarrow$Syn & 6 & +0.47 & [-0.20,+0.73] (0.27) & +0.43 & [-0.20,+0.89] (0.40) \\
SO-Unconstrained & Eng$\leftrightarrow$HPO & 6 & +0.20 & [-0.33,+0.60] (0.72) & +0.43 & [-0.43,+0.75] (0.40) \\
SO-Unconstrained & Syn$\leftrightarrow$HPO & 6 & -0.33 & [-0.73,+0.41] (0.47) & -0.14 & [-0.84,+0.38] (0.79) \\
SO-Constrained & Eng$\leftrightarrow$Syn & 4 & +0.67 & [+0.33,+1.00] (0.33) & +0.80 & [+0.40,+1.00] (0.20) \\
MO-Unconstrained & Eng$\leftrightarrow$Syn & 6 & +0.73 & [+0.47,+1.00] (0.06) & +0.89 & [+0.71,+1.00] (0.02) \\
MO-Constrained & Eng$\leftrightarrow$Syn & 4 & +0.67 & [+0.33,+1.00] (0.33) & +0.80 & [+0.60,+1.00] (0.20) \\
\bottomrule\end{tabular}
\end{table*}

\section{Ablations}

\subsection{\texorpdfstring{Ablation 1: Why $n_{\text{init}}=D$?}{Ablation 1: Why n\_init = D?}}
\begin{figure*}[tp]
    \centering
    \includegraphics[width=1\linewidth]{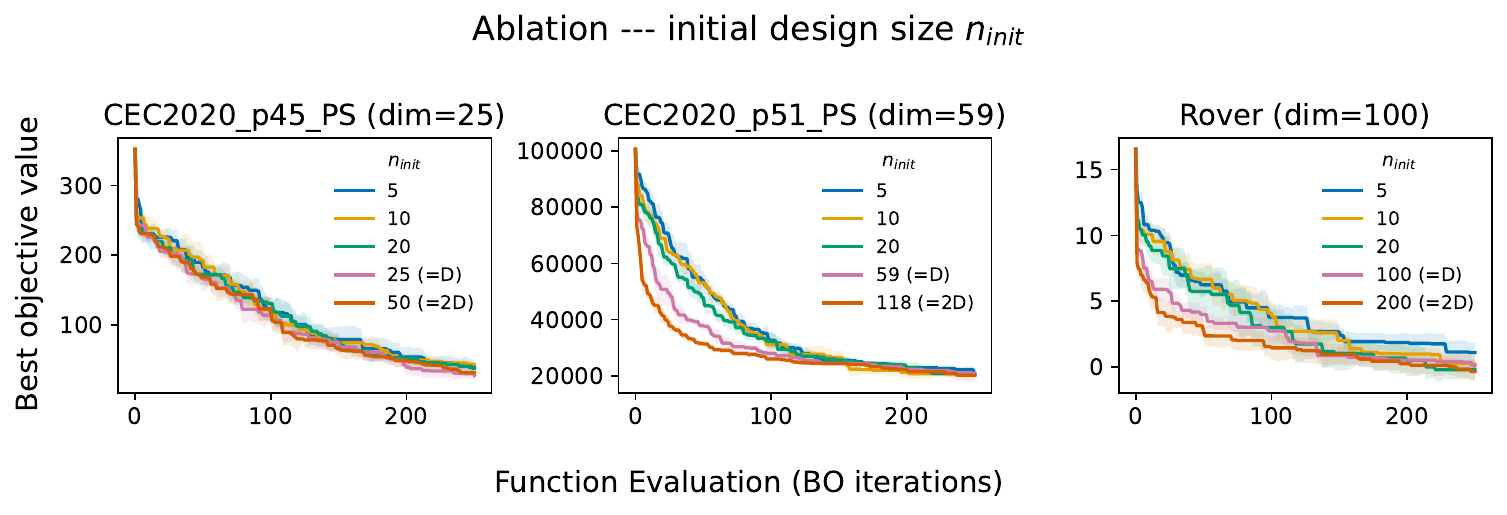}
    \caption{Ablation on $n_{\text{init}}$ for GP-TuRBO on CEC2020\_p45 ($D{=}25$), CEC2020\_p51 ($D{=}59$), and Rover ($D{=}100$). Best objective (lower is better) vs.\ function evaluations, with $n_{\text{init}}\in{5,10,20,D,2D}$ over 15 seeds.}
    \label{fig:ninit_ablation}
\end{figure*}

We set $n_{\text{init}}=D$ following common BO practice \citep{muller2023pfns4bo}. Based on the results in Figure~\ref{fig:ninit_ablation}, we can see that this default is a reasonable but not uniformly optimal with a larger initial design ($n_{\text{init}}=2D$) matches or slightly converge to a better design than $n_{\text{init}}=D$ in two out of three problems. In practice, the user can easily experiment $n_{\text{init}}=D$ parameter in BOCoDe's API. 

\paragraph{Takeaway.} To our knowledge, there is no systematic account of how $n_{\text{init}}$ should scale with dimension. Literature reported choices range from small fixed budgets to $2D{+}1$, with no consensus. Our results show the best value is problem- and dimension-dependent. A principled study of $n_{\text{init}}(D)$ is, we believe, an under-studied question for the BO community.

\subsection{\texorpdfstring{Ablation 2: Why $n_{\text{pop}}=20$ for EA methods?}{Ablation 2: Why n\_pop = 20 for EA methods?}}

\begin{figure*}[tp]
    \centering
    \includegraphics[width=.9\linewidth]{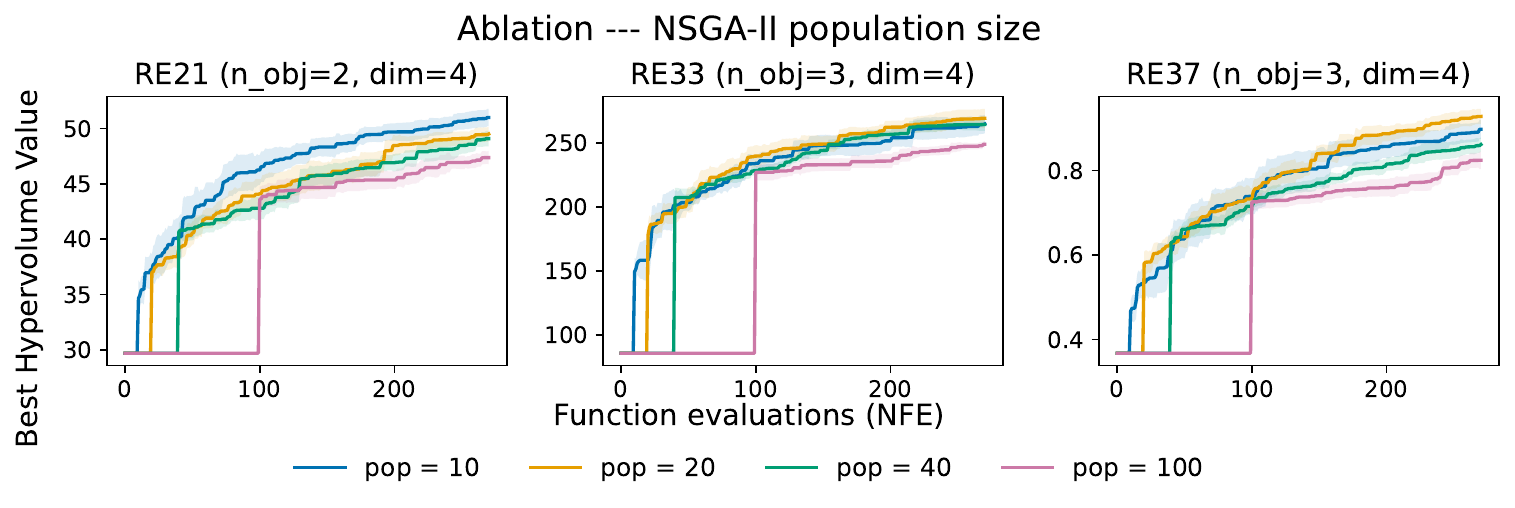}
    \caption{NSGA-II population-size ablation on RE21, RE33, and RE37. Best hypervolume (higher is better) vs.\ function evaluation for $n_{\text{pop}}\in{10,20,40,100}$, 15 seeds.}
    \label{fig:npopulation_ablation}
\end{figure*}

We give the evolutionary baselines the same total function-evaluation budget as the BO methods ($250$ evaluations), following the function-evaluation-equalized, anytime-hypervolume (HV) convention used to compare BO against EA \citep{Lukovic2020Diversity,daulton2020qnehvi}. We empirically show in Figure \ref{fig:npopulation_ablation}, small-to-medium populations (10--40) reach comparable final HV, while a population of $100$ does clearly worse. Therefore, we set our the default population of $20$ in the main experiment with this empirical validated result. 

\paragraph{Takeaway.} A fair, agreed-upon metric for comparing sample-efficient BO against population-based EAs remains an open question, and we see it as a valuable direction for future work. 

\subsection{Ablation 3: Surrogate Ablations}
\label{app:surrogate}

\begin{figure*}[tp]
    \centering
    \includegraphics[width=.9\linewidth]{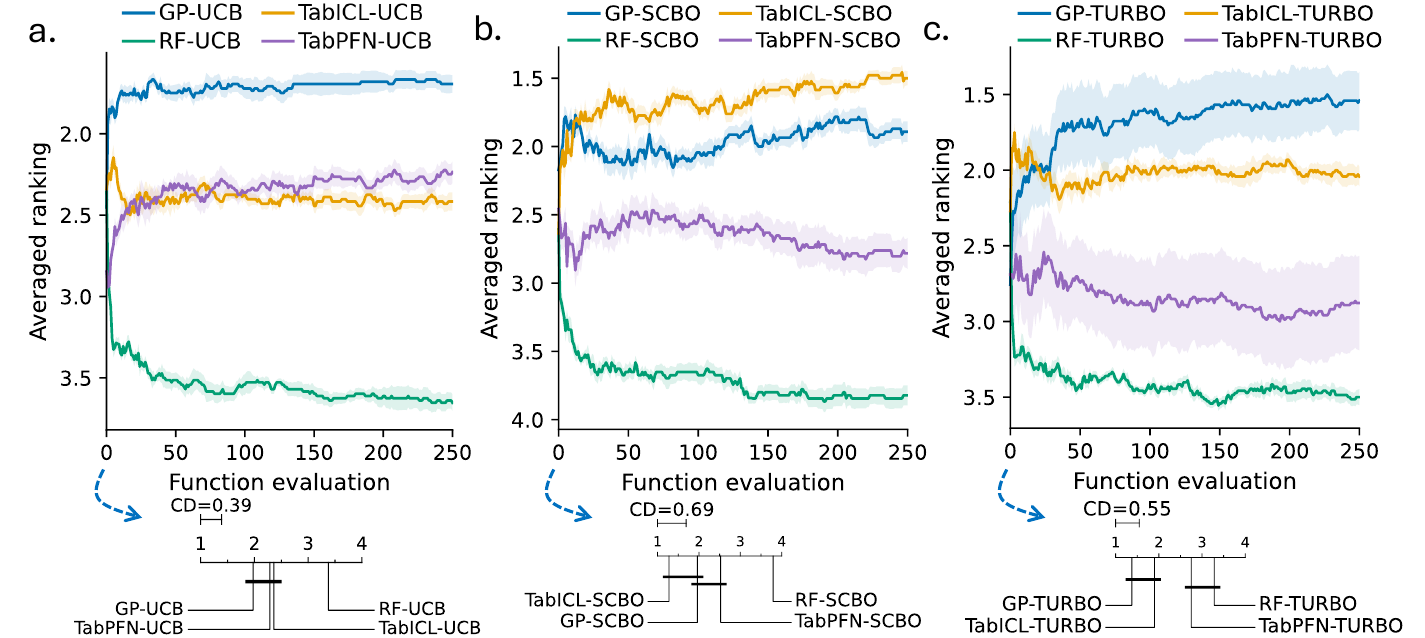}
    \caption{Surrogate ablation under a fixed acquisition. Averaged rank over function evaluations when only the surrogate is swapped (GP, RF, TabPFN, TabICL) under (a) UCB, (b) SCBO, and (c) TuRBO, with final-iteration critical-difference diagrams. Lower is better; bar-connected methods are statistically tied.}
    \label{fig:surrogate_with_UCB_all}
\end{figure*}

\begin{figure*}[tp]
    \centering
    \includegraphics[width=.9\linewidth]{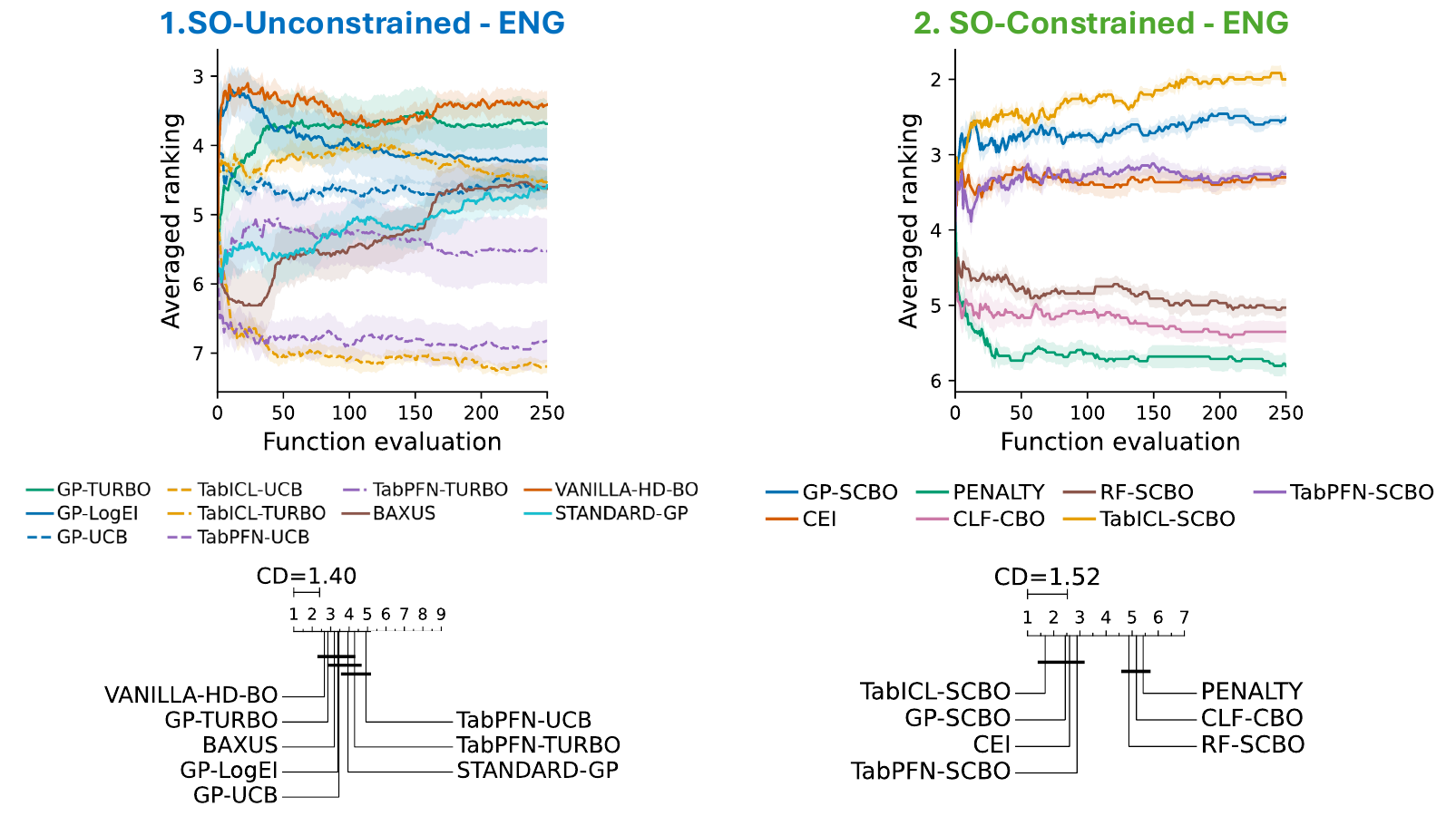}
    \caption{Full per-class rank trajectories including all surrogate variants, for (1) SO-Unconstrained (10 methods) and (2) SO-Constrained (7 methods), with final critical-difference diagrams. Lower is better; bar-connected methods are statistically tied.}
    \label{fig:gp_vs_tfm_rank_iter}
\end{figure*}

We extend the surrogate ablation of the main paper from two acquisitions to three, and from a controlled swap to a full per-class ranking. In Figure~\ref{fig:surrogate_with_UCB_all} we fix the acquisition function and its optimizer and vary only the surrogate (GP, RF, TabPFN, TabICL) under UCB, SCBO, and TuRBO. In Figure~\ref{fig:gp_vs_tfm_rank_iter} we let every surrogate variant compete across the full SO-Unconstrained and SO-Constrained classes.

Across all three acquisitions, the GP is statistically tied for best, and the random forest is consistently worst. Among the foundation-model (FM)  surrogates, TabICL is the strongest: it is statistically indistinguishable from the GP under SCBO and TuRBO, and mid-ranked under UCB, while TabPFN sits below both the GP and TabICL throughout. The full rankings (Figure~\ref{fig:gp_vs_tfm_rank_iter}) tell the same story: GP-based methods and TabICL-SCBO lead their classes, and the RF variants trail.

\paragraph{Takeaway.} To our knowledge, TabICL has not previously been used as a surrogate for BO. That an off-the-shelf TFM is already statistically indistinguishable from a well-tuned GP on several classes makes FM-based surrogates a promising direction for BO that we believe warrants further study.

\section{BOCoDe Problem Directory} \label{app:directory}

This section catalogs every BOCoDe problem with its defining attributes (dimension $D$, number of objectives $M$, number of constraints $G$, variable type, and the cited literature source) organized by optimization class and, within each class, by domain (Engineering, Synthetic, HPO). SO-Unconstrained problems appear in Tables~\ref{tab:probs_SOUnconstrained_Engineering_all}, \ref{tab:probs_SOcontunconstrained_HPO} and~\ref{tab:probs_SOcontunconstrained_Synthetic}; SO-Constrained in Tables~\ref{tab:probs_SOcontconstrained_Synthetic} and~\ref{tab:probs_SOcontconstrained_Engineering}; MO-Unconstrained in Tables~\ref{tab:probs_MOunconstrained_Engineering} and~\ref{tab:probs_MOunconstrained_Synthetic}; MO-Constrained in Tables~\ref{tab:probs_MOconstrained_Engineering} and~\ref{tab:probs_MOconstrained_Synthetic}; and SO-Mixed-Variable in Tables~\ref{tab:probs_Mixedvariable_Engineering}, \ref{tab:probs_Mixedvariable_HPO} and~\ref{tab:probs_Mixedvariable_Synthetic}. Every entry is traceable to its original paper or code through the source column, matching the machine-readable metadata.

\begin{table*}[p]
\centering\tiny\setlength{\tabcolsep}{3pt}
\begin{minipage}[t]{0.49\textwidth}
\centering
\caption{BOCoDe problems in SO-Unconstrained class (single-objective, unconstrained, continuous) and Engineering category (74 problems). \#Obj: number of objectives; \#Con: number of inequality constraints; Dim: design-space dimension.}
\label{tab:probs_SOUnconstrained_Engineering_all}
\begin{tabular}{@{}lrrrl@{}}
\toprule
Problem & \#Obj & \#Con & Dim & Source \\
\midrule
\texttt{HalfCheetahProblem} & 1 & 0 & 6 & \cite{todorov2012mujoco} \\
\texttt{HeatExchanger\_PS} & 1 & 0 & 8 & \cite{yang2012bat} \\
\texttt{HopperPolicy} & 1 & 0 & 33 & \cite{todorov2012mujoco} \\
\texttt{HopperProblem} & 1 & 0 & 3 & \cite{todorov2012mujoco} \\
\texttt{HumanoidProblem} & 1 & 0 & 17 & \cite{todorov2012mujoco} \\
\texttt{HumanoidStandupProblem} & 1 & 0 & 17 & \cite{todorov2012mujoco} \\
\texttt{InvertedDoublePendulumProblem} & 1 & 0 & 1 & \cite{todorov2012mujoco} \\
\bottomrule
\end{tabular}
\end{minipage}\hfill
\begin{minipage}[t]{0.49\textwidth}
\centering
\caption{BOCoDe problems in SO-Unconstrained class (single-objective, unconstrained, continuous) and HPO category (52 problems). \#Obj: number of objectives; \#Con: number of inequality constraints; Dim: design-space dimension.}
\label{tab:probs_SOcontunconstrained_HPO}
\begin{tabular}{@{}lrrrl@{}}
\toprule
Problem & \#Obj & \#Con & Dim & Source \\
\midrule
\texttt{Bayesmark\_ada\_breast} & 1 & 0 & 2 & \cite{turner2019bayesian,pmlr-v133-turner21a} \\
\texttt{Bayesmark\_ada\_digits} & 1 & 0 & 2 & \cite{turner2019bayesian,pmlr-v133-turner21a} \\
\texttt{Bayesmark\_ada\_iris} & 1 & 0 & 2 & \cite{turner2019bayesian,pmlr-v133-turner21a} \\
\texttt{Bayesmark\_ada\_wine} & 1 & 0 & 2 & \cite{turner2019bayesian,pmlr-v133-turner21a} \\
\texttt{Bayesmark\_DT\_breast} & 1 & 0 & 6 & \cite{turner2019bayesian,pmlr-v133-turner21a} \\
\texttt{Bayesmark\_DT\_digits} & 1 & 0 & 6 & \cite{turner2019bayesian,pmlr-v133-turner21a} \\
\texttt{Bayesmark\_DT\_iris} & 1 & 0 & 6 & \cite{turner2019bayesian,pmlr-v133-turner21a} \\
\bottomrule
\end{tabular}
\end{minipage}
\end{table*}

\begin{table*}[p]
\centering\tiny\setlength{\tabcolsep}{3pt}
\begin{minipage}[t]{0.49\textwidth}
\centering
\caption{BOCoDe problems in SO-Unconstrained class (single-objective, unconstrained, continuous) and Synthetic category (24 problems). \#Obj: number of objectives; \#Con: number of inequality constraints; Dim: design-space dimension.}
\label{tab:probs_SOcontunconstrained_Synthetic}
\begin{tabular}{@{}lrrrl@{}}
\toprule
Problem & \#Obj & \#Con & Dim & Source \\
\midrule
\texttt{Ackley} & 1 & 0 & 10 & \cite{hansen2021coco} \\
\texttt{AckleyShifted} & 1 & 0 & 10 & \cite{hansen2021coco} \\
\texttt{Branin} & 1 & 0 & 2 & \cite{hansen2021coco} \\
\texttt{BraninShifted} & 1 & 0 & 2 & \cite{hansen2021coco} \\
\texttt{DixonPrice} & 1 & 0 & 10 & \cite{hansen2021coco} \\
\texttt{DixonPriceShifted} & 1 & 0 & 10 & \cite{hansen2021coco} \\
\texttt{EggHolder} & 1 & 0 & 2 & \cite{hansen2021coco} \\
\texttt{Griewank} & 1 & 0 & 10 & \cite{hansen2021coco} \\
\texttt{GriewankShifted} & 1 & 0 & 10 & \cite{hansen2021coco} \\
\texttt{Hartmann} & 1 & 0 & 6 & \cite{hansen2021coco} \\
\texttt{HartmannShifted} & 1 & 0 & 6 & \cite{hansen2021coco} \\
\texttt{HolderTable} & 1 & 0 & 2 & \cite{hansen2021coco} \\
\texttt{Levy} & 1 & 0 & 10 & \cite{hansen2021coco} \\
\texttt{LevyShifted} & 1 & 0 & 10 & \cite{hansen2021coco} \\
\texttt{Powell} & 1 & 0 & 10 & \cite{hansen2021coco} \\
\texttt{PowellShifted} & 1 & 0 & 10 & \cite{hansen2021coco} \\
\texttt{Rastrigin} & 1 & 0 & 10 & \cite{hansen2021coco} \\
\texttt{RastriginShifted} & 1 & 0 & 10 & \cite{hansen2021coco} \\
\texttt{Rosenbrock} & 1 & 0 & 10 & \cite{hansen2021coco} \\
\texttt{RosenbrockShifted} & 1 & 0 & 10 & \cite{hansen2021coco} \\
\texttt{SixHumpCamel} & 1 & 0 & 2 & \cite{hansen2021coco} \\
\texttt{SixHumpCamelShifted} & 1 & 0 & 2 & \cite{hansen2021coco} \\
\texttt{StyblinskiTang} & 1 & 0 & 10 & \cite{hansen2021coco} \\
\texttt{StyblinskiTangShifted} & 1 & 0 & 10 & \cite{hansen2021coco} \\
\bottomrule
\end{tabular}

\vspace{1.2em}
\caption{BOCoDe problems in SO-Constrained class (single-objective, constrained, continuous) and Synthetic category (11 problems). \#Obj: number of objectives; \#Con: number of inequality constraints; Dim: design-space dimension.}
\label{tab:probs_SOcontconstrained_Synthetic}
\begin{tabular}{@{}lrrrl@{}}
\toprule
Problem & \#Obj & \#Con & Dim & Source \\
\midrule
\texttt{ConstrainedAckley10D} & 1 & 2 & 10 & \cite{gardner2014constrainedei} \\
\texttt{ConstrainedAckley2D} & 1 & 2 & 2 & \cite{gardner2014constrainedei} \\
\texttt{ConstrainedAckley6D} & 1 & 2 & 6 & \cite{gardner2014constrainedei} \\
\texttt{ConstrainedGramacy} & 1 & 2 & 2 & \cite{hansen2021coco} \\
\texttt{ConstrainedHartmann} & 1 & 1 & 6 & \cite{hansen2021coco} \\
\texttt{ConstrainedHartmannSmooth} & 1 & 1 & 6 & \cite{hansen2021coco} \\
\texttt{GKXWC1} & 1 & 1 & 2 & \cite{gardner2014constrainedei} \\
\texttt{GKXWC2} & 1 & 1 & 2 & \cite{gardner2014constrainedei} \\
\texttt{JLH1} & 1 & 1 & 2 & \cite{gardner2014constrainedei} \\
\texttt{JLH2} & 1 & 1 & 2 & \cite{gardner2014constrainedei} \\
\texttt{KeaneBumpFunction} & 1 & 2 & 10 & \cite{hansen2021coco} \\
\bottomrule
\end{tabular}

\vspace{1.2em}
\caption{BOCoDe problems in MO-Unconstrained class (multi-objective, unconstrained, continuous) and Engineering category (17 problems). \#Obj: number of objectives; \#Con: number of inequality constraints; Dim: design-space dimension.}
\label{tab:probs_MOunconstrained_Engineering}
\begin{tabular}{@{}lrrrl@{}}
\toprule
Problem & \#Obj & \#Con & Dim & Source \\
\midrule
\texttt{Penicillin} & 3 & 0 & 7 & \cite{liang2021penicillin} \\
\texttt{RE21} & 2 & 0 & 4 & \cite{tanabe2020re} \\
\texttt{RE22} & 2 & 0 & 3 & \cite{tanabe2020re} \\
\texttt{RE23} & 2 & 0 & 4 & \cite{tanabe2020re} \\
\texttt{RE24} & 2 & 0 & 2 & \cite{tanabe2020re} \\
\texttt{RE25} & 2 & 0 & 3 & \cite{tanabe2020re} \\
\texttt{RE31} & 3 & 0 & 3 & \cite{tanabe2020re} \\
\texttt{RE32} & 3 & 0 & 4 & \cite{tanabe2020re} \\
\texttt{RE33} & 3 & 0 & 4 & \cite{tanabe2020re} \\
\texttt{RE34} & 3 & 0 & 5 & \cite{tanabe2020re} \\
\texttt{RE35} & 3 & 0 & 7 & \cite{tanabe2020re} \\
\texttt{RE36} & 3 & 0 & 4 & \cite{tanabe2020re} \\
\texttt{RE37} & 3 & 0 & 4 & \cite{tanabe2020re} \\
\texttt{RE41} & 4 & 0 & 7 & \cite{tanabe2020re} \\
\texttt{RE42} & 4 & 0 & 6 & \cite{tanabe2020re} \\
\texttt{RE61} & 6 & 0 & 3 & \cite{tanabe2020re} \\
\texttt{RE91} & 9 & 0 & 7 & \cite{tanabe2020re} \\
\bottomrule
\end{tabular}

\end{minipage}\hfill
\begin{minipage}[t]{0.49\textwidth}
\centering
\caption{BOCoDe problems in SO-Constrained class (single-objective, constrained, continuous) and Engineering category (37 problems). \#Obj: number of objectives; \#Con: number of inequality constraints; Dim: design-space dimension.}
\label{tab:probs_SOcontconstrained_Engineering}
\begin{tabular}{@{}lrrrl@{}}
\toprule
Problem & \#Obj & \#Con & Dim & Source \\
\midrule
\texttt{CantileverBeam} & 1 & 11 & 10 & \cite{yang2012bat} \\
\texttt{CEC2020\_p10} & 1 & 3 & 3 & \cite{kumar2020cec} \\
\texttt{CEC2020\_p12} & 1 & 9 & 7 & \cite{kumar2020cec} \\
\texttt{CEC2020\_p13} & 1 & 3 & 5 & \cite{kumar2020cec} \\
\texttt{CEC2020\_p14} & 1 & 10 & 10 & \cite{kumar2020cec} \\
\texttt{CEC2020\_p15} & 1 & 11 & 7 & \cite{kumar2020cec} \\
\texttt{CEC2020\_p16} & 1 & 15 & 14 & \cite{kumar2020cec} \\
\bottomrule
\end{tabular}

\vspace{1.2em}
\caption{BOCoDe problems in MO-Unconstrained class (multi-objective, unconstrained, continuous) and Synthetic category (9 problems). \#Obj: number of objectives; \#Con: number of inequality constraints; Dim: design-space dimension.}
\label{tab:probs_MOunconstrained_Synthetic}
\begin{tabular}{@{}lrrrl@{}}
\toprule
Problem & \#Obj & \#Con & Dim & Source \\
\midrule
\texttt{BraninCurrin} & 2 & 0 & 2 & \cite{hansen2021coco} \\
\texttt{DTLZ1} & 3 & 0 & 7 & \cite{hansen2021coco} \\
\texttt{DTLZ2} & 3 & 0 & 7 & \cite{hansen2021coco} \\
\texttt{DTLZ3} & 3 & 0 & 7 & \cite{hansen2021coco} \\
\texttt{DTLZ4} & 3 & 0 & 7 & \cite{hansen2021coco} \\
\texttt{DTLZ5} & 3 & 0 & 7 & \cite{hansen2021coco} \\
\texttt{ZDT1} & 2 & 0 & 6 & \cite{hansen2021coco} \\
\texttt{ZDT2} & 2 & 0 & 6 & \cite{hansen2021coco} \\
\texttt{ZDT3} & 2 & 0 & 6 & \cite{hansen2021coco} \\
\bottomrule
\end{tabular}
\end{minipage}
\end{table*}
\begin{table*}[p]
\centering\tiny\setlength{\tabcolsep}{3pt}
\begin{minipage}[t]{0.49\textwidth}
\centering
\caption{BOCoDe problems in MO-Constrained class (multi-objective, constrained, continuous) and Engineering category (19 problems). \#Obj: number of objectives; \#Con: number of inequality constraints; Dim: design-space dimension.}
\label{tab:probs_MOconstrained_Engineering}
\begin{tabular}{@{}lrrrl@{}}
\toprule
Problem & \#Obj & \#Con & Dim & Source \\
\midrule
\texttt{CRE22} & 2 & 4 & 4 & \cite{tanabe2020re} \\
\texttt{CRE23} & 2 & 4 & 4 & \cite{tanabe2020re} \\
\texttt{CRE24} & 2 & 11 & 7 & \cite{tanabe2020re} \\
\texttt{CRE25} & 2 & 1 & 4 & \cite{tanabe2020re} \\
\texttt{CRE31} & 3 & 10 & 7 & \cite{tanabe2020re} \\
\texttt{CRE32} & 3 & 9 & 6 & \cite{tanabe2020re} \\
\texttt{CRE51} & 5 & 7 & 3 & \cite{tanabe2020re} \\
\texttt{CS1} & 2 & 7 & 20 & \cite{picard2020modact} \\
\texttt{CS2} & 2 & 8 & 20 & \cite{picard2020modact} \\
\texttt{CT1} & 2 & 7 & 20 & \cite{picard2020modact} \\
\texttt{CT2} & 2 & 8 & 20 & \cite{picard2020modact} \\
\texttt{CTS1} & 3 & 7 & 20 & \cite{picard2020modact} \\
\texttt{CTS2} & 3 & 8 & 20 & \cite{picard2020modact} \\
\texttt{CTSE1} & 4 & 7 & 20 & \cite{picard2020modact} \\
\texttt{CTSE2} & 4 & 8 & 20 & \cite{picard2020modact} \\
\texttt{CTSE3} & 4 & 10 & 20 & \cite{picard2020modact} \\
\texttt{CTSEI1} & 5 & 7 & 20 & \cite{picard2020modact} \\
\texttt{CTSEI2} & 5 & 8 & 20 & \cite{picard2020modact} \\
\texttt{WeldedBeam} & 2 & 4 & 4 & \cite{tanabe2020re} \\
\bottomrule
\end{tabular}

\vspace{1.2em}
\caption{BOCoDe problems in MO-Constrained class (multi-objective, constrained, continuous) and Synthetic category (7 problems). \#Obj: number of objectives; \#Con: number of inequality constraints; Dim: design-space dimension.}
\label{tab:probs_MOconstrained_Synthetic}
\begin{tabular}{@{}lrrrl@{}}
\toprule
Problem & \#Obj & \#Con & Dim & Source \\
\midrule
\texttt{BNH} & 2 & 2 & 2 & \cite{hansen2021coco} \\
\texttt{C2DTLZ2} & 2 & 1 & 12 & \cite{hansen2021coco} \\
\texttt{CONSTR} & 2 & 2 & 2 & \cite{hansen2021coco} \\
\texttt{ConstrainedBraninCurrin} & 2 & 1 & 2 & \cite{hansen2021coco} \\
\texttt{MW7} & 2 & 2 & 2 & \cite{hansen2021coco} \\
\texttt{OSY} & 2 & 6 & 6 & \cite{hansen2021coco} \\
\texttt{SRN} & 2 & 2 & 2 & \cite{hansen2021coco} \\
\bottomrule
\end{tabular}

\vspace{1.2em}
\caption{BOCoDe problems in SO-Mixed-Variable class (single-objective, mixed-variable) and Engineering category (12 problems). \#Obj: number of objectives; \#Con: number of inequality constraints; Dim: design-space dimension.}
\label{tab:probs_Mixedvariable_Engineering}
\begin{tabular}{@{}lrrrl@{}}
\toprule
Problem & \#Obj & \#Con & Dim & Source \\
\midrule
\texttt{Car} & 1 & 10 & 11 & \cite{gandomi2011firefly} \\
\texttt{ColumnBuckling} & 1 & 0 & 4 & \cite{yousefpour2024gpplus} \\
\texttt{CrossedBarrel} & 1 & 0 & 4 & \cite{hase2021olympus} \\
\texttt{EulerBeamMixed} & 1 & 0 & 3 & \cite{cuestaramirez2022mixed} \\
\texttt{GearTrain} & 1 & 0 & 4 & \cite{sandgren1990nonlinear,gandomi2011firefly} \\
\texttt{HelicalSpring} & 1 & 8 & 3 & \cite{gandomi2011firefly} \\
\texttt{HOIP} & 1 & 0 & 3 & \cite{hase2021olympus} \\
\texttt{WeldedBeamCategorical} & 1 & 5 & 5 & \cite{wan2021casmopolitan,dreczkowski2023mcbo} \\
\texttt{PestControl} & 1 & 0 & 25 & \cite{oh2019combinatorial} \\
\texttt{PressureVessel} & 1 & 4 & 4 & \cite{gandomi2011firefly} \\
\texttt{ReinforcedConcreteBeam} & 1 & 2 & 3 & \cite{gandomi2011firefly} \\
\texttt{SpeedReducer} & 1 & 9 & 7 & \cite{yang2012bat} \\
\bottomrule
\end{tabular}
\end{minipage}\hfill
\begin{minipage}[t]{0.49\textwidth}
\centering
\caption{BOCoDe problems in SO-Mixed-Variable class (single-objective, mixed-variable) and HPO category (28 problems). \#Obj: number of objectives; \#Con: number of inequality constraints; Dim: design-space dimension.}
\label{tab:probs_Mixedvariable_HPO}
\begin{tabular}{@{}lrrrl@{}}
\toprule
Problem & \#Obj & \#Con & Dim & Source \\
\midrule
\texttt{HPOB\_4796\_23} & 1 & 0 & 3 & \cite{arango2021hpob} \\
\texttt{HPOB\_5527\_10101} & 1 & 0 & 8 & \cite{arango2021hpob} \\
\texttt{HPOB\_5636\_10101} & 1 & 0 & 6 & \cite{arango2021hpob} \\
\texttt{HPOB\_5859\_125923} & 1 & 0 & 6 & \cite{arango2021hpob} \\
\texttt{HPOB\_5889\_31} & 1 & 0 & 6 & \cite{arango2021hpob} \\
\texttt{HPOB\_5891\_3492} & 1 & 0 & 8 & \cite{arango2021hpob} \\
\texttt{HPOB\_5906\_3889} & 1 & 0 & 16 & \cite{arango2021hpob} \\
\texttt{HPOB\_5965\_10101} & 1 & 0 & 10 & \cite{arango2021hpob} \\
\texttt{HPOB\_5971\_10093} & 1 & 0 & 16 & \cite{arango2021hpob} \\
\texttt{HPOB\_6767\_145804} & 1 & 0 & 18 & \cite{arango2021hpob} \\
\texttt{HPOB\_6794\_10101} & 1 & 0 & 10 & \cite{arango2021hpob} \\
\texttt{HPOB\_7607\_145976} & 1 & 0 & 9 & \cite{arango2021hpob} \\
\texttt{HPOB\_7609\_125923} & 1 & 0 & 9 & \cite{arango2021hpob} \\
\texttt{LCBenchAdult} & 1 & 0 & 7 & \cite{ZimLin2021a} \\
\texttt{LCBenchAirlines} & 1 & 0 & 7 & \cite{ZimLin2021a} \\
\texttt{LCBenchAlbert} & 1 & 0 & 7 & \cite{ZimLin2021a} \\
\texttt{LCBenchAPSFailure} & 1 & 0 & 7 & \cite{ZimLin2021a} \\
\texttt{LCBenchAustralian} & 1 & 0 & 7 & \cite{ZimLin2021a} \\
\texttt{LCBenchBankMarketing} & 1 & 0 & 7 & \cite{ZimLin2021a} \\
\texttt{LCBenchCar} & 1 & 0 & 7 & \cite{ZimLin2021a} \\
\texttt{LCBenchChristine} & 1 & 0 & 7 & \cite{ZimLin2021a} \\
\texttt{LCBenchCovertype} & 1 & 0 & 7 & \cite{ZimLin2021a} \\
\texttt{LCBenchCreditG} & 1 & 0 & 7 & \cite{ZimLin2021a} \\
\texttt{LCBenchFabert} & 1 & 0 & 7 & \cite{ZimLin2021a} \\
\texttt{LCBenchHiggs} & 1 & 0 & 7 & \cite{ZimLin2021a} \\
\texttt{LCBenchJasmine} & 1 & 0 & 7 & \cite{ZimLin2021a} \\
\texttt{LCBenchNomao} & 1 & 0 & 7 & \cite{ZimLin2021a} \\
\texttt{LCBenchVehicle} & 1 & 0 & 7 & \cite{ZimLin2021a} \\
\bottomrule
\end{tabular}

\vspace{1.2em}
\caption{BOCoDe problems in SO-Mixed-Variable class (single-objective, mixed-variable) and Synthetic category (17 problems). \#Obj: number of objectives; \#Con: number of inequality constraints; Dim: design-space dimension.}
\label{tab:probs_Mixedvariable_Synthetic}
\begin{tabular}{@{}lrrrl@{}}
\toprule
Problem & \#Obj & \#Con & Dim & Source \\
\midrule
\texttt{Ackley53} & 1 & 0 & 53 & \cite{wan2021casmopolitan,dreczkowski2023mcbo} \\
\texttt{Ackley5Mixed} & 1 & 0 & 5 & \cite{wan2021casmopolitan,dreczkowski2023mcbo} \\
\texttt{AckleyCat} & 1 & 0 & 20 & \cite{wan2021casmopolitan,dreczkowski2023mcbo} \\
\texttt{BraninCategorical} & 1 & 0 & 3 & \cite{wan2021casmopolitan,dreczkowski2023mcbo} \\
\texttt{BraninLVGP} & 1 & 0 & 2 & \cite{wan2021casmopolitan,dreczkowski2023mcbo} \\
\texttt{CoCaBOFunc2C} & 1 & 0 & 4 & \cite{wan2021casmopolitan,dreczkowski2023mcbo} \\
\texttt{CoCaBOFunc3C} & 1 & 0 & 5 & \cite{wan2021casmopolitan,dreczkowski2023mcbo} \\
\texttt{GoldsteinLVGP} & 1 & 0 & 2 & \cite{wan2021casmopolitan,dreczkowski2023mcbo} \\
\texttt{HartmannCat} & 1 & 0 & 6 & \cite{wan2021casmopolitan,dreczkowski2023mcbo} \\
\texttt{MixedAckley} & 1 & 0 & 5 & \cite{wan2021casmopolitan,dreczkowski2023mcbo} \\
\texttt{RastriginCat} & 1 & 0 & 20 & \cite{wan2021casmopolitan,dreczkowski2023mcbo} \\
\texttt{Rosenbrock5Mixed} & 1 & 0 & 5 & \cite{wan2021casmopolitan,dreczkowski2023mcbo} \\
\texttt{SchwefelCat} & 1 & 0 & 20 & \cite{wan2021casmopolitan,dreczkowski2023mcbo} \\
\texttt{ShekelMixed} & 1 & 0 & 4 & \cite{wan2021casmopolitan,dreczkowski2023mcbo} \\
\texttt{StyblinskiTangCat} & 1 & 0 & 10 & \cite{wan2021casmopolitan,dreczkowski2023mcbo} \\
\texttt{StyblinskiTangMixed} & 1 & 0 & 5 & \cite{wan2021casmopolitan,dreczkowski2023mcbo} \\
\texttt{MaxSAT} & 1 & 0 & 60 & \cite{pmlr-v206-deshwal23a} \\
\bottomrule
\end{tabular}
\end{minipage}
\end{table*}

\section{Full Results}

\subsection{Single-Objective Per-problem Best Searched Value}

Tables~\ref{tab:socont_final_eng_v1}$\sim$\ref{tab:mixed_final_syn} report, for
every single-objective problem, the final best value; the constrained classes 
report the best feasible objective. Pooling over methods makes these problem-level 
reference values: they record what the campaign attained on each problem at the 
shared budget and give future work an absolute anchor to compare against, complementing
the rank-based analyses.

\subsection{Multi-Objective Per-problem Best Hypervolume Value}

Tables~\ref{tab:mounc_final_eng}$\sim$\ref{tab:mocon_final_syn} report the
same summary for the multi-objective problems: the final best dominated
hypervolume under each problem's fixed reference point,
pooled over methods and seeds, with feasible hypervolume only for the
constrained class.

\begin{table*}[p]
\centering\tiny\setlength{\tabcolsep}{3pt}
\renewcommand{\cellset}{\renewcommand{\arraystretch}{0.8}}
\begin{minipage}[t]{0.4\textwidth}
\centering
\caption{SO-Unconstrained Engineering (part 1): final best objective at BO iteration 250, pooled over the 6 methods and seeds (sign matches the convergence figures).}
\label{tab:socont_final_eng_v1}
\begin{tabular}{lcc}
\toprule
Problem & mean$\pm$std & median$\pm$95\%CI \\
\midrule
AgNP & $0.1664 \pm 0.03129$ & $0.1693 \pm 0.002889$ \\
Allison & $0.5723 \pm 0.011$ & $0.57 \pm 0.0001068$ \\
AntPolicySearchProblem & $-182.7 \pm 412.1$ & $18.43 \pm 0.7427$ \\
AntProblem & $-1004 \pm 0.8469$ & $-1004 \pm 0.1803$ \\
AutoAM & $-0.8934 \pm 0.07527$ & $-0.9365 \pm 0.005064$ \\
Borehole & $7.823 \pm 0.004738$ & $7.82 \pm 0.000814$ \\
CEC2020\_p11\_PS & $23.71 \pm 4.531$ & $21.13 \pm 0.6197$ \\
\bottomrule
\end{tabular}
\end{minipage}\hfill
\begin{minipage}[t]{0.49\textwidth}
\centering
\caption{SO-Unconstrained Engineering (part 2): final best objective at BO iteration 250, pooled over the 6 methods and seeds (sign matches the convergence figures).}
\label{tab:socont_final_eng_v2}
\begin{tabular}{lcc}
\toprule
Problem & mean$\pm$std & median$\pm$95\%CI \\
\midrule
HalfCheetahPolicySearchProblem & $-1291 \pm 499.6$ & $-1242 \pm 76.18$ \\
HalfCheetahProblem & $-0.3104 \pm 0.04538$ & $-0.3229 \pm 0.006361$ \\
HeatExchanger\_PS & \makecell{$1.809e+04$\\$\pm 1.33e+04$} & \makecell{$1.586e+04$\\$\pm 2637$} \\
HopperPolicySearchProblem & $-1274 \pm 512.3$ & $-1052 \pm 26.44$ \\
HopperProblem & $-322.4 \pm 102.1$ & $-310.2 \pm 17.36$ \\
HumanoidProblem & $-958.9 \pm 232.9$ & $-963.7 \pm 39.3$ \\
HumanoidStandupProblem & \makecell{$-1.448e+05$\\$\pm 8431$} & \makecell{$-1.469e+05$\\$\pm 1159$} \\
InvertedDoublePendulumProblem & $-91.4 \pm 1.135$ & $-91.55 \pm 0.0001105$ \\
InvertedPendulumProblem & $-52.85 \pm 18.96$ & $-51.5 \pm 3.821$ \\
Mazda\_PS & $19.67 \pm 7.574$ & $14.52 \pm 1.767$ \\
Mazda\_SCA\_PS & $21.39 \pm 4.966$ & $17.82 \pm 1.241$ \\
MiniAeroWing & $243 \pm 1.424$ & $242.4 \pm 0.0987$ \\
P3HT & $-727.9 \pm 111.4$ & $-770.4 \pm 7.43$ \\
PD4CartPole & $-143.9 \pm 56.5$ & $-167.8 \pm 1.158$ \\
PID4Acrobot & $79.53 \pm 1.212$ & $79.9 \pm 0.1652$ \\
Perovskite & \makecell{$3.8e+04$\\$\pm 3.209e+04$} & \makecell{$2.712e+04$\\$\pm 1621$} \\
PusherProblem & $51.88 \pm 1.959$ & $50.82 \pm 0.1614$ \\
QPowerModel & $0.0002482 \pm 0.0002033$ & \makecell{$0.0001805$\\$\pm 4.359e-05$} \\
ReacherProblem & $5.823 \pm 0.113$ & $5.774 \pm 0.008518$ \\
ReactivityModel & \makecell{$7.272e-06$\\$\pm 9.398e-06$} & \makecell{$4.666e-06$\\$\pm 1.007e-06$} \\
RobotPush & $-6.416 \pm 1.846$ & $-6.615 \pm 0.191$ \\
Rover & $-0.3132 \pm 2.009$ & $-0.6649 \pm 0.2126$ \\
SwimmerProblem & $-33.82 \pm 2.279$ & $-33.11 \pm 0.5072$ \\
Walker2DPolicySearchProblem & $-1025 \pm 256.4$ & $-1027 \pm 5.773$ \\
Walker2DProblem & $-871.4 \pm 226.8$ & $-989.3 \pm 46.02$ \\
Wing & $123.5 \pm 0.339$ & $123.3 \pm 0.04799$ \\
\bottomrule
\end{tabular}

\vspace{1.2em}
\caption{SO-Unconstrained Synthetic: final best objective at BO iteration 250, pooled over the 6 methods and seeds (sign matches the convergence figures).}
\label{tab:socont_final_syn}
\begin{tabular}{lcc}
\toprule
Problem & mean$\pm$std & median$\pm$95\%CI \\
\midrule
Ackley & $2.885 \pm 1.877$ & $2.695 \pm 0.3199$ \\
AckleyShifted & $5.419 \pm 4.691$ & $3.732 \pm 0.4893$ \\
Branin & $0.3982 \pm 0.0004936$ & \makecell{$0.398$\\$\pm 4.118e-05$} \\
BraninShifted & $0.3997 \pm 0.004107$ & $0.3981 \pm 0.0002032$ \\
DixonPrice & $166.1 \pm 202.3$ & $86.77 \pm 39.7$ \\
DixonPriceShifted & $658.4 \pm 631.6$ & $495.3 \pm 133.4$ \\
EggHolder & $-890.4 \pm 74.4$ & $-894.6 \pm 9.019$ \\
Griewank & $0.8065 \pm 0.3523$ & $0.9507 \pm 0.02123$ \\
GriewankShifted & $0.9713 \pm 0.09811$ & $0.9873 \pm 0.01399$ \\
Hartmann & $-3.215 \pm 0.2798$ & $-3.321 \pm 0.01749$ \\
HartmannShifted & $-3.311 \pm 0.09041$ & $-3.322 \pm 0.0001792$ \\
HolderTable & $-18.76 \pm 1.436$ & $-19.21 \pm 0.001541$ \\
Levy & $1.497 \pm 2.744$ & $0.6086 \pm 0.1346$ \\
LevyShifted & $4.171 \pm 3.233$ & $3.411 \pm 0.5949$ \\
Powell & $45.2 \pm 39.66$ & $42.97 \pm 10.09$ \\
PowellShifted & $83.55 \pm 72.31$ & $72.34 \pm 16.92$ \\
Rastrigin & $44.84 \pm 24.76$ & $50.03 \pm 3.623$ \\
RastriginShifted & $47.42 \pm 18.29$ & $50.26 \pm 4.68$ \\
Rosenbrock & $869.5 \pm 914.1$ & $598.4 \pm 133.1$ \\
RosenbrockShifted & $1221 \pm 1169$ & $954.6 \pm 197.6$ \\
SixHumpCamel & $-1.031 \pm 0.000546$ & \makecell{$-1.032$\\$\pm 4.066e-05$} \\
SixHumpCamelShifted & $-1.031 \pm 0.001228$ & $-1.031 \pm 0.0001227$ \\
StyblinskiTang & $-349.9 \pm 25.48$ & $-349.2 \pm 4.331$ \\
StyblinskiTangShifted & $-323.2 \pm 42.07$ & $-330.3 \pm 10.61$ \\
\bottomrule
\end{tabular}
\end{minipage}
\end{table*}
\begin{table*}[p]
\centering\tiny\setlength{\tabcolsep}{3pt}
\renewcommand{\cellset}{\renewcommand{\arraystretch}{0.8}}
\begin{minipage}[t]{0.49\textwidth}
\centering
\caption{SO-Unconstrained HPO: final best objective at BO iteration 250, pooled over the 6 methods and seeds (sign matches the convergence figures).}
\label{tab:socont_final_hpo}
\begin{tabular}{lcc}
\toprule
Problem & mean$\pm$std & median$\pm$95\%CI \\
\midrule
Bayesmark\_DT\_breast & $-0.9227 \pm 0.006898$ & $-0.9231 \pm 0.001519$ \\
Bayesmark\_DT\_digits & $-0.7937 \pm 0.08449$ & $-0.8079 \pm 0.0001919$ \\
Bayesmark\_DT\_iris & $-0.9499 \pm 0.0007281$ & $-0.95 \pm 0$ \\
Bayesmark\_DT\_wine & $-0.9383 \pm 0.01365$ & $-0.9367 \pm 0.001899$ \\
Bayesmark\_MLPadam\_breast & $-0.9276 \pm 0.003682$ & $-0.9275 \pm 0.0006053$ \\
Bayesmark\_MLPadam\_digits & $-0.9795 \pm 0.002033$ & $-0.9798 \pm 0.0004302$ \\
Bayesmark\_MLPadam\_iris & $-0.9852 \pm 0.004877$ & $-0.9833 \pm 0.001147$ \\
\bottomrule
\end{tabular}
\end{minipage}\hfill
\begin{minipage}[t]{0.49\textwidth}
\centering
\caption{SO-Constrained Engineering: final best feasible objective at BO iteration 250, pooled over the 4 methods and seeds (sign matches the convergence figures).}
\label{tab:socon_final_eng}
\begin{tabular}{lcc}
\toprule
Problem & mean$\pm$std & median$\pm$95\%CI \\
\midrule
CEC2020\_p10 & $1.168 \pm 0.1137$ & $1.106 \pm 0.04675$ \\
CEC2020\_p12 & $5.821 \pm 3.825$ & $4.642 \pm 0.5677$ \\
CEC2020\_p13 & \makecell{$2.797e+04$\\$\pm 1813$} & \makecell{$2.689e+04$\\$\pm 264.1$} \\
CEC2020\_p14 & \makecell{$8.101e+04$\\$\pm 3.368e+04$} & \makecell{$6.674e+04$\\$\pm 2368$} \\
CEC2020\_p15 & $3058 \pm 99.79$ & $3003 \pm 20.55$ \\
CEC2020\_p17 & $0.03712 \pm 0.05045$ & $0.01324 \pm 0.004763$ \\
CEC2020\_p18 & \makecell{$1.422e+04$\\$\pm 1.33e+04$} & $7379 \pm 1833$ \\
\bottomrule
\end{tabular}

\vspace{1.2em}
\caption{SO-Constrained Synthetic: final best feasible objective at BO iteration 250, pooled over the 4 methods and seeds (sign matches the convergence figures).}
\label{tab:socon_final_syn}
\begin{tabular}{lcc}
\toprule
Problem & mean$\pm$std & median$\pm$95\%CI \\
\midrule
ConstrainedAckley10D & $3.098 \pm 1.883$ & $2.645 \pm 0.4249$ \\
ConstrainedAckley2D & $0.2582 \pm 0.4997$ & $0.04749 \pm 0.03434$ \\
ConstrainedAckley6D & $2.055 \pm 1.386$ & $2.08 \pm 0.3232$ \\
ConstrainedGramacy & $0.6102 \pm 0.02443$ & $0.5999 \pm 0.0005497$ \\
ConstrainedHartmann & $-2.998 \pm 0.7047$ & $-3.318 \pm 0.008165$ \\
ConstrainedHartmannSmooth & $-2.965 \pm 0.7257$ & $-3.319 \pm 0.08125$ \\
GKXWC1 & $-1.997 \pm 0.01106$ & \makecell{$-2$\\$\pm 1.296e-06$} \\
GKXWC2 & $0.2697 \pm 0.06804$ & $0.2535 \pm 0.0006924$ \\
JLH1 & $0.03254 \pm 0.009304$ & \makecell{$0.03127$\\$\pm 7.041e-05$} \\
JLH2 & $-1.799 \pm 0.08495$ & $-1.825 \pm 0.002507$ \\
KeaneBumpFunction & $-0.2745 \pm 0.1071$ & $-0.2295 \pm 0.01764$ \\
\bottomrule
\end{tabular}
\end{minipage}
\end{table*}
\begin{table*}[p]
\centering\tiny\setlength{\tabcolsep}{3pt}
\renewcommand{\arraystretch}{0.88}
\renewcommand{\cellset}{\renewcommand{\arraystretch}{0.8}}
\begin{minipage}[t]{0.49\textwidth}
\centering
\caption{SO-Mixed-Variable HPO: final best objective at BO iteration 250, pooled over the 5 methods and seeds (sign matches the convergence figures).}
\label{tab:mixed_final_hpo}
\begin{tabular}{lcc}
\toprule
Problem & mean$\pm$std & median$\pm$95\%CI \\
\midrule
HPOB\_4796\_23 & $-0.5596 \pm 0.0006587$ & $-0.5601 \pm 0.0002131$ \\
HPOB\_5527\_10101 & $-0.7888 \pm 0.002443$ & $-0.7888 \pm 0.0004198$ \\
HPOB\_5636\_10101 & $-0.7942 \pm 0.001545$ & $-0.7941 \pm 0$ \\
HPOB\_5859\_125923 & $-0.8638 \pm 0.0003158$ & $-0.8638 \pm 0$ \\
HPOB\_5889\_31 & $-0.7749 \pm 0.0004197$ & $-0.775 \pm 0$ \\
HPOB\_5906\_3889 & $-0.9953 \pm 0.0002487$ & \makecell{$-0.9953$\\$\pm 7.63e-05$} \\
HPOB\_5965\_10101 & $-0.7933 \pm 0.002064$ & $-0.7941 \pm 0$ \\
HPOB\_5971\_10093 & $-0.9971 \pm 0.0009445$ & $-0.9971 \pm 0.0001143$ \\
HPOB\_6767\_145804 & $-0.9929 \pm 0.01771$ & $-0.9958 \pm 0.0008407$ \\
HPOB\_6794\_10101 & $-0.7981 \pm 0.001524$ & $-0.7981 \pm 0.0002099$ \\
HPOB\_7607\_145976 & $-0.7797 \pm 0.001446$ & $-0.7799 \pm 0$ \\
HPOB\_7609\_125923 & $-0.8743 \pm 0.001006$ & $-0.8739 \pm 0.0002334$ \\
LCBenchAPSFailure & $-98.7 \pm 0.214$ & $-98.72 \pm 0.04745$ \\
LCBenchAdult & $-80.83 \pm 0.4866$ & $-80.68 \pm 0.02685$ \\
LCBenchAirlines & $-62.29 \pm 1.058$ & $-62.6 \pm 0.1482$ \\
LCBenchAlbert & $-65.97 \pm 0.2999$ & $-66 \pm 0.02147$ \\
LCBenchAustralian & $-90.96 \pm 0.2772$ & $-90.85 \pm 0$ \\
LCBenchBankMarketing & $-88 \pm 1.204$ & $-88.23 \pm 0.02916$ \\
LCBenchCar & $-95.75 \pm 2.393$ & $-96.87 \pm 0$ \\
LCBenchChristine & $-74 \pm 0.3037$ & $-74.04 \pm 0.01359$ \\
LCBenchCovertype & $-73.4 \pm 3.914$ & $-74.83 \pm 1.061$ \\
LCBenchCreditG & $-74.24 \pm 0.7273$ & $-74.32 \pm 0.07373$ \\
LCBenchFabert & $-66.73 \pm 0.167$ & $-66.67 \pm 0.04719$ \\
LCBenchHiggs & $-71.45 \pm 0.7464$ & $-71.77 \pm 0.06092$ \\
LCBenchJasmine & $-81.86 \pm 0.1925$ & $-81.85 \pm 0.049$ \\
LCBenchNomao & $-96.05 \pm 0.2683$ & $-96.16 \pm 0$ \\
LCBenchVehicle & $-80.83 \pm 1.513$ & $-81.38 \pm 0$ \\
LCBenchVolkert & $-62.15 \pm 2.601$ & $-63.76 \pm 0.4603$ \\
\bottomrule
\end{tabular}

\vspace{0.6em}
\caption{SO-Mixed-Variable Engineering: final best objective at BO iteration 250, pooled over the 5 methods and seeds (sign matches the convergence figures).}
\label{tab:mixed_final_eng}
\begin{tabular}{lcc}
\toprule
Problem & mean$\pm$std & median$\pm$95\%CI \\
\midrule
Car & $20.35 \pm 1.356$ & $19.83 \pm 0.006961$ \\
ColumnBuckling & \makecell{$-2.966e+05$\\$\pm 0.06631$} & \makecell{$-2.966e+05$\\$\pm 0$} \\
CrossedBarrel & $-45.7 \pm 0.989$ & $-44.94 \pm 0.2773$ \\
EulerBeamMixed & $1287 \pm 0.03384$ & $1287 \pm 0.001419$ \\
GearTrain & \makecell{$2.28e-06$\\$\pm 5.837e-06$} & \makecell{$8.362e-08$\\$\pm 2.159e-07$} \\
HOIP & $-16.74 \pm 0.1666$ & $-16.79 \pm 0$ \\
HelicalSpring & $2.112 \pm 0.4573$ & $2.045 \pm 0.09399$ \\
PestControl & $12.46 \pm 0.4496$ & $12.24 \pm 0.1244$ \\
PressureVessel & $1355 \pm 1529$ & $704.1 \pm 130.3$ \\
ReinforcedConcreteBeam & $266.3 \pm 1.939$ & \makecell{$265.9$\\$\pm 4.747e-05$} \\
SpeedReducer & $2382 \pm 71.6$ & $2353 \pm 2.953$ \\
WeldedBeamCategorical & $2.518 \pm 0.8279$ & $2.437 \pm 0.1133$ \\
\bottomrule
\end{tabular}

\vspace{0.6em}
\caption{SO-Mixed-Variable Synthetic: final best objective at BO iteration 250, pooled over the 5 methods and seeds (sign matches the convergence figures).}
\label{tab:mixed_final_syn}
\begin{tabular}{lcc}
\toprule
Problem & mean$\pm$std & median$\pm$95\%CI \\
\midrule
Ackley53 & $9.338 \pm 7.005$ & $9.113 \pm 2.215$ \\
Ackley5Mixed & $0.07492 \pm 0.1373$ & $0.01565 \pm 0.004732$ \\
AckleyCat & $3.271 \pm 4.479$ & \makecell{$4.441e-16$\\$\pm 1.216$} \\
BraninCategorical & $0.5956 \pm 0.5044$ & \makecell{$0.3979$\\$\pm 2.851e-05$} \\
BraninLVGP & $3.111 \pm 0.7486$ & $2.791 \pm 0.006349$ \\
CoCaBOFunc2C & $-0.1956 \pm 0.04151$ & $-0.206 \pm 0.0005037$ \\
CoCaBOFunc3C & $-0.6381 \pm 0.1636$ & $-0.7172 \pm 0.01092$ \\
GoldsteinLVGP & $3.305 \pm 0.9291$ & $3 \pm 0.003386$ \\
HartmannCat & $-3.126 \pm 0.2715$ & $-3.291 \pm 0.09017$ \\
MaxSAT & \makecell{$-1.542e+04$\\$\pm 260.3$} & \makecell{$-1.554e+04$\\$\pm 79.21$} \\
MixedAckley & $3.406 \pm 3.836$ & $2.712 \pm 0.8415$ \\
RastriginCat & $2.451 \pm 6.351$ & $0 \pm 0$ \\
Rosenbrock5Mixed & $0.7807 \pm 0.7305$ & $0.5619 \pm 0.09858$ \\
SchwefelCat & $7398 \pm 249.4$ & $7257 \pm 17.63$ \\
ShekelMixed & $-5.576 \pm 3.398$ & $-4.07 \pm 1.184$ \\
StyblinskiTangCat & $-356.5 \pm 22.92$ & $-367.2 \pm 0$ \\
StyblinskiTangMixed & $-150.1 \pm 9.524$ & $-155.4 \pm 1.615$ \\
\bottomrule
\end{tabular}
\end{minipage}\hfill
\begin{minipage}[t]{0.49\textwidth}
\centering
\caption{MO-Unconstrained Engineering: final best hypervolume value at BO iteration 250, pooled over the 6 methods and seeds (sign matches the convergence figures).}
\label{tab:mounc_final_eng}
\begin{tabular}{lcc}
\toprule
Problem & mean$\pm$std & median$\pm$95\%CI \\
\midrule
Penicillin & \makecell{$1.804e+06$\\$\pm 5.272e+04$} & \makecell{$1.793e+06$\\$\pm 1.026e+04$} \\
RE21 & $52.2 \pm 2.27$ & $53.24 \pm 0.4861$ \\
RE22 & \makecell{$3.968e+04$\\$\pm 6443$} & \makecell{$4.175e+04$\\$\pm 1084$} \\
RE23 & \makecell{$8.244e+09$\\$\pm 9.755e+08$} & \makecell{$8.583e+09$\\$\pm 4.881e+07$} \\
RE24 & \makecell{$2.173e+04$\\$\pm 82.24$} & \makecell{$2.174e+04$\\$\pm 17.01$} \\
RE25 & \makecell{$8.249e+05$\\$\pm 7.42e+04$} & \makecell{$8.496e+05$\\$\pm 1.02e+04$} \\
RE31 & \makecell{$1.057e+17$\\$\pm 1.48e+15$} & \makecell{$1.061e+17$\\$\pm 6.723e+13$} \\
RE32 & \makecell{$3.728e+14$\\$\pm 6.358e+12$} & \makecell{$3.745e+14$\\$\pm 1.376e+11$} \\
RE33 & $284 \pm 15.49$ & $284.1 \pm 3.723$ \\
RE34 & $33.18 \pm 3.303$ & $34.25 \pm 0.7688$ \\
RE35 & \makecell{$2.173e+09$\\$\pm 5.173e+07$} & \makecell{$2.186e+09$\\$\pm 1.059e+07$} \\
RE36 & $84.97 \pm 9.026$ & $90.37 \pm 1.763$ \\
RE37 & $1.017 \pm 0.09634$ & $1.072 \pm 0.0218$ \\
RE41 & $317.1 \pm 48.94$ & $342.3 \pm 12.79$ \\
RE42 & \makecell{$3.118e+10$\\$\pm 2.254e+10$} & \makecell{$4.596e+10$\\$\pm 5.79e+09$} \\
RE61 & \makecell{$9.44e+25$\\$\pm 8.135e+24$} & \makecell{$9.502e+25$\\$\pm 2.369e+24$} \\
RE91 & $1.535 \pm 0.5341$ & $1.343 \pm 0.1822$ \\
\bottomrule
\end{tabular}

\vspace{0.6em}
\caption{MO-Unconstrained Synthetic: final best hypervolume value at BO iteration 250, pooled over the 6 methods and seeds (sign matches the convergence figures).}
\label{tab:mounc_final_syn}
\begin{tabular}{lcc}
\toprule
Problem & mean$\pm$std & median$\pm$95\%CI \\
\midrule
BraninCurrin & $56.6 \pm 4.84$ & $57.44 \pm 0.4917$ \\
DTLZ1 & \makecell{$6.384e+07$\\$\pm 1.62e+05$} & \makecell{$6.392e+07$\\$\pm 2.154e+04$} \\
DTLZ2 & $0.5733 \pm 0.1871$ & $0.6786 \pm 0.04297$ \\
DTLZ3 & \makecell{$9.999e+11$\\$\pm 3.621e+08$} & \makecell{$1e+12$\\$\pm 1.807e+06$} \\
DTLZ4 & $0.5741 \pm 0.1856$ & $0.6777 \pm 0.04279$ \\
DTLZ5 & $987.5 \pm 2.715$ & $987.9 \pm 0.2201$ \\
ZDT1 & $118 \pm 3.307$ & $119.8 \pm 0.6841$ \\
ZDT2 & $113.2 \pm 7.321$ & $113.3 \pm 1.817$ \\
ZDT3 & $124.3 \pm 3.697$ & $125.3 \pm 1.09$ \\
\bottomrule
\end{tabular}

\vspace{0.6em}
\caption{MO-Constrained Engineering: final best feasible hypervolume value at BO iteration 250, pooled over the 4 methods and seeds (sign matches the convergence figures).}
\label{tab:mocon_final_eng}
\begin{tabular}{lcc}
\toprule
Problem & mean$\pm$std & median$\pm$95\%CI \\
\midrule
CRE22 & \makecell{$7.246e+05$\\$\pm 5.077e+04$} & \makecell{$7.485e+05$\\$\pm 8905$} \\
CRE23 & $8.105 \pm 0.5161$ & $8.391 \pm 0.1536$ \\
CRE24 & \makecell{$3.997e+06$\\$\pm 9.134e+05$} & \makecell{$4.585e+06$\\$\pm 2.002e+05$} \\
CRE25 & $194.3 \pm 41.41$ & $223.5 \pm 10.15$ \\
CRE31 & $27.34 \pm 5.641$ & $31.73 \pm 1.817$ \\
CRE32 & \makecell{$7.905e+08$\\$\pm 7.432e+08$} & \makecell{$1.431e+09$\\$\pm 2.599e+08$} \\
CRE51 & \makecell{$3.373e+25$\\$\pm 2.62e+24$} & \makecell{$3.31e+25$\\$\pm 9.8e+23$} \\
CS1 & $15.76 \pm 10.39$ & $15.86 \pm 3.014$ \\
CS2 & $30.54 \pm 19.9$ & $35.08 \pm 5.376$ \\
CT1 & $0.8123 \pm 0.318$ & $0.8069 \pm 0.1005$ \\
CT2 & $0.7602 \pm 0.3216$ & $0.7897 \pm 0.1007$ \\
CTS1 & $208.2 \pm 56.16$ & $214.3 \pm 14.11$ \\
CTS2 & $199.3 \pm 59.32$ & $207.2 \pm 16.34$ \\
CTSE1 & $15.99 \pm 8.349$ & $14.9 \pm 2.359$ \\
CTSE2 & $14.42 \pm 8.236$ & $14.21 \pm 2.164$ \\
CTSEI1 & $7451 \pm 3732$ & $5757 \pm 1262$ \\
CTSEI2 & $7080 \pm 4048$ & $6038 \pm 1485$ \\
\bottomrule
\end{tabular}

\vspace{0.6em}
\caption{MO-Constrained Synthetic: final best feasible hypervolume value at BO iteration 250, pooled over the 4 methods and seeds (sign matches the convergence figures).}
\label{tab:mocon_final_syn}
\begin{tabular}{lcc}
\toprule
Problem & mean$\pm$std & median$\pm$95\%CI \\
\midrule
C2DTLZ2 & $0.2293 \pm 0.1295$ & $0.2203 \pm 0.0476$ \\
CONSTR & $91.31 \pm 0.6223$ & $91.36 \pm 0.1887$ \\
ConstrainedBraninCurrin & $568.6 \pm 50.5$ & $595.3 \pm 14.03$ \\
MW1 & $0.6973 \pm 0.09379$ & $0.7043 \pm 0.02625$ \\
MW3 & $0.8662 \pm 0.04254$ & $0.8782 \pm 0.003395$ \\
MW7 & $0.4593 \pm 0.04913$ & $0.4733 \pm 0.01053$ \\
OSY & $5823 \pm 4029$ & $6443 \pm 1411$ \\
\bottomrule
\end{tabular}
\end{minipage}
\end{table*}

\subsection{Convergence Plot}

\begin{figure*}[htpb!]
    \centering
    \includegraphics[width=.9\linewidth]{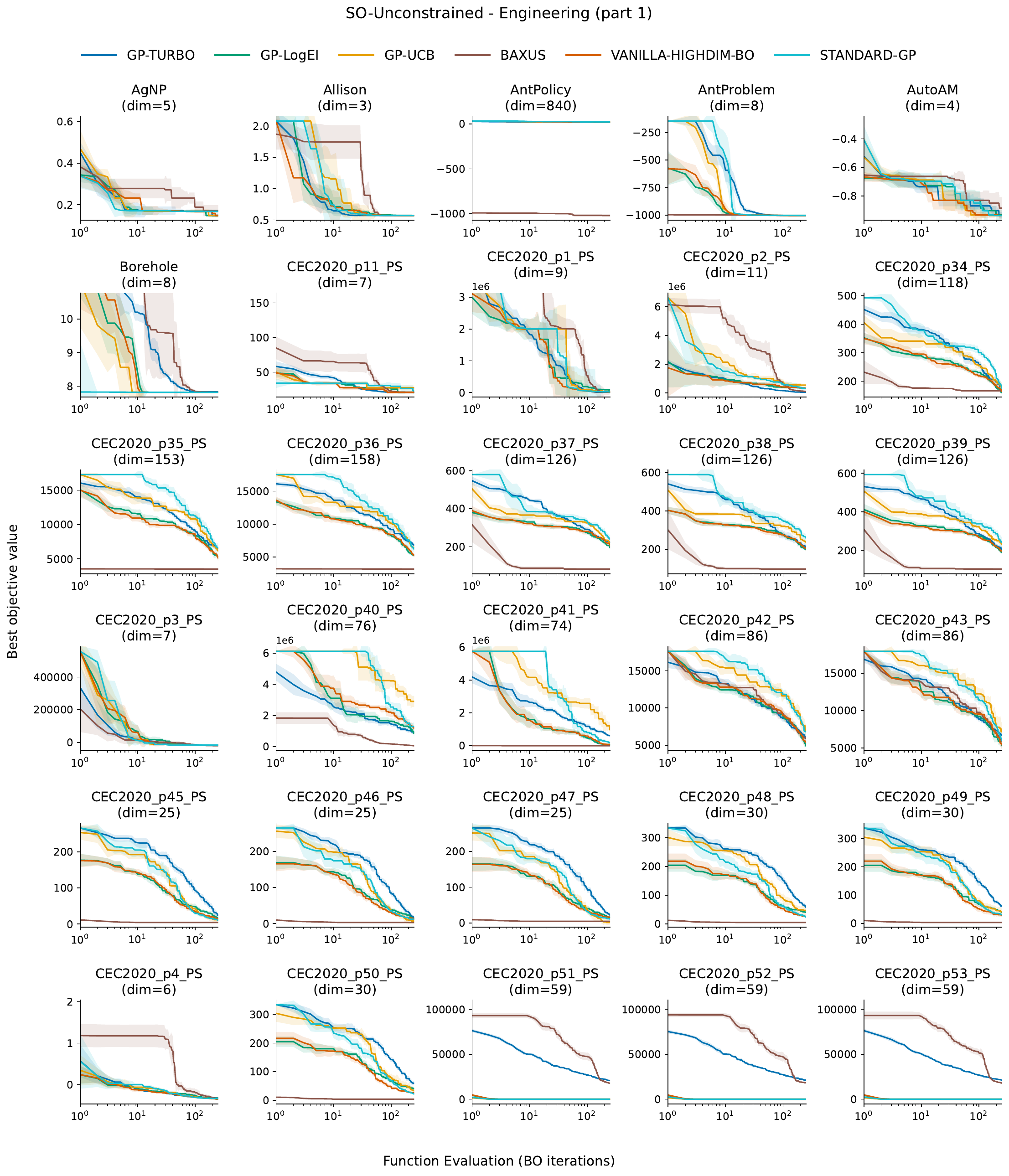}
    \caption{Per-problem optimization convergence plot of SO-Unconstrained Engineering problems (1).}
    \label{fig:SOUN_Eng_01}
\end{figure*}

Figures~\ref{fig:SOUN_Eng_01}$\sim$\ref{fig:SOMixed_syn} show the
per-problem convergence of the campaign, one panel per problem, grouped by
optimization class and domain. Single-objective panels track the best
objective value found so far in the minimization convention, so lower is
better; multi-objective panels track the best dominated hypervolume of the
evaluated points under the problem's fixed reference point, so higher is
better. The constrained classes plot feasible values only. Solid lines are
the median over the $25$ seeds, and shaded bands are $95\%$ confidence
intervals.

\begin{figure*}[htpb!]
    \centering
    \includegraphics[width=.9\linewidth]{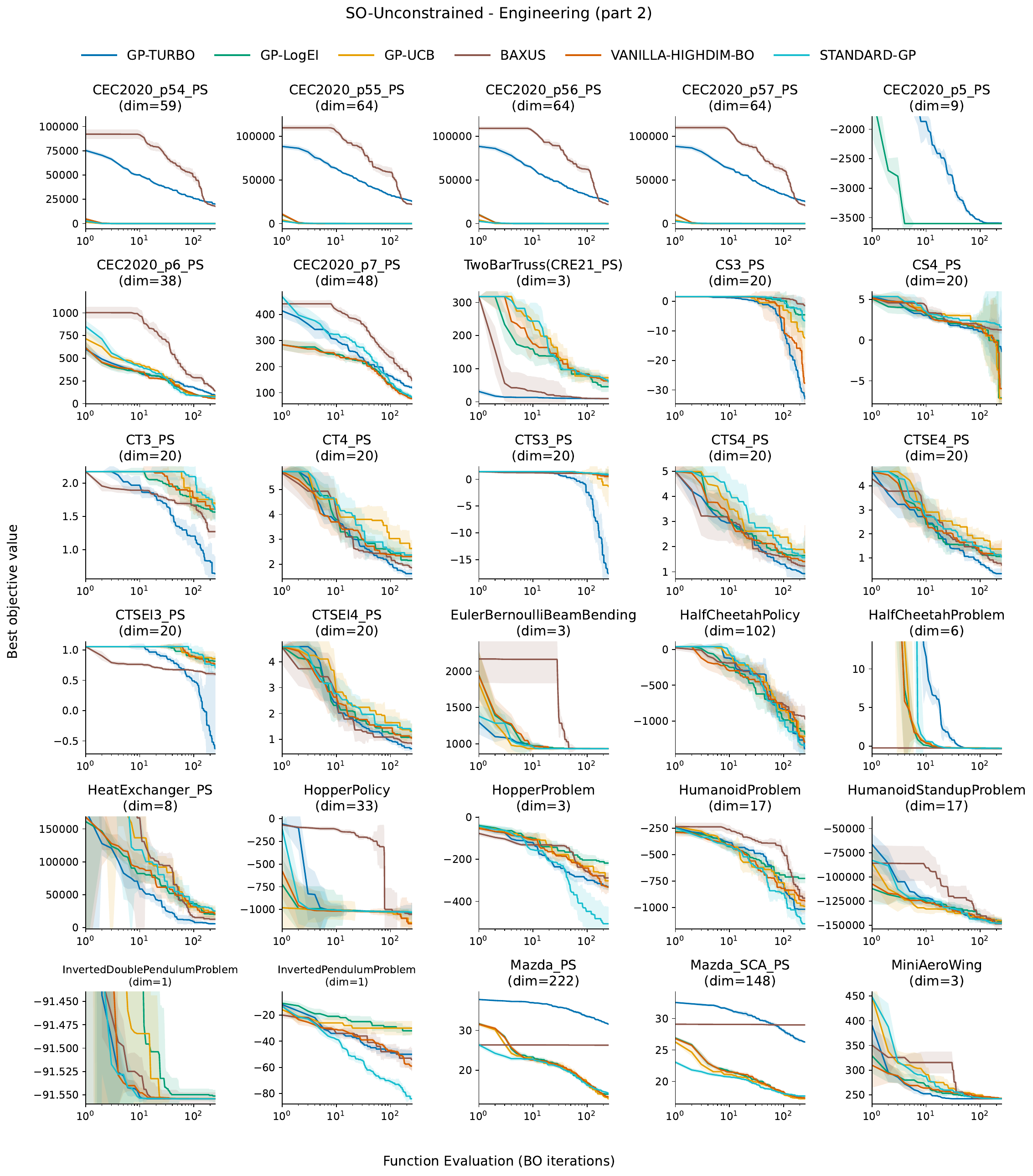}
    \caption{Per-problem optimization convergence plot of SO-Unconstrained Engineering problems (2).}
    \label{fig:SOUN_Eng_02}
\end{figure*}

\begin{figure*}[htpb!]
    \centering
    \includegraphics[width=.9\linewidth]{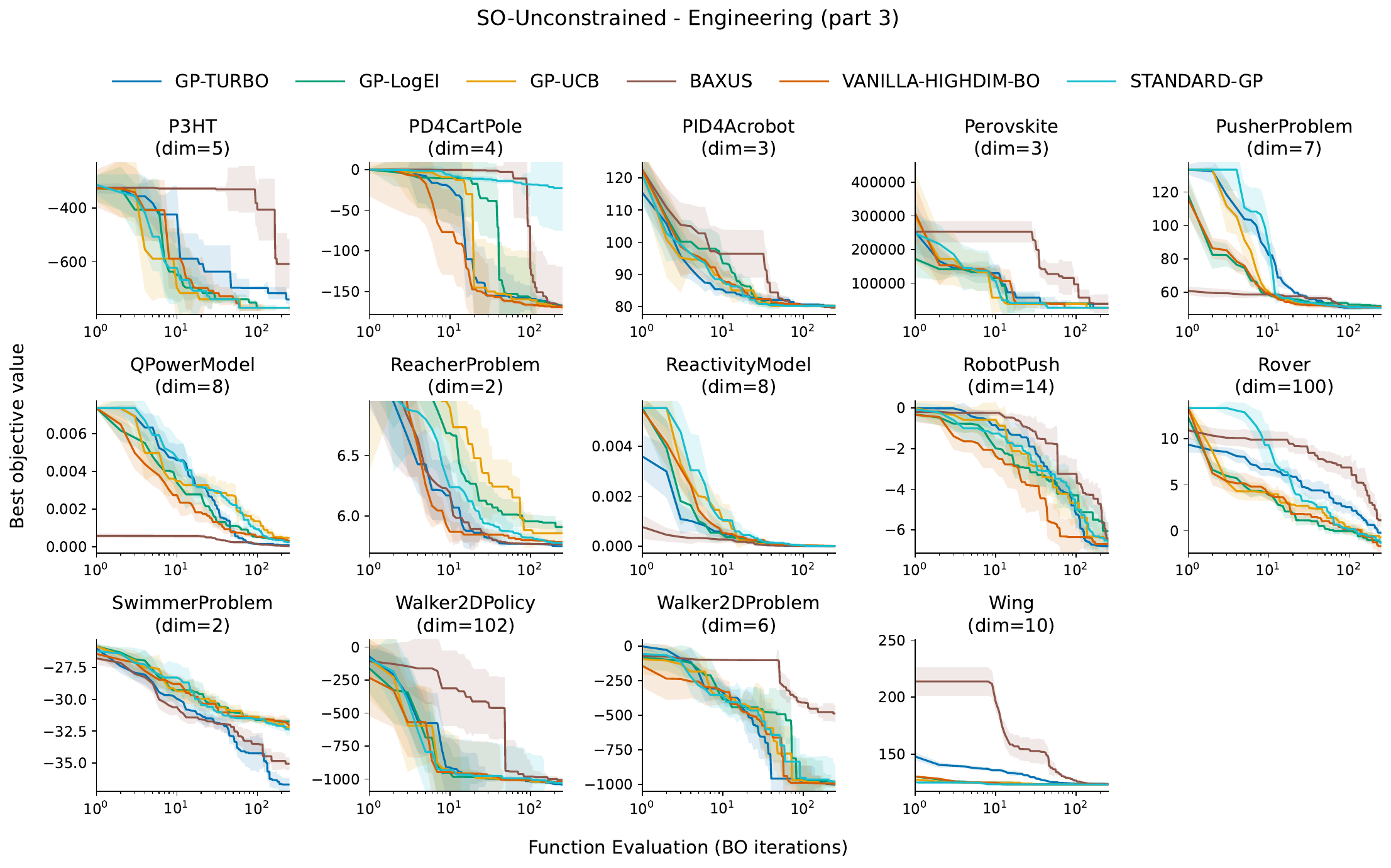}
    \caption{Per-problem optimization convergence plot of SO-Unconstrained Engineering problems (3).}
    \label{fig:SOUN_Eng_03}
\end{figure*}

\begin{figure*}[htpb!]
    \centering
    \includegraphics[width=.9\linewidth]{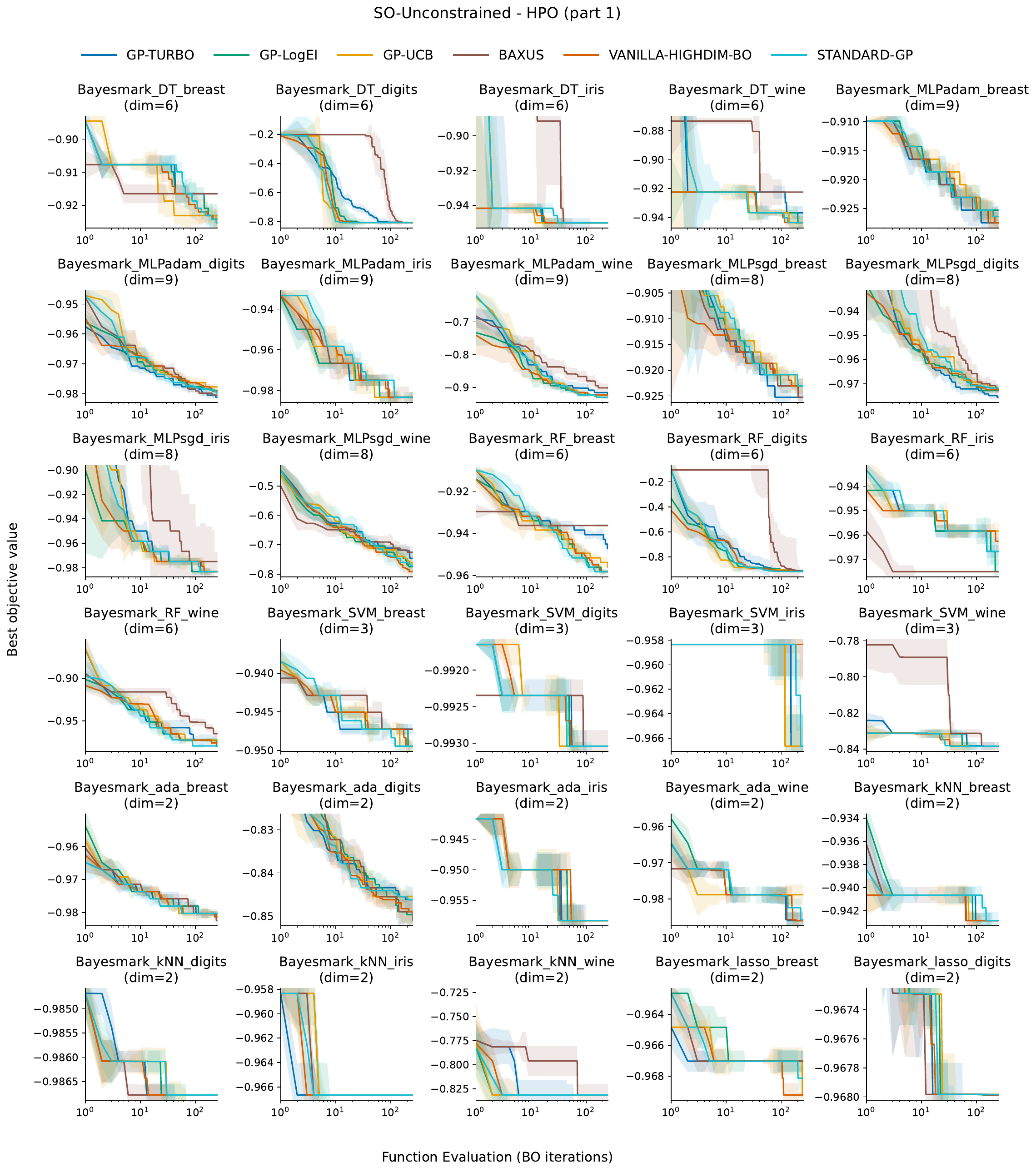}
    \caption{Per-problem optimization convergence plot of SO-Unconstrained HPO problems.}
    \label{fig:SOUN_HPO_1}
\end{figure*}

\begin{figure*}[htpb!]
    \centering
    \includegraphics[width=.9\linewidth]{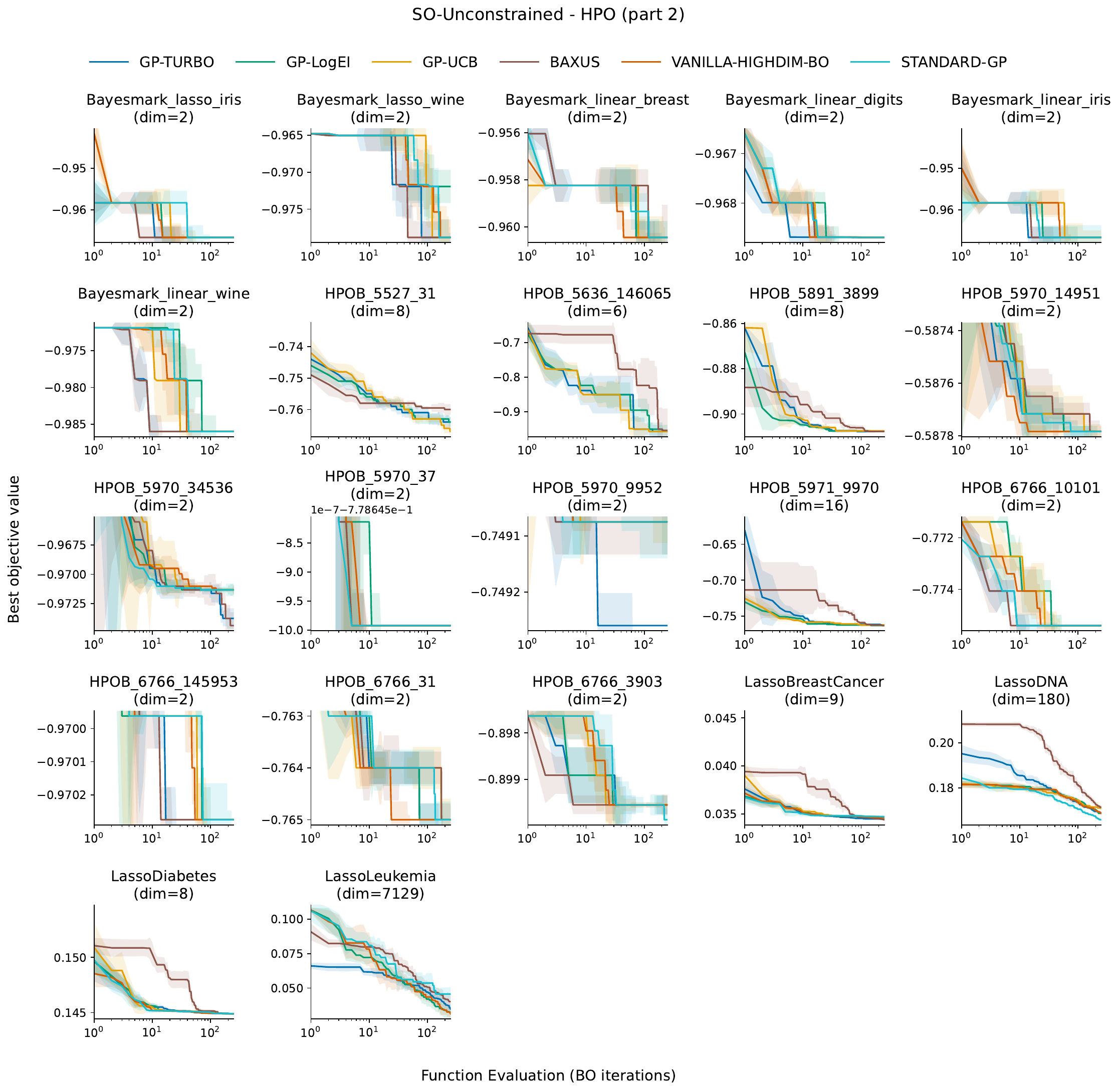}
    \caption{Per-problem optimization convergence plot of SO-Unconstrained HPO problems.}
    \label{fig:SOUN_HPO_2}
\end{figure*}

\begin{figure*}[htpb!]
    \centering
    \includegraphics[width=.9\linewidth]{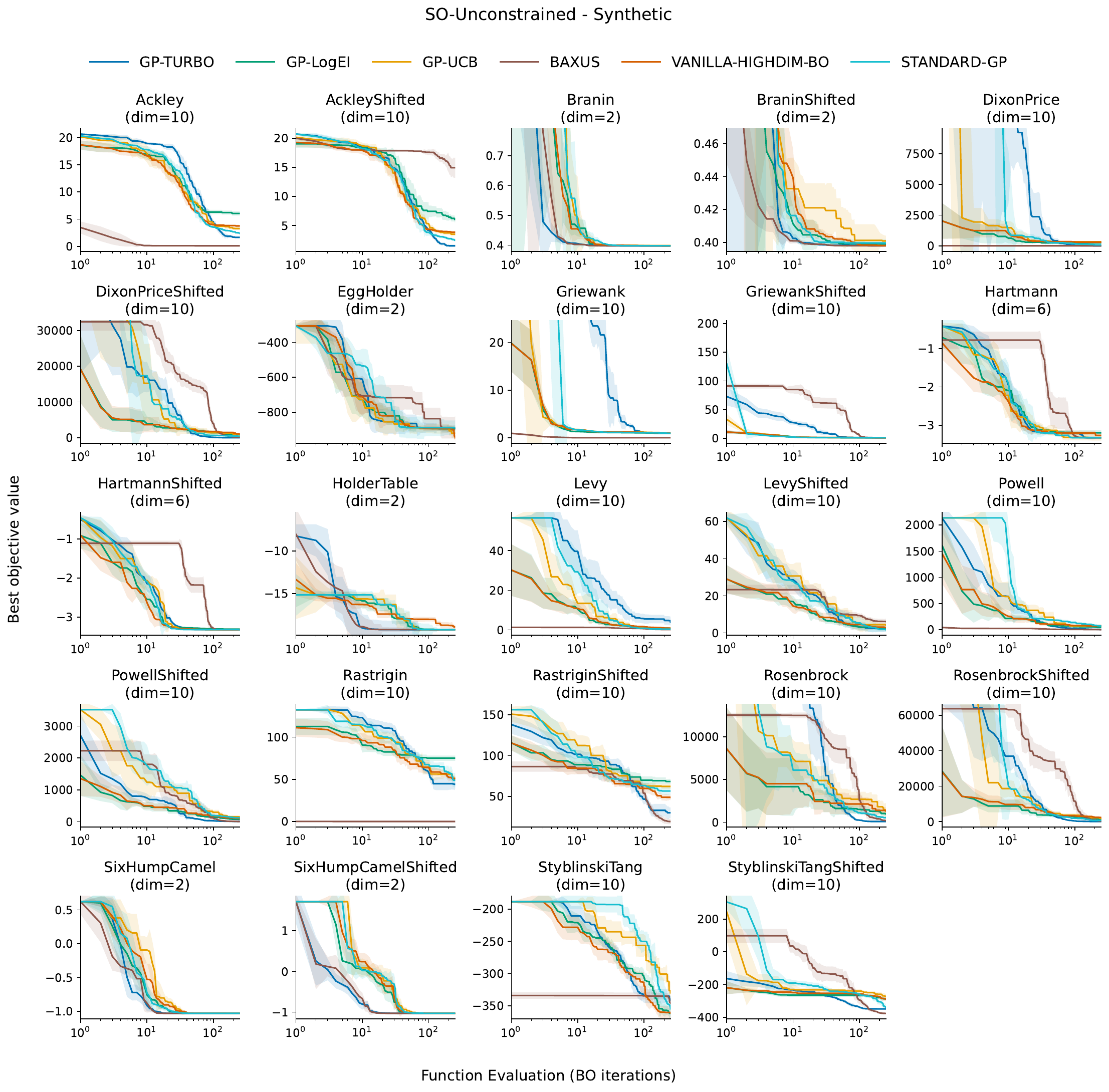}
    \caption{Per-problem optimization convergence plot of SO-Unconstrained Synthetics problems.}
    \label{fig:SOUN_SYN}
\end{figure*}

\begin{figure*}[htpb!]
    \centering
    \includegraphics[width=.9\linewidth]{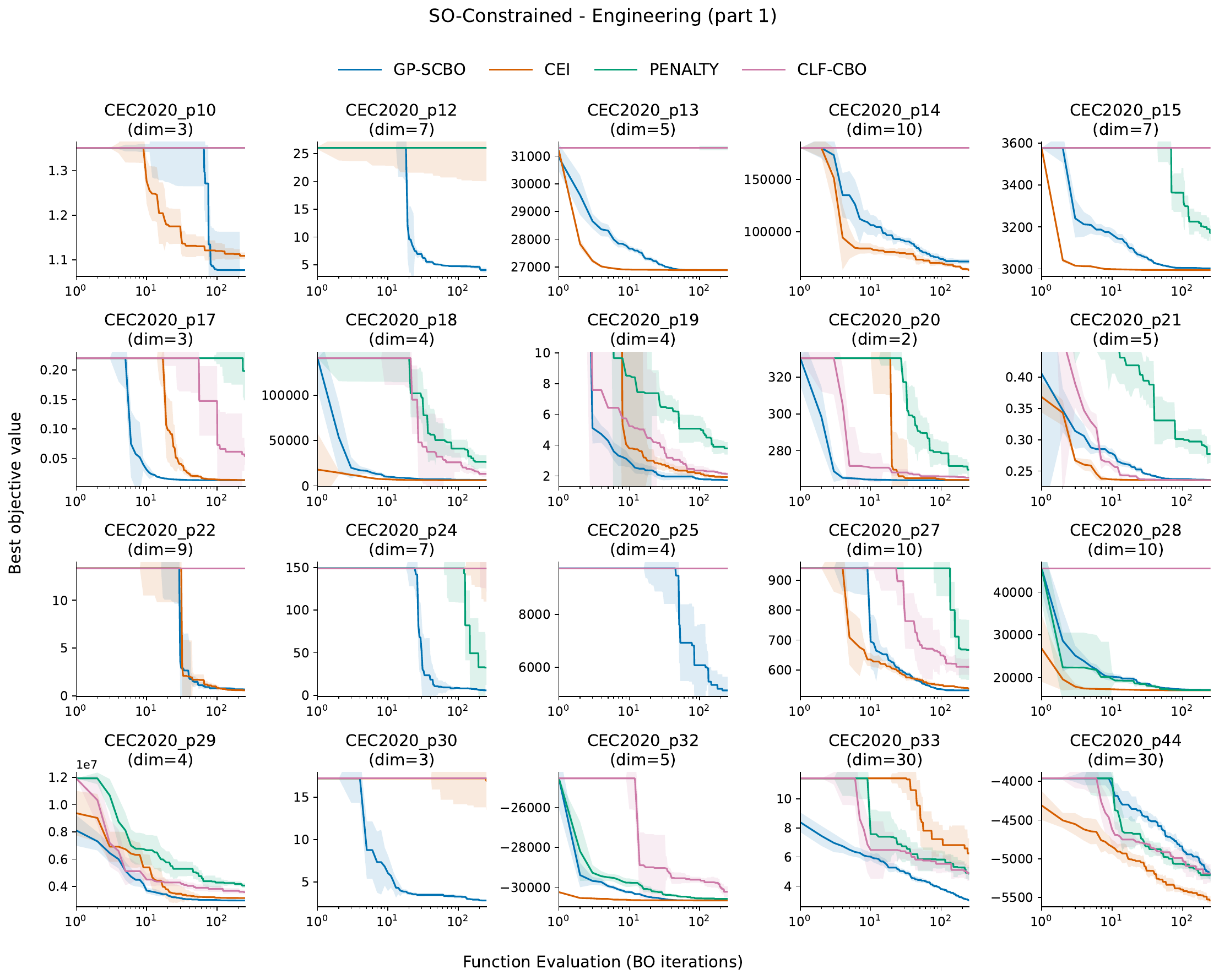}
    \caption{Per-problem optimization convergence plot of SO-Constrained Engineering problems.}
    \label{fig:SOConstrained_eng_01}
\end{figure*}

\begin{figure*}[htpb!]
    \centering
    \includegraphics[width=.9\linewidth]{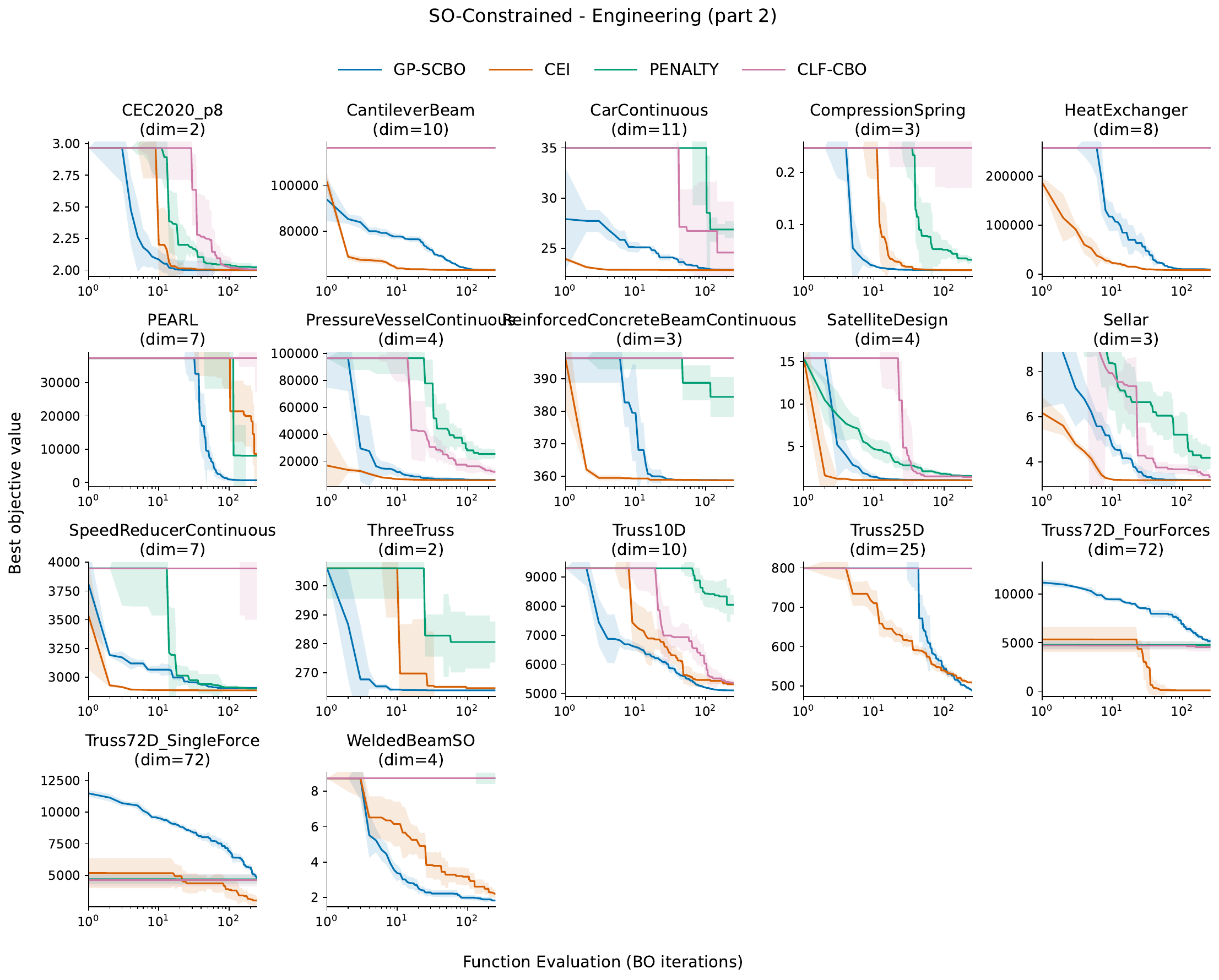}
    \caption{Per-problem optimization convergence plot of SO-Constrained Engineering problems.}
    \label{fig:SOConstrained_eng_02}
\end{figure*}

\begin{figure*}[htpb!]
    \centering
    \includegraphics[width=.9\linewidth]{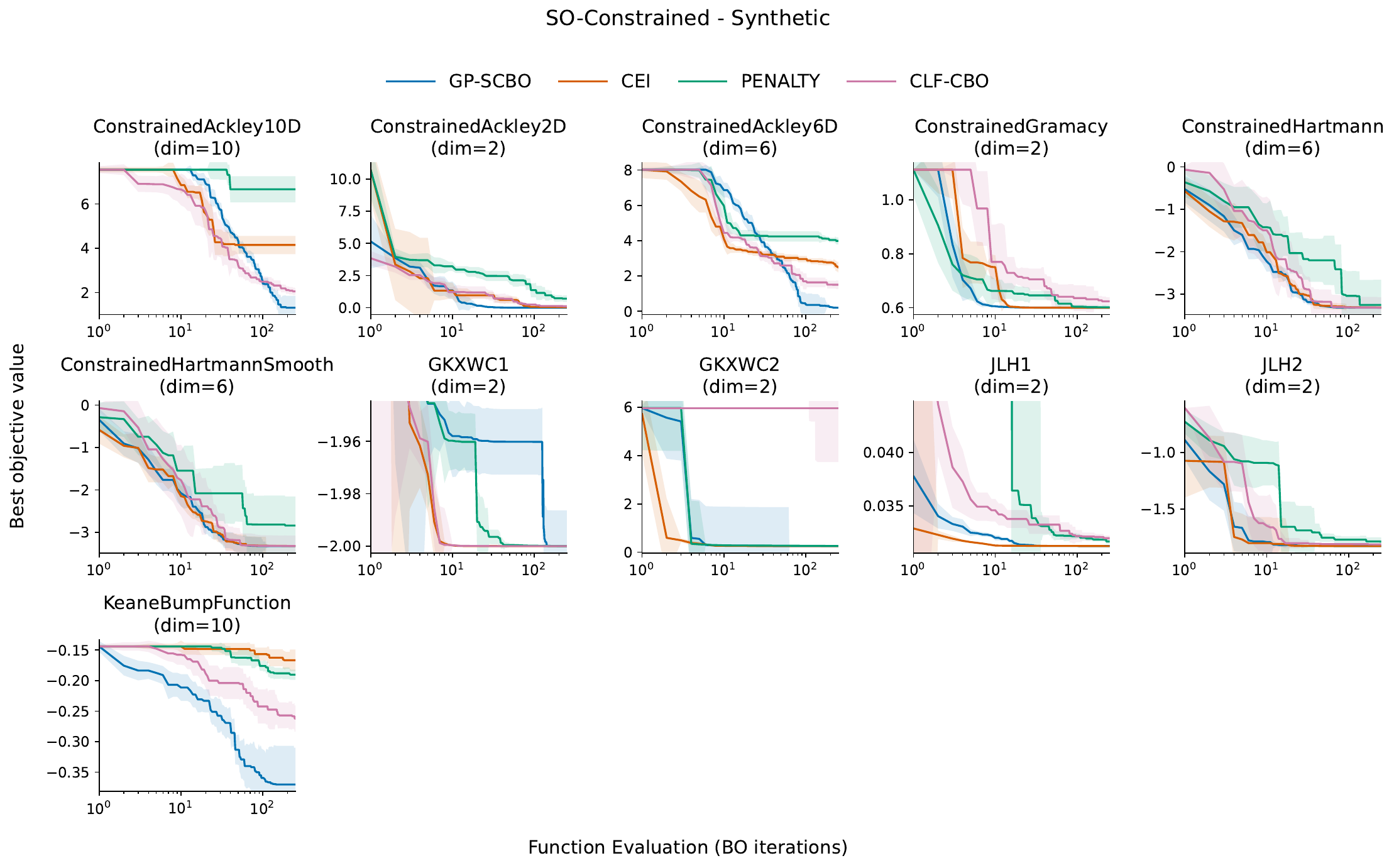}
    \caption{Per-problem optimization convergence plot of SO-Constrained Synthetics problems.}
    \label{fig:SOConstrained_SYN}
\end{figure*}

\begin{figure*}[htpb!]
    \centering
    \includegraphics[width=.9\linewidth]{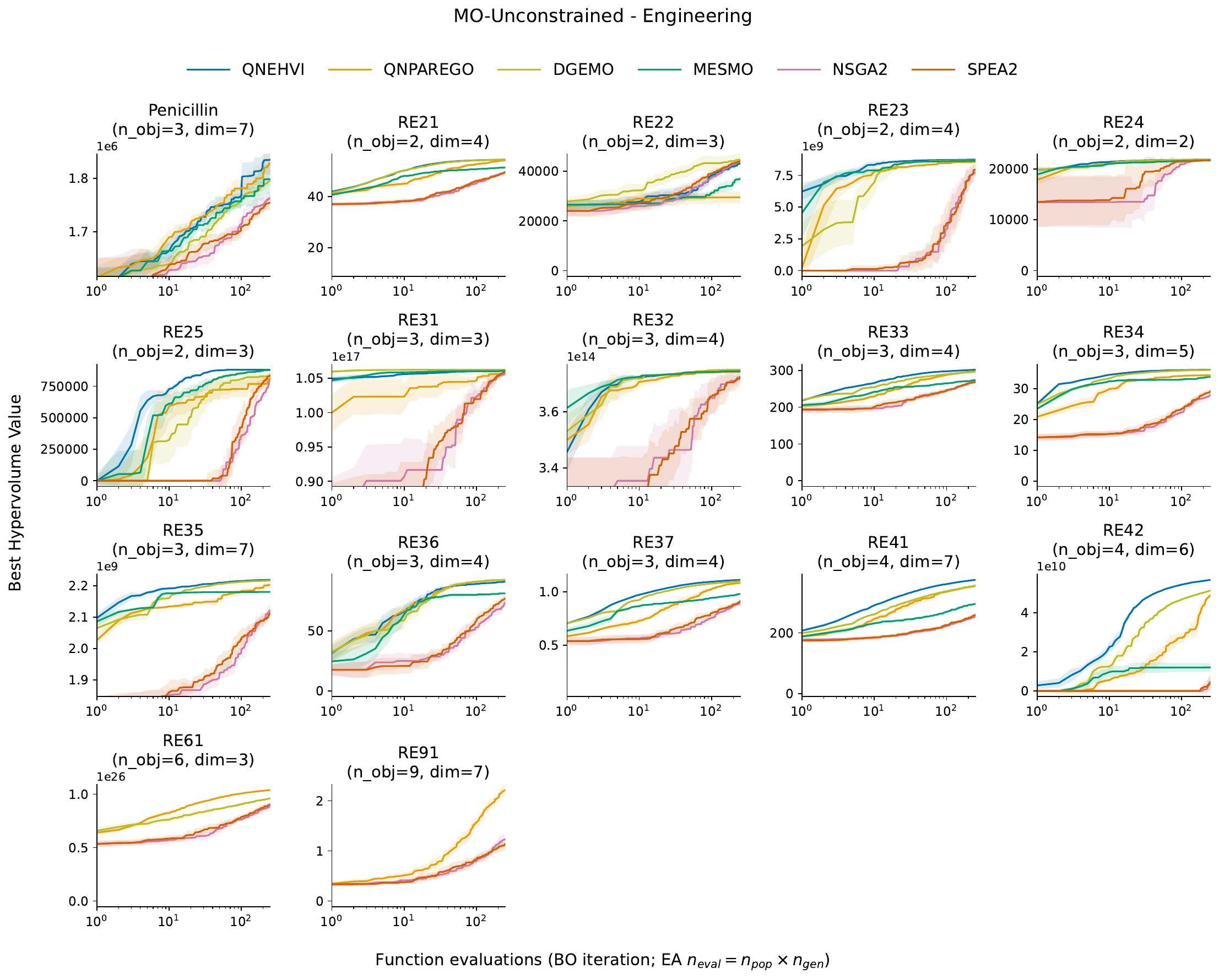}
    \caption{Per-problem optimization convergence plot of MO-Unconstrained Engineering problems.}
    \label{fig:MOUN_eng}
\end{figure*}

\begin{figure*}[htpb!]
    \centering
    \includegraphics[width=.9\linewidth]{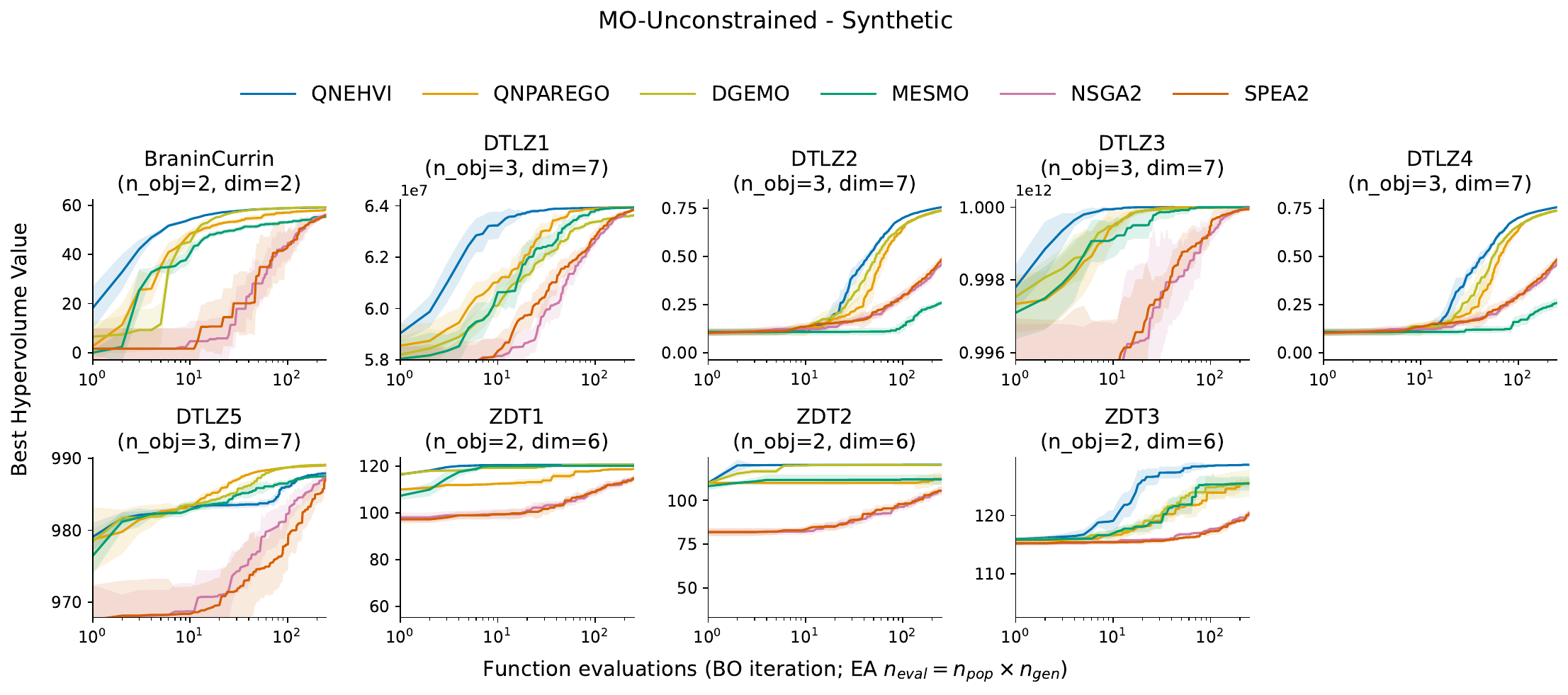}
    \caption{Per-problem optimization convergence plot of MO-Unconstrained Synthetic problems.}
    \label{fig:MOUN_syn}
\end{figure*}

\begin{figure*}[htpb!]
    \centering
    \includegraphics[width=.9\linewidth]{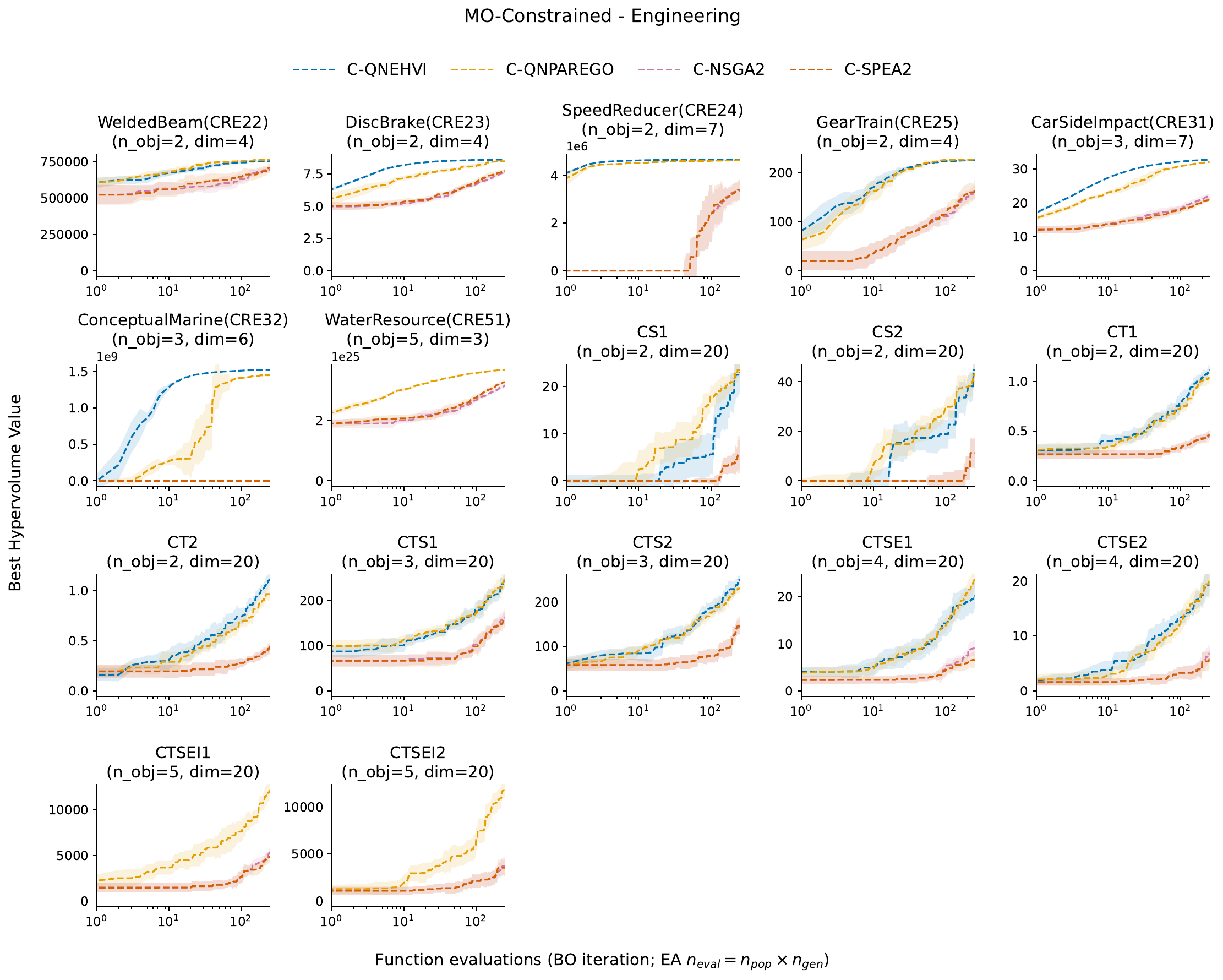}
    \caption{Per-problem optimization convergence plot of MO-Constrained Engineering problems.}
    \label{fig:MOCON_Eng}
\end{figure*}

\begin{figure*}[htpb!]
    \centering
    \includegraphics[width=.9\linewidth]{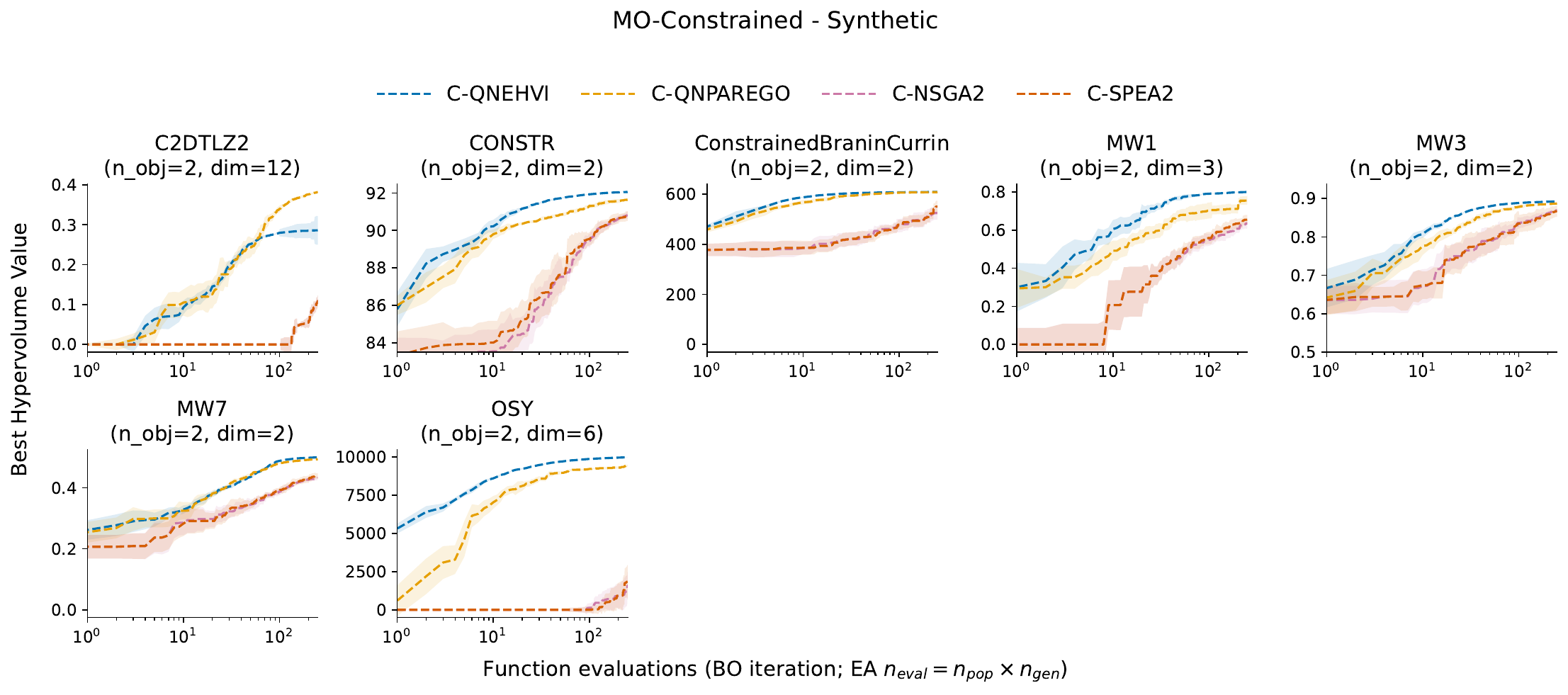}
    \caption{Per-problem optimization convergence plot of MO-Constrained Synthetic problems.}
    \label{fig:MOCON_synthetics}
\end{figure*}

\begin{figure*}[htpb!]
    \centering
    \includegraphics[width=.9\linewidth]{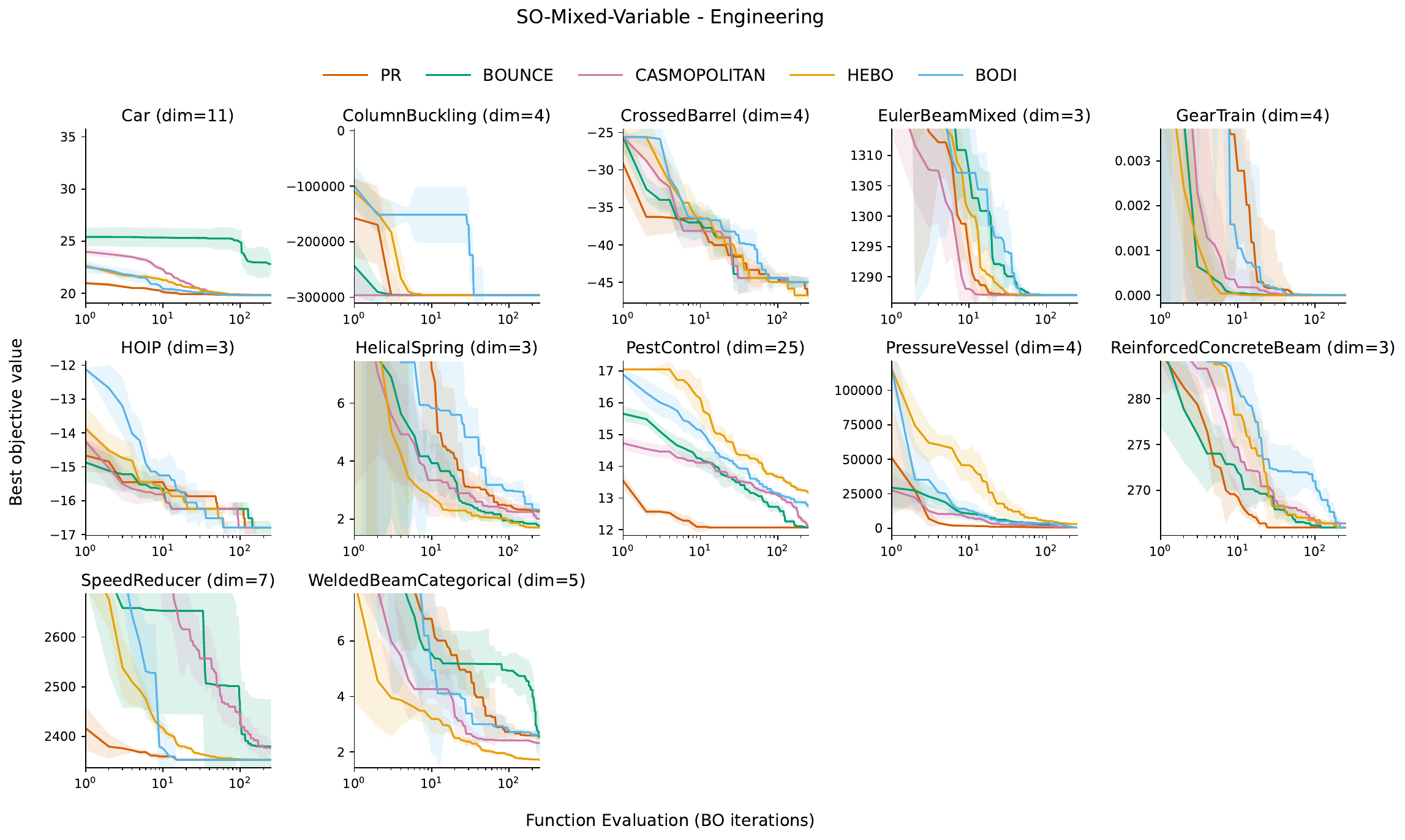}
    \caption{Per-problem optimization convergence plot of SO-Mixed-Variable Engineering problems.}
    \label{fig:SOMixed_eng}
\end{figure*}

\begin{figure*}[htpb!]
    \centering
    \includegraphics[width=.9\linewidth]{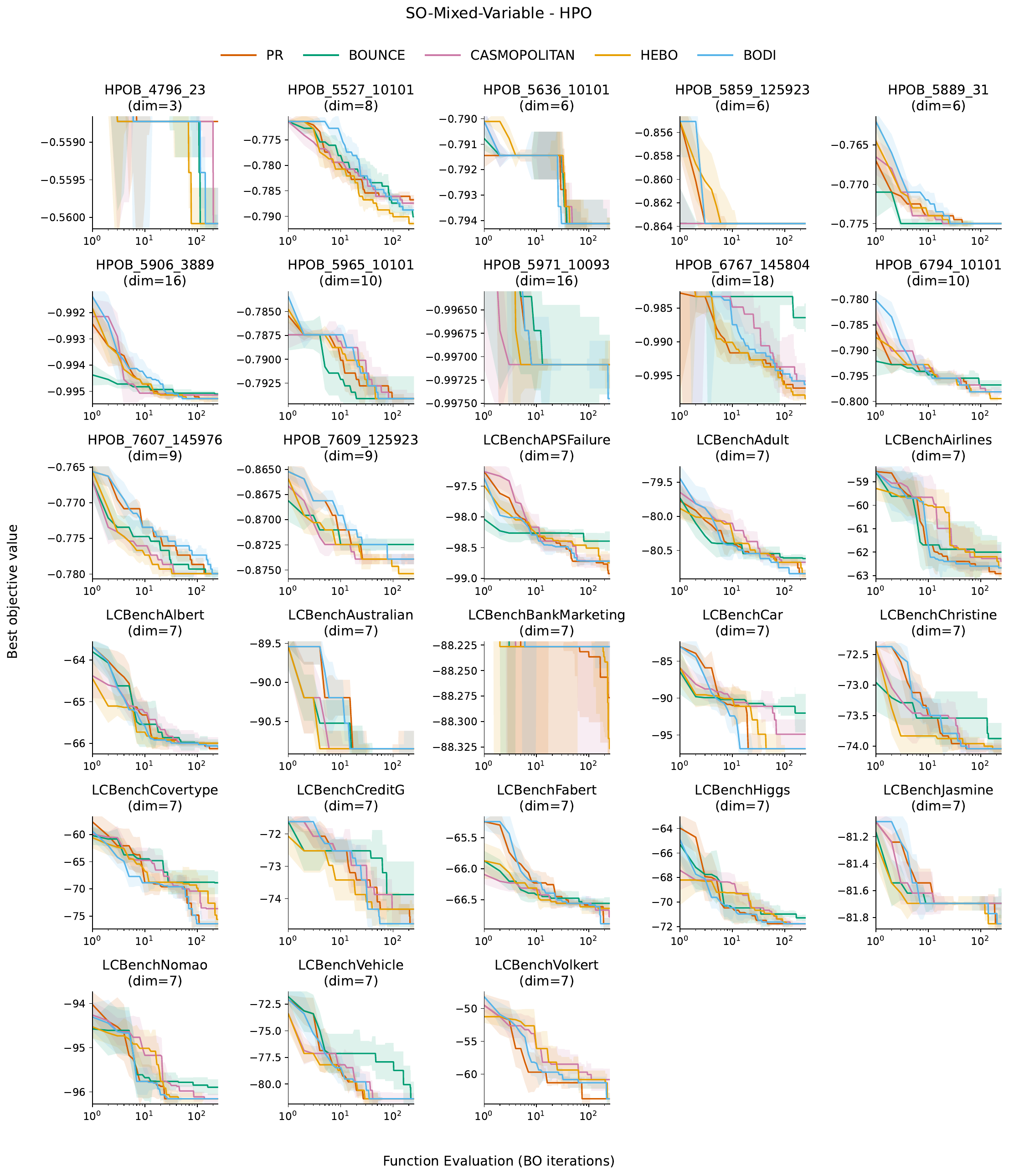}
    \caption{Per-problem optimization convergence plot of SO-Mixed-Variable HPO problems.}
    \label{fig:SOMixed_hpo}
\end{figure*}

\begin{figure*}[htpb!]
    \centering
    \includegraphics[width=.9\linewidth]{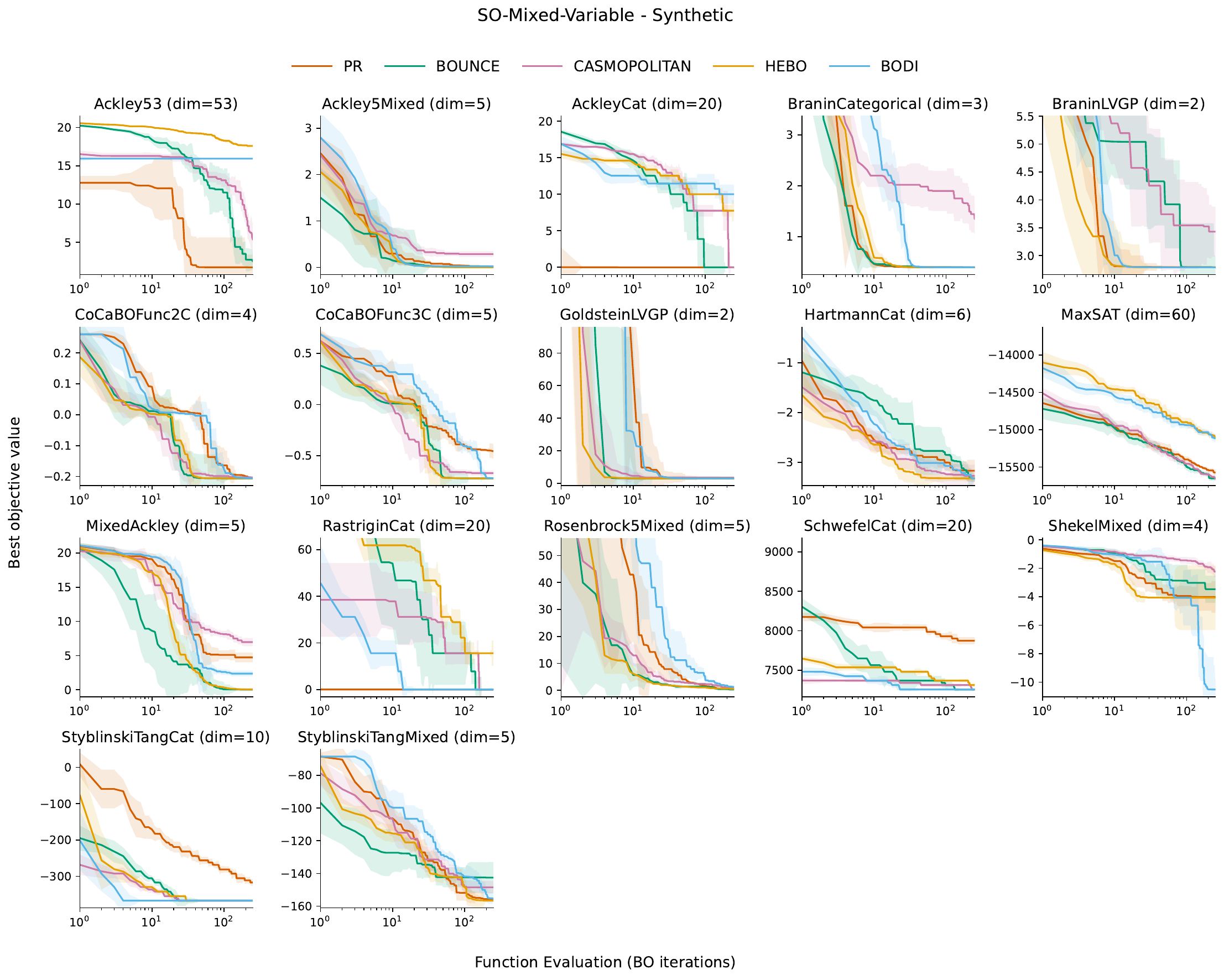}
    \caption{Per-problem optimization convergence plot of SO-Mixed-Variable Synthetic problems.}
    \label{fig:SOMixed_syn}
\end{figure*}

\end{document}